\documentclass[structabstract]{aa} 

\usepackage{amsmath}
\usepackage{rotating}
 \usepackage{color}
\usepackage{graphicx} 
\usepackage{natbib}

\usepackage{overpic}
\usepackage{mathptmx}
\usepackage{mathrsfs}

\usepackage{appendix}

\usepackage{lscape} \usepackage{longtable, needspace} \usepackage{ulem}
\def\new#1 {{\bf #1} } \def\cut#1 {\sout{#1}}

\usepackage{subfigure}
\usepackage{txfonts}
\usepackage{multirow}
\usepackage{url}
\usepackage{xcolor}
\definecolor{darkgreen}{RGB}{0 128 128}

\usepackage[labelfont=bf,singlelinecheck=false]{caption}

\usepackage{tikz}
\newcommand\encircle[1]{%
  \tikz[baseline=(X.base)] 
    \node (X) [draw, shape=circle, inner sep=0] {\strut #1};}

\usepackage{lipsum}
\usepackage{arydshln}

\begin{document}

\title{Are infrared dark clouds  really quiescent?} \author{S.   Feng\inst{1, 2}, 
H.  Beuther\inst{2},   Q.  Zhang\inst{3}, Th.  Henning\inst{2},  H. Linz \inst{2}, S. Ragan  \inst{4}, R. Smith  \inst{5}}
\institute{1.  Max-Planck-Institut f\"ur Extraterrestrische Physik, 
Gie{\ss}enbachstra{\ss}e 1,  D-85748,  Garching bei M\"unchen, Germany\\
2. Max-Planck-Institut f\"ur Astronomie, K\"onigstuhl 17,  D-69117,  Heidelberg, Germany \\
3. Harvard-Smithsonian Center for Astrophysics, 60 Garden Street, Cambridge MA 02138, USA\\
4. School of Physics and Astronomy, The University of Leeds, Leeds, LS2 9JT, UK\\
5. Jodrell Bank Centre for Astrophysics, School of Physics and Astronomy, University of Manchester, Oxford Road, Manchester, M13 9PL, UK \\
}

\offprints{syfeng@mpe.mpg.de} 
\date{\today}

 \abstract {The dense, cold regions where high-mass stars form are poorly characterized, yet they represent an ideal opportunity to learn more about the initial conditions of high-mass star formation (HMSF) since high-mass starless cores (HMSCs) lack the violent feedback seen at later evolutionary stages.
 }
{We investigate the initial conditions of HMSF by studying the dynamics and chemistry of HMSCs.}
{ We present continuum maps obtained from the Submillimeter Array (SMA) interferometry at 1.1 mm for four infrared dark clouds (IRDCs, G28.34S, IRDC\,18530, IRDC\,18306, and IRDC\,18308). For these clouds, we also present line surveys at 1 mm/3 mm obtained from IRAM 30\,m single-dish  observations. 
} 
{ (1) At an angular resolution of 2\arcsec~($\sim 10^4$ AU at an average distance of 4 kpc), the 1.1\,mm SMA observations resolve each source into several fragments. The mass of each fragment is on average $\rm >10\,M_{\odot}$, which exceeds the predicted thermal Jeans mass of the entire clump by a factor of up to 30, indicating that thermal pressure  does not dominate the fragmentation process. Our measured velocity dispersions in the lines obtained by 30\,m imply that non-thermal motion provides the extra support against gravity in the fragments.
(2)  Both non-detection of high-J transitions and the hyperfine multiplet fit of $\rm N_2H^+~(J=1\rightarrow0)$, $\rm C_2H~(N=1\rightarrow0)$, $\rm HCN~(J=1\rightarrow0)$, and $\rm H^{13}CN~(J=1\rightarrow0)$ indicate that our sources are cold and young. However, the obvious detection of SiO and the asymmetric line profile of $\rm HCO^{+}~(J=1\rightarrow0)$ in  G28.34\,S indicate a potential protostellar object and probable  infall motion.
(3)  With a large number of N-bearing species, the existence of carbon rings and molecular ions, and the anti-correlated spatial distributions between $\rm N_2H^+$/$\rm NH_2D$ and CO, our  large-scale high-mass clumps exhibit similar chemical features to small-scale low-mass prestellar objects.  
}
{This study of a small sample of IRDCs  illustrates that thermal Jeans instability alone cannot explain the fragmentation of the clump into cold ($\rm T\sim15$ K), dense ($\rm >10^5~cm^{-3}$) cores and that these IRDCs are not completely quiescent.}

 \titlerunning{Are the Infrared Dark Clouds  Really Quiescent?} \authorrunning{Feng et al.} \maketitle

\section{Introduction}

{ 
The high temperatures ($\rm >50\,K$) and densities  ($\rm >10^5\,cm^{-3}$) in high-mass protostellar objects (HMPOs; \citealp{beuther02}) give rise to a rich spectrum of molecular lines and so HMPOs are fruitful objects for study. Their formation mechanism, however, is still under debate.
Before the protostellar objects exist, 
the so-called high-mass starless (prestellar) cores (HMSCs, \citealt{beuther09}), form in the densest clumps within the infrared-dark clouds (IRDCs). 
Following the literatures, we define the following structure hierarchy: IRDCs are large-scale filamentary structures (several tens of \,pc; \citealt{jackson10,beuther11,wangk14}); clumps are dense gas on the order of   $\rm \sim1$\,pc \citep{zhang09}; HMSCs usually have high density ($\rm n\ge10^3\text{--}10^5~cm^{-3}$, \citealt{teyssier02,rathborne06,butler09,vasyunin09,ragan09})  and small sizes (on a scale  of $\rm \sim 0.1\,pc$). On the smallest scale ($\rm \sim 0.01$ pc), HMSCs harbor  some internal structures known as   condensations.
 The prestellar phase of high-mass sources is not observationally well constrained  owing to its short duration,  to the sources large average distance (several kpc), and to the characteristically low temperatures ($\rm T<20\,K$, \citealt{carey98,sridharan05,pillai06,wangy08,wienen12,chira13}), which   lead to little excitation. In fact, even the existence of HMSCs  has been questioned (e.g., \citealt{motte07}) until large survey samples were able to constrain the collapse timescale (on the order of $\rm 5\times10^4$\,yrs, e.g., \citealt{russeil10,tackenberg12}).
 }
 \\

The Herschel guaranteed time key programme ``The Earliest Phases of Star Formation (EPoS)'' surveyed a sample of 45 IRDCs \citep{ragan12b}. This survey revealed several instances of $\rm 70\,\mu$m dark regions corresponding to peaks in submillimeter  dust emission from the ATLASGAL\footnote{``The APEX Telescope Large Area Survey of the Galaxy" (ATLASGAL) is a large area survey
of the galaxy at $\rm 870\,\mu$m \citep{schuller09}.}. These sources { are ideal locations to seek for HMSCs}  (e.g., \citealt{beuther10b,henning10,linz10,beuther12b,ragan12b}) and are prime targets with which to address several outstanding questions about the starless phase:  What is the structure of HMSCs? What governs their fragmentation (e.g., \citealp{zhang09,bontemps10,zhang11,beuther13b,zhang15})?  Is thermal pressure sufficient to support HMSCs against collapse (e.g., \citealt{wangk11,wangk14})?  What are the chemical properties of HMSCs, and how do they compare to more advanced phases  (e.g.,  \citealt{miettinen11,vasyunin11})?
\\

To investigate these questions, we select a sample of four ``quiescent"
\footnote{ IRDCs are classified as ``active'' if they host protostellar dominated cores and ``quiescent'' if they host prestellar dominated cores, depending on whether both 4.5 and 24\,$\mu$m emission is detected \citep{chambers09,rathborne10}. } { IRDC candidates} from the EPoS sample: G28.34\,S, IRDC\,18530, IRDC\,18306, and IRDC\,18308. The kinematic distances ({\it D}) range from 3.6 to 4.8\,kpc. We describe our Submillimeter Array (SMA) and IRAM 30-meter telescope observations in Section  \ref{irdc:obs} and present the maps in Section \ref{irdc:result}. Our analyses of the kinematics and chemistry are presented in Section \ref{irdc:analysis}, and we summarize in Section  \ref{irdc:conclusion}.\\

\section{Observations}\label{irdc:obs}
\subsection{Submillimeter Array }
We carried out four-track observations on our sample with the SMA at 260/270\,GHz (1.1\,mm)  using the extended (EXT) and compact (COMP) configurations (summarized in Table~\ref{irdc:SMAconfI}). From May to July 2013, we observed two sources per track, which share  bandpass and flux calibrators.
For all the observations, phase and amplitude calibrations were performed via frequent
switch (every 20 min) on quasars 1743-038 and 1751+096. 
The primary beam size is 48\arcsec.  The phase center for each target  is pinned down to the $\rm 870\,\mu$m ATLASGAL continuum peak \citep{schuller09}, which is listed in Table~\ref{irdc:SMAconfII}.
The baselines (BL) range from 16\,m to 226\,m with six or seven antennas ($\rm N_{ant}$) { on different days}, making structure with extent {$\ge 10^{''}$}  being filtered out. 
Bandpass calibrations were done with BL\,Lac  (EXT), 3C\,279 (EXT, COMP), and 3C\,84 (COMP).
Flux calibrations were estimated  using Neptune (EXT) and Uranus (COMP).
The zenith opacities,  measured with water vapor monitors mounted on the Caltech Submillimeter Observatory (CSO) or the James Clerk Maxwell Telescope (JCMT),  were satisfactory
during all tracks with $\tau (\rm 225~ GHz)\sim0.01\text{--}0.3$.
Further technical descriptions of the SMA and its calibration schemes can be found in \citet{ho04}.\\

The double sideband receiver was employed, and we tuned the center of   spectral band 22 in the
lower sideband  as the rest frequency of the $\rm H^{13}CO^+$
$(3\rightarrow2)$ line, 260.255\,GHz. { Separated by 10\,GHz, the frequency coverages are 258.100--261.981\,GHz (LSB) and 269.998--273.876\,GHz (USB). }
Each sideband has a total bandwidth of 4\,GHz
and a native frequency resolution of 0.812\,MHz (a velocity resolution of 0.936$\rm
~km\,s^{-1}$).  
The frequency resolutions in the spectral windows of 260.221--260.299\,GHz and 271.679--271.757\,GHz 
are 0.406\,MHz (corresponding to
a velocity resolution of 0.468$\rm ~km\,s^{-1}$); frequency resolution in the
spectral windows of  260.041--260.221\,GHz and 271.757--271.937\,GHz
 is 1.625\,MHz (velocity resolution of 1.872$\rm ~km\,s^{-1}$).\\

The  flagging and calibration was done with the 
MIR package\footnote{The MIR package  was originally developed for the Owens Valley Radio Observatory, and is now adapted for the SMA,
http://cfa-www.harvard.edu/$\sim$cqi/mircook.html. }.  The imaging and
data analysis was conducted with MIRIAD package
\citep{sault95}. 
The synthesized beams and 1$\sigma$ rms  of the continuum image  from dual-sidebands  at each configuration are listed in Table~\ref{irdc:SMAconfII}, along with the results obtained by combining configurations with natural weighting.  We also list the  1$\sigma$ rms  of the { primary beam averaged spectrum of each source.} \\

\begin{table*}
\small
\caption{SMA observation toward four IRDCs: I. configurations \label{irdc:SMAconfI}}
\centering
 \scalebox{1}{
\begin{tabular}{cc|p{2.5cm}cc|cc|cc}
\hline
\hline
Configuration  &Date  &Source  &$\rm N_{ant}$  &BL &\multicolumn{2}{c|}{Calibrator}  &$\rm T_{sys}$  &$\rm \tau_{225\,GHz}$ \\
                       &(yyyymmdd) & &                    &(m)   &Bandpass     &Flux      &(K)  &   \\
\hline
EXT     &20130530 & IRDC\,18306 &7  &44-226    &BL\,Lac   &Neptune   &200-400        &0.05-0.2\\
            &                 & IRDC\,18308  &   &               & 3C\,279                     &                    &                  &\\                     
\hline
EXT      &20130611  & G28.34\,S    &6   &68-226  &BL\,Lac   &Neptune   &200-600        &0.2-0.3\\
&&IRDC\,18530  &   &                                   &  3C\,279                   & Titan        &         &\\ 
\hline
COMP    &20130717 & IRDC\,18306  &6   &16-69  &3C\,279  &Uranus   &200-500  &0.1-0.25\\
&&IRDC\,18308  &   &                                     & 3C\,84                   &     &             &\\ 
\hline
COMP   &20130719  & G28.34\,S  &6   &16-69  &3C\,279  &Uranus   &100-300  &0.01-0.2\\
&& IRDC\,18530  &   &                                    & 3C\,84                   &       &           &\\ 
\hline

\end{tabular}
}
\end{table*}

\begin{table*}

\centering
\caption{SMA observations toward four IRDCs: II. phase center of each source and the synthetic image quality at different configurations \label{irdc:SMAconfII}}
 \scalebox{0.85}{
\begin{tabular}{p{2cm}c|c|c|c|c}
\hline\hline
Source  &               &G28.34\,S  &IRDC\,18530   &IRDC\,18306   &IRDC\,18308\\
\hline
R.A. &[J2000]         &$\rm 18^h42^m47^s$    &$\rm 18^h55^m30^s$      &$\rm 18^h33^m32^s$       &$\rm 18^h33^m35^s$      \\
Dec. &[J2000]         &$\rm -04^{\circ}04^{'}07^{''}$  &$\rm 02^{\circ}17^{'}06^{''}$    &$\rm -08^{\circ}32^{'}27^{''}$    &$\rm -08^{\circ}35^{'}53^{''}$ \\   
\it D  &($\rm kpc$)    &4.8  &4.6  &3.6  &4.4  \\
$\rm V_{lsr}$  &($\rm km\,s^{-1}$)   &78.4   &75.9   &54.8    &73.7   \\
\hline
\multirow{2}{*}{COMP$^a$}  &($\rm Maj\times Min$, P.A)   &$\rm 3.47^{''} \times 1.79^{''},~32.8^\circ$   &$\rm 3.29^{''} \times 1.77^{''},~34.8^\circ$ &$\rm 3.28^{''} \times 2.17^{''},~-7.7^\circ$    &$\rm 3.26^{''} \times 2.14^{''},~9.6^\circ$ \\
                                    &${\rm \sigma_{conti}} ^{\it d}$ (mJy/beam)    &1.00     &0.86    &1.28    &1.40\\
                                     &${\rm \sigma_{line}} ^{\it e}$ (mK)    &78     &83    &146    &162\\
\hline
\multirow{2}{*}{EXT$^b$}    &($\rm Maj\times Min$, P.A)  &$\rm 0.88^{''} \times 0.69^{''},~-82.2^\circ$   &$\rm 0.88^{''} \times 0.67^{''},~-78.9^\circ$  &$\rm 0.93^{''} \times 0.79^{''},~-85.4^\circ$      &$\rm 0.93^{''} \times 0.79^{''},~-88.6^\circ$  \\
                                    &${\rm \sigma_{conti}} ^{\it d}$  (mJy/beam)  &1.26    &1.29     &1.31   &1.29  \\
                                    &${\rm \sigma_{line}} ^{\it e}$ (mK)    &161     &167   &141    &160\\
\hline
\multirow{2}{*}{COMP+EXT$^c$}  &($\rm Maj\times Min$, P.A)  &$\rm 2.55^{''} \times 1.51^{''},~33.5^\circ$   &$\rm 2.44^{''} \times 1.49^{''},~35.1^\circ$   &$\rm 1.79^{''} \times 1.56^{''},~5.9^\circ$     &$\rm 1.67^{''} \times 1.50^{''},~9.6^\circ$\\
                                         &${\rm \sigma_{conti}} ^{\it d}$  (mJy/beam)       &0.97    &0.77  & 0.99    &0.98\\
                                         &${\rm \sigma_{line}} ^{\it e}$ (mK)    &54     &65    &268   &324\\
\hline\hline
\multicolumn{6}{l}{{\bf Note.} {\it a}.  Synthetic beam of compact configuration. }\\
\multicolumn{6}{l}{~~~~~~~~~~{\it b}.  Synthetic beam of extended configuration. }\\
\multicolumn{6}{l}{~~~~~~~~~~{\it c}.  Synthetic beam of compact+extended configuration, combined with 
 the ``natural weighting''. }\\
\multicolumn{6}{l}{~~~~~~~~~~{\it d}.  1$\sigma$ rms measured from  the continuum map of each source. }\\
\multicolumn{6}{l}{~~~~~~~~~~{\it e}.   1$\sigma$ rms measured from the  { primary beam averaged spectrum  of each source}, which is converted into brightness temperature.  }\\

\end{tabular}
}

\end{table*}

\begin{table*}
\small
\centering
\caption{IRAM 30\,m observations on the four IRDCs\label{irdc:source30m}}
 \scalebox{0.95}{
\begin{tabular}{cp{2.5cm}p{2.5cm}c|cp{2.5cm}p{2.5cm}c}
\hline
\hline
Source &R.A. &Dec.  &abbrev.  &Source &R.A. &Dec.  &abbrev. \\
       &[J2000] &[J2000]     &         &        &[J2000] &[J2000]  &   \\

\hline
G28.34\,S    &$\rm 18^h42^m46^s.597$      &$\rm -04^{\circ}04^{'}11^{''}.940$    &G28.34\,S
&IRDC\,18530    &$\rm 18^h55^m30^s.128$      &$\rm 02^{\circ}17^{'}09^{''}.300$    &18530\\
                                                                                                                          
IRDC\,18306    &$\rm 18^h33^m32^s.044$      &$\rm -08^{\circ}32^{'}28^{''}.620$    &18306
&IRDC\,18308      &$\rm 18^h33^m35^s.090$      &$\rm -08^{\circ}36^{'}00^{''}.120$    &18308\\
\hline\hline
\end{tabular}
}

\vspace{0.3cm}

\scalebox{0.95}{
\begin{tabular}{cccc|ccc}
\hline\hline
Date$^a$  &Source  &Pointing &Focus  &\multicolumn{2}{c}{$\rm T_{sys}\it^b$  }  &$\rm \tau_{225\,GHz}\it ^b$\\  
(yyyymmdd) &   &                 &        &1mm (K)   &3mm (K)    &\\
\hline
20140528   &G28.34\,S, 18530  &Saturn, K3-50A, Venus, 1749+096   &Saturn, Venus         &450-550             &115-130    &0.5-0.6\\
20140529   &18306, 18530  &Saturn, K3-50A,1749+096    &Saturn          &270-310             &90-110      &0.2-0.4\\
20140530   &18306, 18308 &Saturn, 1749+096    &Saturn         &275-285             &90-100   &0.2-0.4\\
$\rm 20140531^*$   &G28.34\,S, 18306, 18308  &W3OH, 1749+096   &W3OH          &210-280            &90-110   &0.1-0.3\\
\hline\hline
\multicolumn{7}{l}{{\bf Note.} {\it a.} ``*'' denotes the date when  observational bands were tuned from 85.8--93.6\,GHz to 85.2--92.9\,GHz, and from 215.1--222.8\,GHz to 217.0--224.8\,GHz.}
\\
\multicolumn{7}{l}{~~~~~~~~~~{\it b.} `$\rm T_{sys}$ and $\rm \tau$ are averaged system temperature and precipitable water vapor  in each observation.}
\end{tabular}
}
\end{table*}

\subsection{Single-dish observations with the IRAM 30~m telescope}
In addition to the high spatial  resolution observations performed with the SMA, we conducted an imaging  line survey of our sample
with the IRAM 30~m telescope at 1\,mm/3\,mm. Observations were performed in the on-the-fly mode from May 28  to May 30, 2014, mapping a $1.5'\times1.5'$ area of each source. A broad bandpass (8\,GHz  bandwidth for each sideband) of EMIR covers the range of 85.819--93.600\,GHz { (E0)} with a velocity resolution of 0.641\,$\rm km\,s^{-1}$, and 215.059--222.841\,GHz  { (E2)} with a velocity resolution of 0.265\,$\rm km\,s^{-1}$.  After achieving the expected sensitivity,  we tuned the band center on May 31 and observed an ``extra band'' of 85.119--92.900\,GHz and 217.059--224.841\,GHz to search for more lines.
The phase center, the  weather conditions, focus, and pointing information\footnote{The SMA continuum peak of each source has some shift from its observational phase center. Therefore, we set the 30\,m mapping center according to the SMA continuum peak and use Gildas software to ``reproject'' the SMA offset  accordingly.}  are  listed in   Table~\ref{irdc:source30m}. 
Using a forward efficiency ($\rm F_{eff}$, 94\% at 1mm and 95\% at 3\,mm) and a main beam efficiency ($\rm B_{eff}$, 63\% at 1mm and 81\% at 3\,mm)\footnote{http://www.iram.es/IRAMES/mainWiki/Iram30\,mEfficiencies},  we converted the data from
antenna temperature ($\rm T_{A}$) to main beam brightness temperature ($\rm T_{mb}=F_{eff}/B_{eff}\times T_{A}$).   The
beam of the 30~m telescope is
$\sim12''$  at 1\,mm and $\sim30''$  at 3\,mm. 
We used the Gildas\footnote{http://www.iram.fr/IRAMFR/GILDAS} software for data reduction and the  first step line identification\footnote{The ``Weeds'' is an extension of Gildas for line identification  \citep{maret11}.} (Table~\ref{irdc:tab:line}). The $\rm1\sigma$ rms $\rm T_{mb}$ in the line free channels are 6-8\,mK at 3\,mm and 26-33\,mK at 1\,mm.

\section{Observational results}\label{irdc:result}

\subsection{Fragmentation on small scale from SMA observations}\label{irdc:continuum}
\begin{figure*}
\centering
\begin{tabular}{p{4cm}p{8cm}p{5cm}p{4cm}}
&APEX+{\it Herschel}  & &SMA+IRAM 30\,m
\end{tabular}
\begin{tabular}{p{0.1cm}p{19.9cm}}
\rotatebox{90}{~~~~~~~~~~~~~~~~~~~~~~~~~~~~~~~~~~~~~~~~~~~~~~~~~~~~~~~~~~~~~~~~~~~~~~~~~~~~~~~~~~~~~~~~~~~~~~~~~~~~~~\small Dec [J2000]}
&\scalebox{0.85}{
\includegraphics[width=6.66cm]{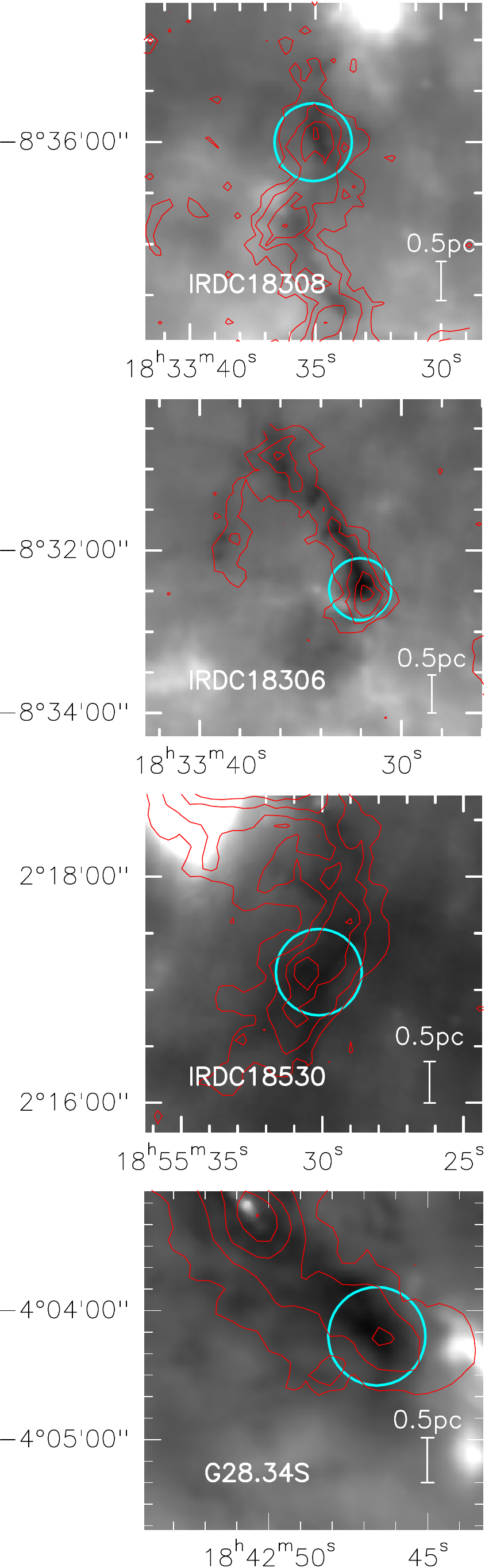}
\includegraphics[width=13.0cm]{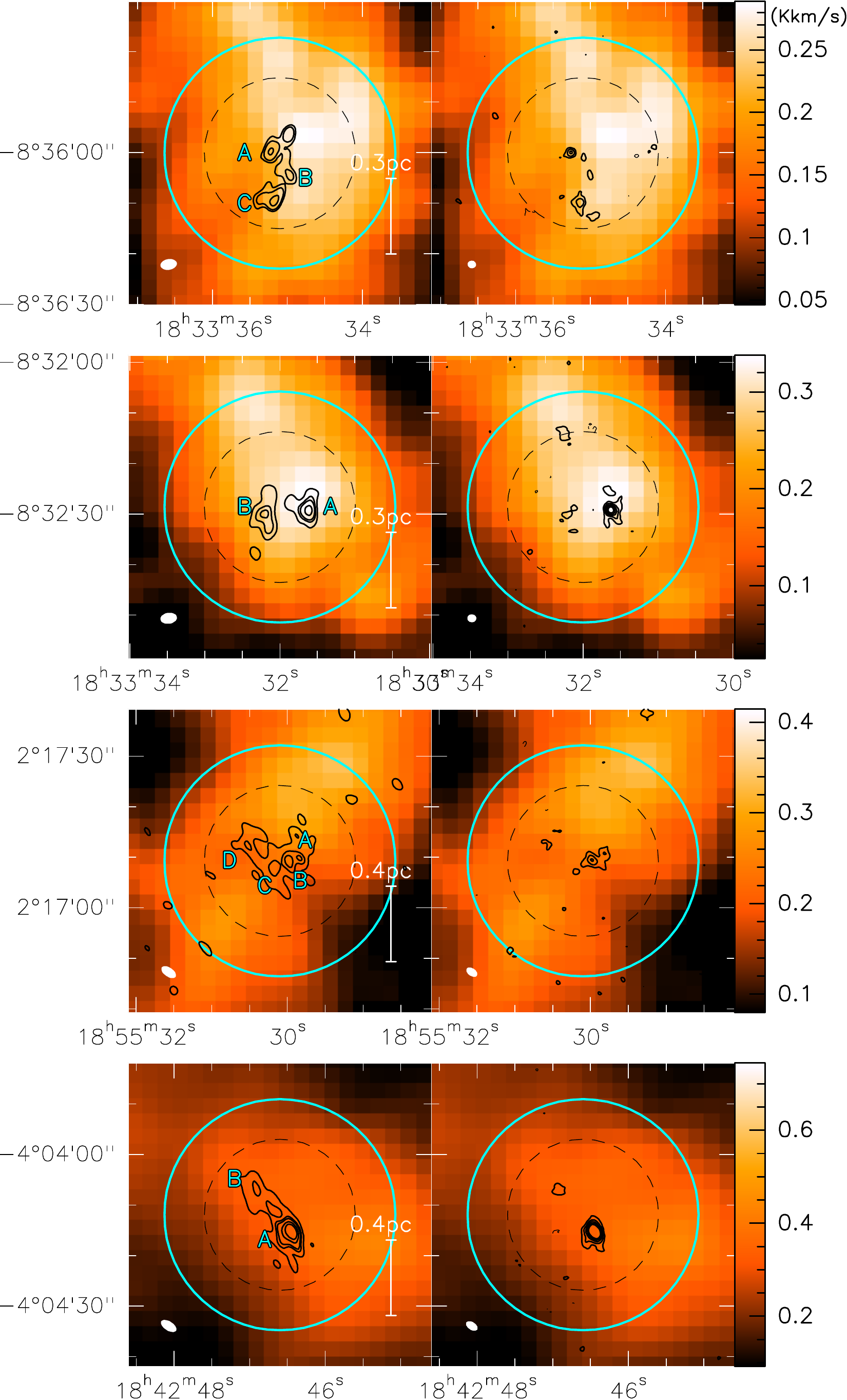}
}
\end{tabular}
\begin{tabular}{c}
\small RA [J2000]
\end{tabular}
\caption[Compilation of the continuum data from $70\,\mu$m to 1.1\,mm wavelength for 4 G28.34\,S, IRDC\,18530, IRDC\,18306 and IRDC 18308]{Compilation of the continuum data from $70\,\mu$m to 1.2\,mm wavelength for G28.34\,S, IRDC\,18530, IRDC\,18306, and IRDC 18308. Left column: Graymaps of the  dust emission observed by {\it Herschel}  at $70\,\mu$m \citep{ragan12}.  The red contours  show continuum emission observed by ATLASGAL at $870\,\mu$m  \citep{schuller09}, {starting from 10$\sigma$ rms  and continuing in 10$\sigma$ rms steps. }
Middle and Right columns: Black contours show continuum observed by SMA COMP configuration (middle) and COMP+EXT combination (right), overlaying  the moment 0 colormaps of $\rm H^{13}CO^+(1\rightarrow0)$ (integrated through its velocity dispersion) from IRAM 30\,m observations.  COMP contours start from 5$\sigma$ rms and continue in 5$\sigma$ rms steps, while COMP+EXT contours start from 3$\sigma$ and continue in 3$\sigma$ steps. $\rm 1 \sigma$ rms of each source configuration is  listed in Table~\ref{irdc:source30m}. Cyan letters mark the fragments having $>3\sigma$ rms continuum emission in both COMP and COMP+EXT maps. In each panel of the middle and right columns, the SMA synthesized beam is in the bottom left.  The cyan circles show the { primary} beam of SMA at 1.1\,mm, and the black dashed circles show the beam of 30\,m at 3\,mm.
\label{irdc:fig:conti}}
\end{figure*}

 As shown in Figure~\ref{irdc:fig:conti}, our sources locate on the ATLASGAL $870\,\mu$m continuum peaks of the pc-scale filaments  (shown as red contours in the left column). All targets are  embedded in the   {\it Herschel} $70\,\mu$m dark clouds \citep{ragan12}, and each of them is away from the  $70\,\mu$m  bright region at a projected distance of $\sim$ 1--2\,pc (1\arcmin--2\arcmin). 
At an angular resolution of $\sim$2\arcsec~(corresponding to  $\rm 10^4$ AU on linear scales), the 1.1\,mm SMA observations resolve each $870\,\mu$m continuum peak into several  compact substructures (hereafter ``fragments'',  plotted as  black contours in the middle and right columns of Figure~\ref{irdc:fig:conti}).  Combined with  ``natural weighting'', observations with the COMP+EXT configuration achieve a relatively high angular resolution, but high sensitivity only to the compact structures. Observations with the COMP configuration, however,  achieve a higher signal-to-noise ratio and recover some extended structures of the envelope, so they are essential for studying the morphology of the fragments, e.g., size, mass. \\

We find 2--4 fragments per source  with continuum emission levels $>3\sigma$ rms  in both the COMP and COMP+EXT maps. These fragments  have typical sizes of 0.05--0.2\,pc and masses of 4--62 $\rm M_{\sun}$. They are mostly separated at a projected distance of 0.07--0.24\,pc (see Table~\ref{irdc:fragI} and discussion in Section ~\ref{fragmentation}). In particular, fragments in G28.34\,S and IRDC\,18308 are well aligned along the large-scale filamentary direction. The spatial resolution of our observations only allow  the ``core'' structure to be resolved, 
so we cannot tell whether the cores contain further hierarchic fragments into condensations or stay as monolithic objects in this paper.

\subsection{Line survey from IRAM 30\,m at 1\,mm/3\,mm}\label{irdc:spectra}
{ Our target lines in the SMA observations included} cold, dense gas tracers such as $\rm H^{13}CO^{+}~(3\rightarrow2,~E_u/k_B=25~K)$, $\rm H^{13}CN~(3\rightarrow2,~E_u/k_B=25~K)$, $\rm HN^{13}C~(3\rightarrow2,~E_u/k_B=25~K)$, and  $\rm HNC~(3\rightarrow2,~E_u/k_B=26~K)$. However, none of them has  $>4\sigma$ rms emission at 1.1\,mm.  
{
Possible explanations for this include the following: (1) These species were tracing the more extended structures which our observations are not sensitive to; (2) Their emission was too faint to detect, given the sensitivity of our SMA observations; or (3) these high-J lines were not yet excited in these young, cold HMSCs. Our new IRAM 30\,m line survey now recovers the extended structure information with 3--10 times more sensitivity than the SMA observations. We now detect emission from lower-J transitions of the above species at 3\,mm.
}

\subsubsection{Line identification}\label{irdc:id}
\begin{figure*}
\centering
~~~~~~~~~~~~~~~3\,mm\\
\includegraphics[width=16cm]{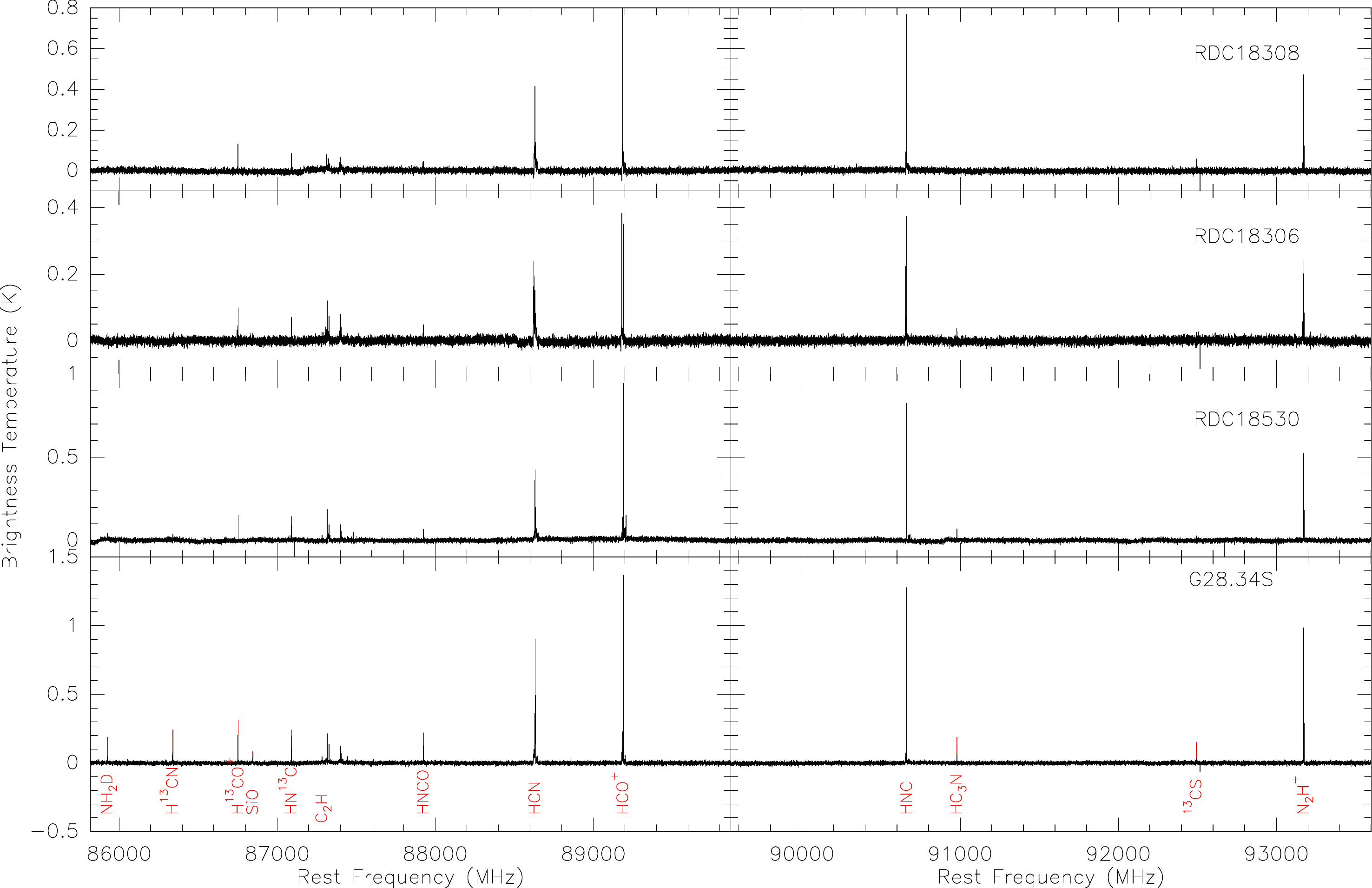}\\
~~~~~~~~~~~~~~~1\,mm\\
\includegraphics[width=16cm]{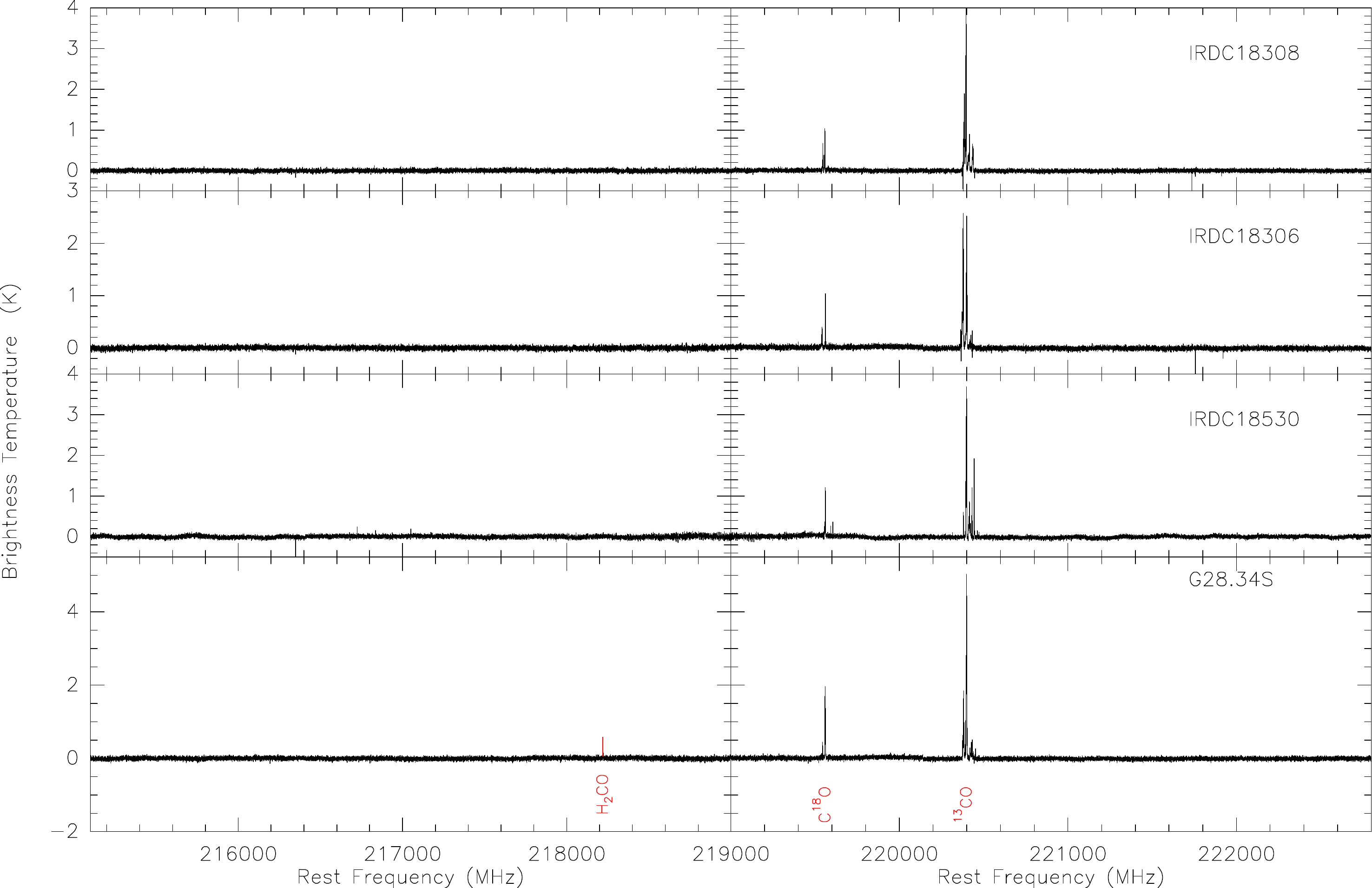}
\caption[Spectra of G28.34\,S, IRDC\,18530, IRDC\,18306 and IRDC 18308 from  IRAM 30\,m line survey at 1\,mm \& 3\,mm]{Averaged spectra from  IRAM 30\,m line survey at { 3\,mm/1\,mm}, which are extracted from a square region ([20\arcsec, 20\arcsec] to [-20\arcsec, -20\arcsec] offset)  centered on  the SMA continuum peak. The spectral resolution is 0.194 MHz ($ \rm 0.265~km\,s^{-1}$ at 1\,mm and $ \rm 0.641~km\,s^{-1}$ at 3\,mm). All  detected lines are labeled.  The ``extra band" is not shown here because of less integration time and thus there is  a lower signal-to-noise ratio. 
}\label{irdc:spec}
\end{figure*}

We present the spectra from each 30\,m targeted source  in Figure~\ref{irdc:spec}. 
To increase the signal-to-noise ratio for the weak emission in the dark clouds, 
each spectrum is averaged  from a square region ([20\arcsec, 20\arcsec] to [-20\arcsec, -20\arcsec] offset)  centered on the SMA continuum peak (the mapping center of the 30\,m observations).
Using the Splatalogue database\footnote{A compilation of the Jet Propulsion Laboratory (JPL, http://spec.jpl.nasa.gov,  \citealp{pickett98}),   Cologne Database for Molecular Spectroscopy  (CDMS, http://www.astro.uni-koeln.de/cdms/catalog,  \citealp{muller05}),  and Lovas/NIST catalogues  
 \citep{lovas04}, http://www.splatalogue.net.}, we identified 32 lines from  15 species  (including  20 isotopologues) in the 16\,GHz-broad 1\,mm/3\,mm band (Table~\ref{irdc:tab:line},  the lines detected only in the extra band are in blue).  Most of them are  ``late depleters'', which have low critical densities, low $\rm E_u/k_B$, and low  binding energies, and thus are hard to deplete onto grain surfaces \citep{bergin03}.  Although IRDC\,18306 is the nearest source in our sample, its lines  have the lowest brightness temperatures. In contrast, G28.34\,S has the largest number of detected lines with  the highest brightness temperatures in our sample, indicating that this source may be  chemically more evolved than the others in our sample.

\subsubsection{Molecular spatial distribution}\label{distribution}

When a high-mass star-forming region (HMSFR) is in the early chemical stages, different species trace gas with different temperatures and densities. 
Therefore, the spatial distribution diversity from different species can tell us the chemical status of a particular host source.\\

From Figure~\ref{irdc:spec}, G28.34\,S has the largest number of emission lines, so we use this source for line identification and list all the detections (having emission $\rm >4\sigma$ rms) in Table \ref{irdc:tab:line}. The line profiles as plotted in Figure~\ref{irdc:velpro}  show that some lines have blended multiplets. { To account for this multiplicity when mapping the distribution of all identified species in each source, we integrate each line according to its detection and multiplet using the  rules below. \\
$\bullet$ If a line that has $\rm \ge4\sigma$ detection in a particular source contains a multiplet, we fit the strongest multiplet transitions using the hyperfine-structure (HFS) fit; otherwise, we fit it using  a Gaussian profile\footnote{The HFS fit and the Gaussian fit are both based on the Gildas software package, http://www.iram.fr/IRAMFR/GILDAS/doc/html/class- html/class.html}   (as discussed in Section \ref{chemipro} and listed in Table~\ref{tab:lineprofile})\footnote{For HCN in IRDC18306, we image the  $\rm F=1\rightarrow1$ line. For the rest, we image the line with the strongest relative intensity (according to CDMS/JPL line intensity at 300 K, $\ell$(300 K) in Table~\ref{irdc:tab:line}).}.
 Subsequently, we map the integrated  intensities of these lines over their  velocity range (down to where the line goes into the noise, Table~\ref{irdc:tab:intrange}).\\
$\bullet$ If a G28.34\,S-detection species has $\rm <4\sigma$ detection  in another particular source, we integrate a total of  three channels around the systematic $\rm V_{lsr}$ at the rest frequency of its strongest transition { (marked with ``*'' in Table~\ref{irdc:tab:intrange}) in that source.} \\
 }
 
 { Figure \ref{irdc:molint} presents the spatial distributions of all the detected species in our sample.}
We note that { the intensity of some lines in the $\rm 70\,\mu$m bright sources in our field of view ($\sim$1\arcmin~ NE to G28.34S, IRDC\,18530, and NW to IRDC\,18308) are stronger than in our target $\rm 70\,\mu$m dark sources, which makes the spatial origin of these species hard to judge. Therefore, we plot the contour that marks the half maximum  integrated intensity of a particular species (the black contours in Figure~\ref{irdc:molint}). }
In general, most species  show the strongest emission towards the 870 $\mu$m continuum peak. We note that molecular emission  peaks in IRDC\,18306 and IRDC\,18308 have a systematic offset of  $\rm \sim5\arcsec\text{--}10\arcsec$ north of the continuum peak. This may be caused by the pointing uncertainties between 30\,m and APEX. { Nevertheless, comparison among the sources reveals some similarities and special features of the species that  we  outline in greater detail for different types of chemistry. 
Figure~\ref{irdc:velpro} shows the line profiles and Figure \ref{irdc:molint} the distributions\footnote{ Spatial distributions (Figure \ref{irdc:molint})  and line profiles (Figure~\ref{irdc:velpro}) are ordered by different groups and listed from the bottom left to the upper right.}.} \\

\onecolumn
 \begin{figure*}
\includegraphics[width=16cm]{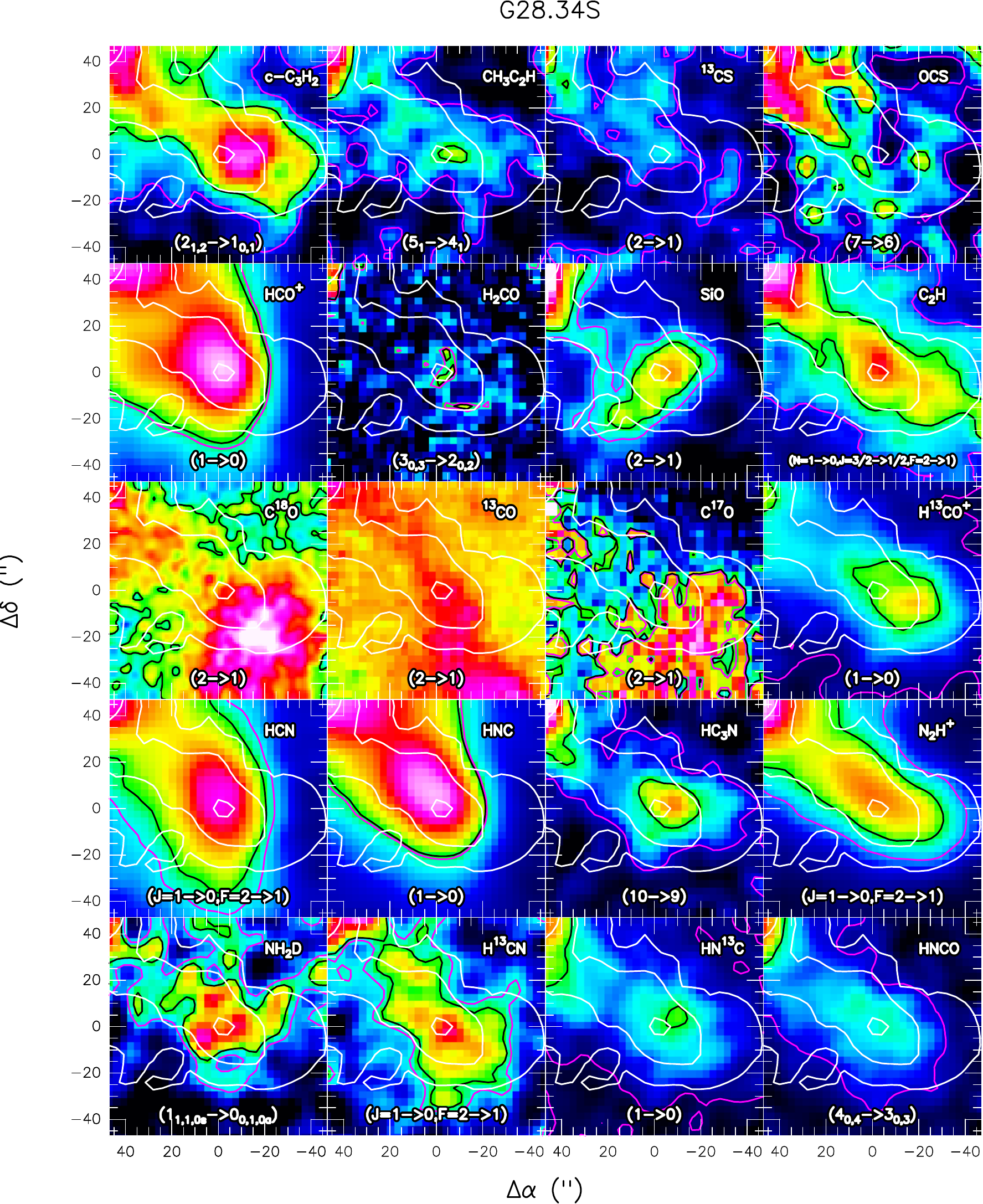}
\caption[Molecular spatial distribution in G28.34\,S, IRDC\,18530, IRDC\,18306 and IRDC 18308]{Molecular line integrated intensity maps overlaid with the $\rm 870\mu$m continuum contours in G28.34\,S, IRDC\,18530, IRDC\,18306, and IRDC 18308. The maps in each source are { ordered in different groups according to line rest frequencies and the leading isotopes}. Colormaps show the intensity integrations ($\rm Kkm\,s^{-1}$)  over the  velocity dispersion of each line. { For the lines with $\rm <4\sigma$ detections, we only integrate a total of three channels around the system $\rm V_{lsr}$ at their rest frequencies. }The purple contour in each panel indicates the area with integrated intensity $>4\sigma$ rms. { The black contour indicates  the area within which integrated intensity is over the half maximum in the field of view.} The white contours show the continuum emission from $870\,\mu$m ATLASGAL data, starting from $10\sigma$ and increasing by $10\sigma$.
}\label{irdc:molint}
\end{figure*}

\begin{figure*}
\ContinuedFloat
\includegraphics[width=16cm]{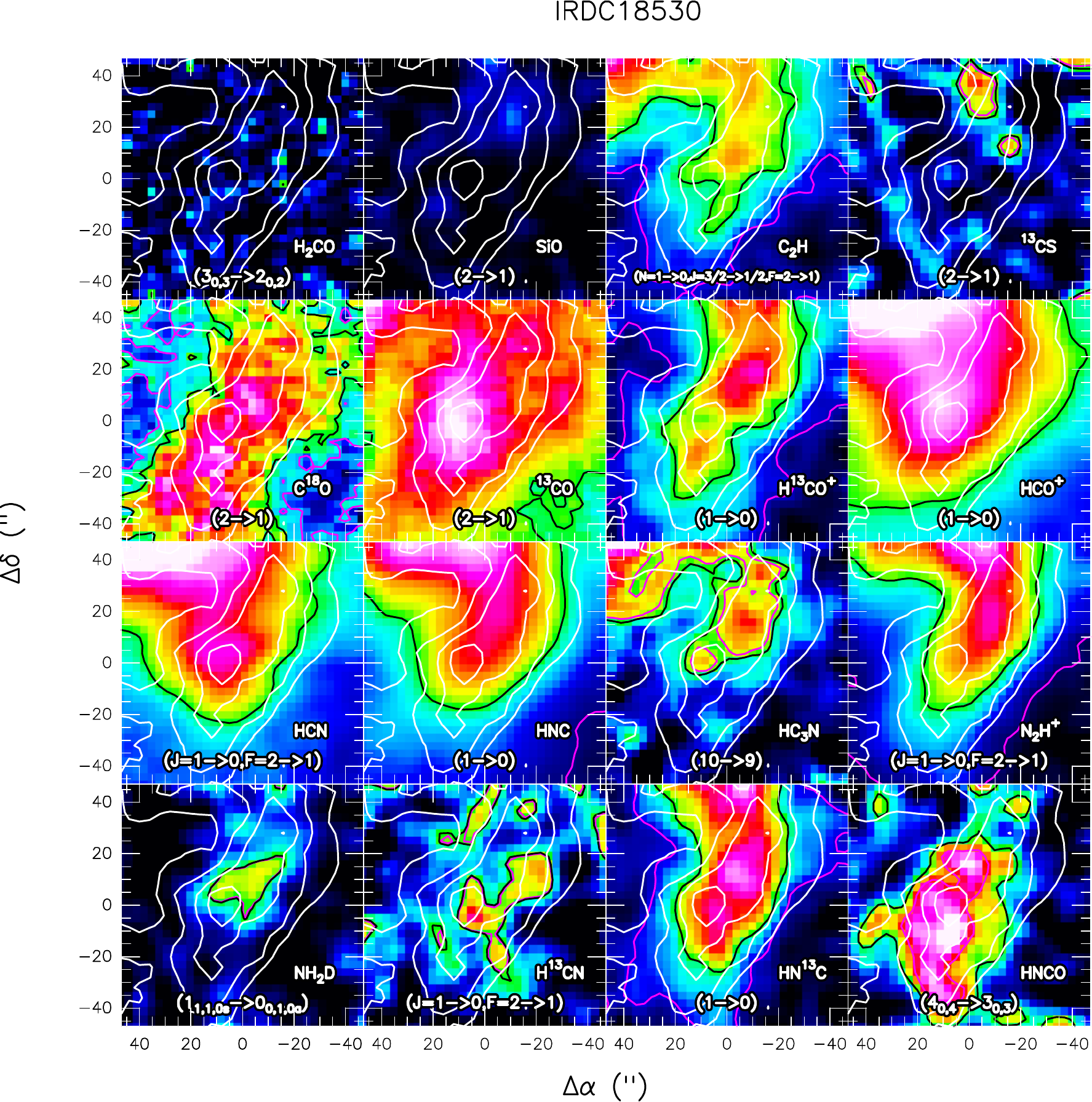}
\caption{(continued)}
\end{figure*}

\begin{figure*}
\ContinuedFloat
\includegraphics[width=16cm]{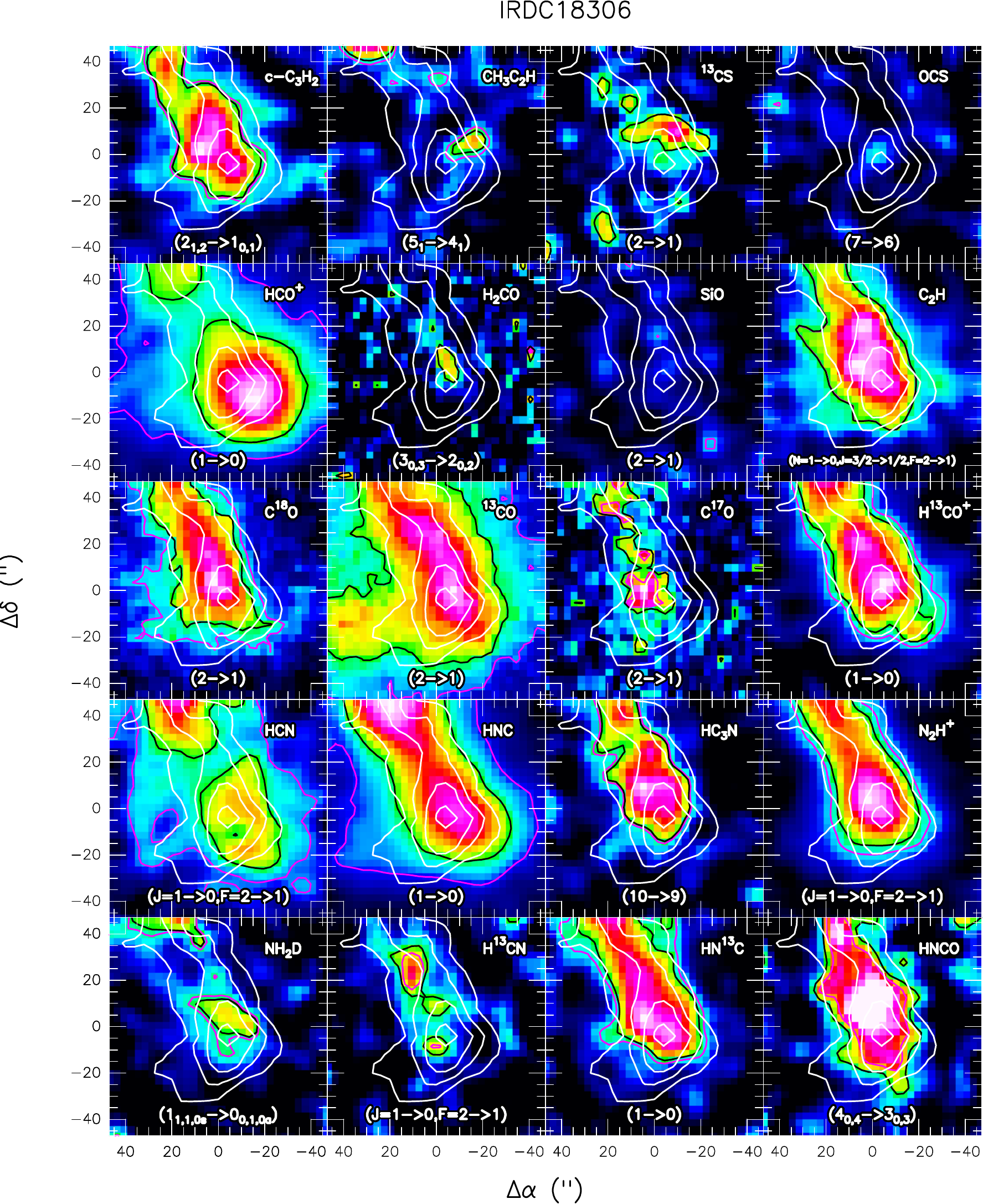}
\caption{(continued)}
\end{figure*}

\begin{figure*}
\ContinuedFloat
\includegraphics[width=16cm]{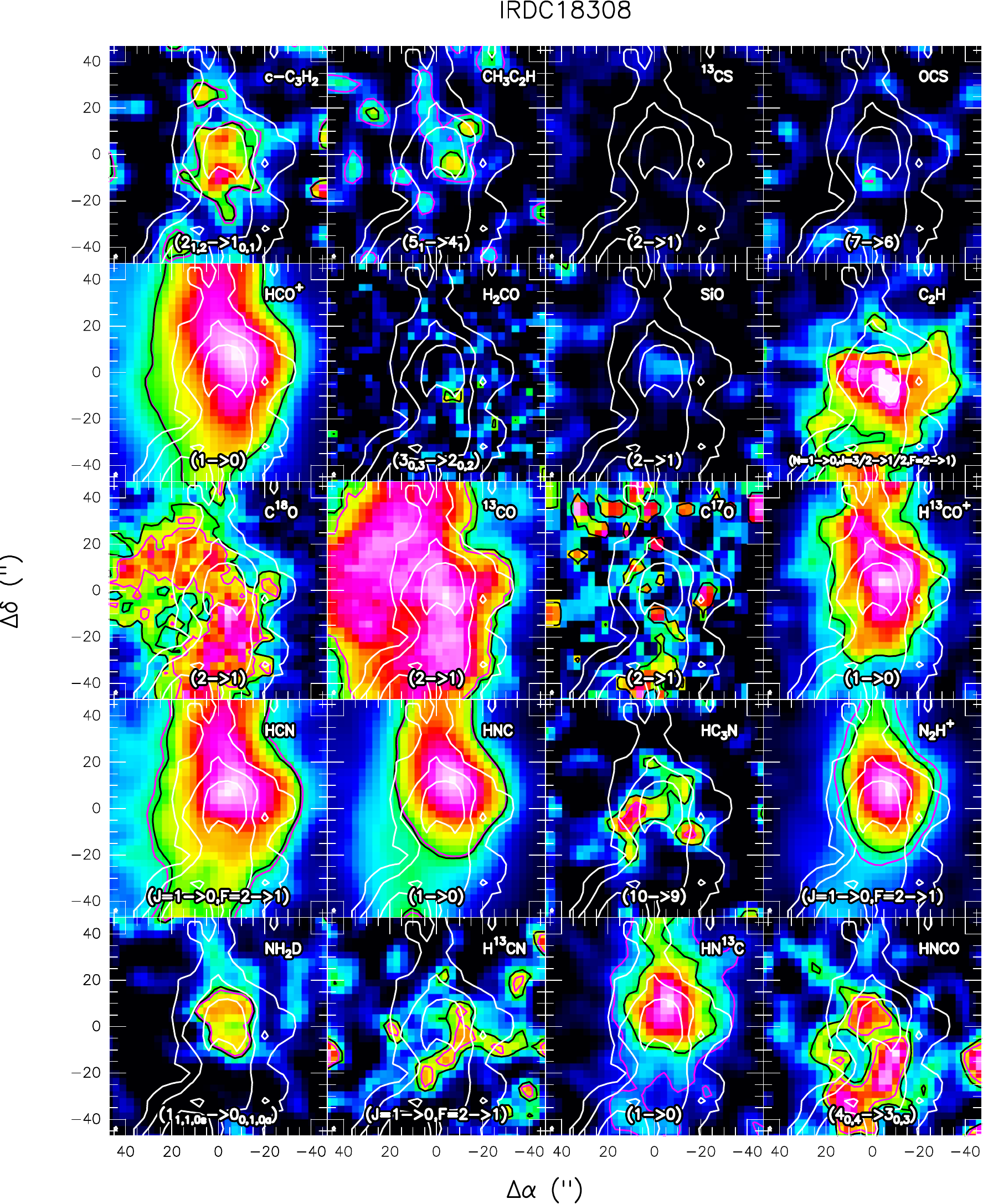}
\caption{(continued)}
\end{figure*}

\twocolumn

\begin{itemize}
\item {\bf Nitrogen (N-) bearing} species have the largest number of lines detected in our 30\,m observations.   Previous studies especially towards low-mass star-forming regions  (e.g., \citealt{lee01,pirogov03,fuller05,ragan06,sanhueza12,fontani14}) show that these species  usually trace the cold, dense gas because of their large dipole moments,  relatively low optical depths, or simple linear structures (e.g., $\rm HC_3N$, \citealp{bergin96}). Among them,  the abundance of $\rm NH_2D$ can be enhanced  by deuteration from $\rm NH_3$ ($\rm H_3^++HD\rightarrow H_2D^++H_2+\Delta E$)  when CO has frozen out onto the dust grains ($\rm T<20$ K, \citealp{crapsi05,chenhr10,pillai11}); HCN ($\rm H^{13}CN$)  is a well-known infall  tracer in low-mass prestellar environment (e.g., \citealt{sohn07}); the ratio of HCN/HNC  ($\rm H^{13}CN/HN^{13}C$)  is strongly temperature dependent, high in the warmer region like Orion GMC \citep{goldsmith86,schilke92}  and approximately around unity \citep{sarrasin10} in the cold environment (see discussion in Section \ref{irdc:hcn}); multiplets of $\rm N_2H^+$ are commonly used to  reveal  the excitation temperature of this species (e.g., \citealp{caselli02c}) 
 in the dark, quiescent regions.  All these species have  emission peaks or strong emission coincident with the 870  $\mu$m continuum peaks, and the offsets  between them are less than the pointing uncertainty between 30\,m and APEX. In particular, isotopologues of HNC and HCN are co-spatial in sources except for HCN ($\rm H^{13}CN$) in IRDC\,18308. This exception may be explained by the abnormal enhancement or suppression of the  individual hyperfine line discussed in Section \ref{chemipro}, indicating non-LTE (local thermal equilibrium) excitations of HCN \citep{loughnane12}. \\

In addition to 
the simple N-bearing species, HNCO is the simplest organic species containing C, H, O, and N elements. Moreover,  it has been suggested as a shock tracer, 
 based on its correlated spatial distribution with  SiO (e.g.,  \citealt{zinchenko00}) and $\rm CH_3OH$ (e.g., \citealt{meier05}). We found extended emission of HNCO in all our dark, cold clumps. The velocity dispersion of this line is similar to the other lines ($\rm 1-3\,km\,s^{-1}$) on the scale of 0.8\,pc, so we are  not able to tell whether the shocks are from cloud-cloud collision (e.g., \citealp{jimenez09,jimenez10,nguyen13}) or from protostellar objects deeply embedded.\\

\item {\bf Carbon oxidized and  hydride species } are important indicators of the chemical status of a particular star-forming region. Because they are one of the most abundant molecules in the gas phase,  possessing a low dipole moment and low critical density, CO isotopologues ($\rm ^{13}CO$, $\rm  C^{18}O,$ and $\rm C^{17}O$) can be used to indicate the gas temperature (e.g., \citealt{tafalla98,pineda08}, see also Appendix \ref{appen:cotemp}). $\rm HCO^+$ is the typical dense gas tracer produced from CO, and its abundance is enhanced  in regions with a high ionization fraction and/or shocked by outflows (e.g., \citealt{codella01,hofner01,rawlings00,rawlings04}). $\rm H_2CO$ is an important carbon hydride which forms $\rm CH_3OH$ and other more complex organics later in the gas phase (e.g., \citealp{horn04,garrod08}); sublimation of this species needs a warmer environment, so the detection of this species indicates the host cloud may be in a more evolved phase.\\

These species show the following features in G28.34S: (1) a clear anti-correlated distribution between CO isotopologues (especially the optically thin $\rm 2\rightarrow1$ lines of $\rm C^{18}O$ and $\rm C^{17}O$) and  $\rm N_2H^+$/$\rm NH_2D$/$\rm  H^{13}CO^+$, indicating strong  depletion and ionization even on the scale of 0.8\,pc (see discussion in Section \ref{irdc:codepletion}); (2) a blue asymmetric line profile of the optically thick $\rm HCO^+~(1\rightarrow0)$ comparing to   $\rm  H^{13}CO^+~(1\rightarrow0)$, indicating possible large velocity of  infall motion (\citealt{myers96,mardones97}, see Figure~\ref{irdc:h13co+} and Section ~\ref{sec:col} for detailed discussion); and (3) $>4\sigma$ line emission from $\rm H_2CO$ coincident with the continuum peak, indicating warmer gas there. Despite a tentative anti-correlated distribution between $\rm C^{17}O$ and  $\rm N_2H^+$/$\rm  H^{13}CO^+$ on the scale of 0.5\,pc in IRDC\,18308,  a 10\arcsec~ (0.2\,pc) offset between CO and $\rm NH_2D$ emission peak in IRDC\,18308, and a tentative detection of $\rm H_2CO$ ($\rm 2\sigma$ emission) in IRDC\,18306, the other sources do not have the same  features found in G28.34\,S. \\

{ We note that} the emission peak of $\rm HCO^+~(1\rightarrow0)$ and   $\rm  H^{13}CO^+~(1\rightarrow0)$ are not strictly co-spatial in any source (offsets are $\rm 5\arcsec-10\arcsec$), and this may come from the large opacity of $\rm HCO^+$. However, $\rm H^{13}CO^+$ shows a ``deficiency'' at the continuum peak of IRDC\,18530 where $\rm ^{13}CO$ is abundant, which is probably due to the dissociative recombination in this environment (see discussion in Section \ref{irdc:iosec}).\\

\item {\bf SiO} is a typical shocked gas tracer, and  is commonly associated with one or more embedded, energetic young outflows. We  detect  SiO ($\rm 2\rightarrow1$) only in G28.34\,S, where this line shows strong emission and broad line wings (Figure~\ref{irdc:velpro}). Coincident with the $\rm H_2O$ maser in G28.34\,S \citep{wangy06}, its emission may be from outflow and thus implies that a protostar is already embedded in this source. However, we cannot rule out the possibility that the  SiO  emission may also come from the  bright infrared  source which is 40\arcsec~south of our targeted region, or  from G28.34\,P1 which is 1\arcmin~to the north and hosts several jet-like outflows \citep{wangk11,zhang15}. Therefore, the spatial origin of SiO  is crucial to diagnose the evolutionary status of G28.34\,S, and thus further kinetic study with higher sensitivity data\footnote{The SiO $\rm (6\rightarrow5)$ line has only $2.5\sigma$ detection in our SMA data.}  is needed to confirm its  spatial origin. \\

\item { {\bf Carbon chains/rings} are a mysterious group of species whose gas/grain origin is still not clear. }Among the species we detected,   $\rm
C_2H$  is thought to reside (predicted by chemical models) 
in the source center in the early stage,
before transforming into other species (e.g., CO, OH, $\rm H_2O$), and it can have high abundance in the outer shells even at later stages \citep{beuther08a}; the carbon ring  {c-$\rm C_3H_2$}, first detected in the cold dark cloud TMC-1 by \citet{matthews85}, is reported to survive even  in a hot region where $\rm NH_3$ is photon-destroyed  \citep{palau14}; and $\rm CH_3C_2H$ is considered  a good probe of temperature for early stages of star-forming regions \citep{bergin94}. We detect the above-mentioned carbon chains/ring towards the continuum peak of all our sources, except in IRDC\,18530 where only $\rm C_2H$ is observed. However, since these species also show stronger or roughly the same emission towards the nearby $\rm 70\,\mu$m bright sources (according to the half maximum intensity contour), we cannot tell whether these species are good cold gas tracers or not.\\

\item {\bf Sulfur (S-) bearing species} also have unknown chemistry. Two S-bearing lines, $\rm  ^{13}CS (2\rightarrow1)$  and OCS  $\rm (7\rightarrow6)$ are covered in our observations (except for the OCS line in IRDC\,18530). Owing to their high dipole moments, CS isotopologues are used as  dense gas tracers (e.g., \citealt{bronfman96}) in  both low-mass  (e.g., \citealt{tafalla02}) and high-mass  prestellar cores (e.g., \citealt{jones08}), but they show high depletions in the center of high-mass dense cores (e.g.,  \citealt{beuther09b}). OCS forms via surface chemistry and is the only S-bearing molecule that has actually been observed on
 interstellar ices \citep{palumbo97}. Both $\rm ^{13}CS$ and OCS are only detected in G28.34\,S with $\rm>4\sigma$ emission, but these emissions have been shown to  originate from the NE bright source, 
 according to their half maximum intensity contours. Therefore, we do not claim their spatial origin from the cold gas.

\end{itemize}

In short, G28.34\,S appears chemically more evolved than the other sources in our sample, thanks to  the sole detections of  $\rm H_2CO$ and OCS,  the strong emission of HNCO, the  $\rm HCO^+$ asymmetric line profile implying significant infall, and SiO coinciding with previous $\rm H_2O$ maser detection. All these signatures  indicate that G28.34\,S  is the host of potential protostellar objects.

\section{Analysis and discussion}\label{irdc:analysis}

\subsection{Fragmentation}\label{fragmentation}
Figure \ref{irdc:fig:conti} presents  the continuum maps of {\it Herschel}  $70\,\mu$m,  APEX $870\,\mu$m (ATLASGAL), and  SMA  1.1\,mm. Because the COMP+EXT configuration of SMA filters out some extended structures, in the following we calculate parameters using data obtained with  the COMP configuration only.\\

We assume that the dust emission is optically thin, and that  dust is fully coupled with gas with the same temperature. Therefore, we estimate the mass  $\rm M_{fr}$  and column density  $\rm N_{fr}$ in the resolved fragments from the SMA continuum maps with \citep{hildebrand83,schuller09}:
 \begin{eqnarray}\label{irdc:gas-SMA}
 \rm M_{fr}
                 &=&\rm \frac{S_{1.1mm}R{\it D}^2}{B_{1.1mm}(T_{SMA}) \kappa_{1.1mm}} ~~~~~~~~~~~~~~~~~~~~~~~~~~~~~~~~~~~~~~~~~~(g)\\
  \rm N_{fr}
                 &=&\rm \frac{I_{1.1mm}R}{B_{1.1mm}(T_{SMA})\Omega_{SMA} \kappa_{1.1mm}\mu m_H}~~~~~~~~~~~~~~~~~~~(cm^{-2})  
 \end{eqnarray}
The averaged column density of the entire clump (in the $\rm 40\arcsec\times 40\arcsec$ region) $\rm N_{c}$  can be estimated from  each ATLASGAL continuum map as well,
 \begin{eqnarray}\label{irdc:gas-ATLAS}
  \rm N_{c}
                 &=& \rm \frac{I_{0.87mm}R}{B_{0.87mm}(T_{ATLAS})\Omega_{ATLAS} \kappa_{0.87mm}\mu m_H}~~(cm^{-2})  \label{irdc:atlasgas}
 ,\end{eqnarray}
   
\noindent  where  {\it D} is the source kinematic distance (from the sun), $\rm S_{1.1mm}$ is the total continuum flux density at 1.1\,mm (measured from each fragment  after a 2-D Gaussian fit\footnote{We fit the 2-D Gaussian structure with CASA, http://casa.nrao.edu}, in units of Jy); $\rm I_{1.1mm}$ is the specific intensity for the continuum  peak of each fragment (in $\rm Jy~beam^{-1}$); $\rm I_{0.87mm}$ is the averaged specific intensity in the $\rm 40\arcsec\times 40\arcsec$ region (in $\rm Jy~beam^{-1}$);  
R is the isothermal gas-to-dust mass ratio (taken to be 150, from \citealt{draine11}),  $\rm B_{1.1mm}(T_{SMA})$ and $\rm B_{0.87mm}(T_{ATLAS})$ are the Planck functions for a dust temperature $\rm T_{SMA}$ or $\rm T_{ATLAS}$; $\rm \Omega_{SMA}$ and $\rm \Omega_{ATLAS}$ are the solid angles of the SMA and ATLASGAL beam  (in $\rm rad^2$); $\rm \kappa_{1.1 mm}=1.0~\rm cm^2g^{-1} $ and $\rm \kappa_{0.87 mm}=1.8~\rm cm^2g^{-1} $ are the dust absorption coefficients at 1.1\,mm and 870 $\mu$m (assuming a model of agglomerated grains with thick ice mantles for densities {\color{black} $\rm 10^5\text{--}10^6~cm^{-3}$}, extrapolated from \citealt{ossenkopf94});
 $\mu$ is the mean molecular weight of the ISM, which is assumed to be 2.33;  and $\rm m_H$ is the mass of a hydrogen atom  ($\rm 1.67 \times 10^{-24}$ g).
\\

As described in Table~\ref{irdc:fragI},   the sizes of fragments (from 2-D Gaussian fitting) are 2--4 times larger than the SMA beam size on small scales. On larger scales,  the sizes of filamentary structures (projected area with $\rm >10\sigma$ emission) seem to be more extended than the primary beam of ATLASGAL  { ($\rm \theta_{ALTLASGAL}=18\arcsec$)}. Therefore, we take the filling factors as unity in both spatial scales.\\

{ 
Firstly, we compare the gas concentrations on different spatial scales. Following \citet{kirk06}, assuming Bonner-Ebert (BE) spheres for both  large-scale clumps and small-scale cores,  the gas concentration can be estimated in terms of observable parameters as
 
 \begin{eqnarray}
\rm Con_{ATLAS}&=&\rm 1-\frac{1.44\Omega_{ATLAS}}{I_{0.87mm}}\overline{\frac{S_{0.87mm}}{A_{0.87mm}}}  \label{irdc:conc_atlas}     \\
\rm Con_{SMA}&=&\rm 1-\frac{1.44\Omega_{SMA}S_{1.1mm}}{A_{1.1mm}I_{1.1mm}}   \label{irdc:conc_sma}
 \end{eqnarray}
 
 Here $\rm A_{1.1mm}$ is the size of each small-scale fragment derived from 2-D Gaussian fits (almost the same size as the area within the $5\sigma$ continuum contour). Since ATLASGAL filaments are not suitable to fit with a 2-D Gaussian, we measure the total flux per unit projection size on each ATLASGAL clumps ($\rm \overline{S_{0.87mm}/A_{0.87mm}}$) with the following approach. 
{ Following \citet{wangk15},} we extract a flux intensity profile from a cut perpendicular to the filament elongated direction, which is centered on the mapping center of the 30\,m observations\footnote{We assume that the local temperature in the cut is the same at different radii, so the flux intensity profile is the first-order estimate instead of the density profile in \citet{wangk15}.}. Then we measure the  FWHM width of the filament locally around our source. For each filament, we set three rectangles along the filamentary orientation. These rectangles all have the same centers as the previous cut and the same width as the local FWHM width. The length of the rectangles along the filament are one, two, and three times of the FWHM width. Then, we measure the total flux and projection size within each rectangle and derive the mean.\\

We find that the gas concentration { (listed in Tables~\ref{irdc:fragI}--\ref{irdc:fragII}) on  large scales  ($ \rm Con_{ATLAS}\sim0.6$) is lower than that in the small-scale fragments  ($\rm Con_{SMA}\sim0.8$). } Moreover, the  large-scale concentration is on the verge of requiring additional non-thermal support mechanism (critical value of BE sphere as 0.72 from \citealp{walawender05,kirk06}).\\

 }

Secondly, assuming that all  small-scale fragments and large-scale cloud clumps are spherically symmetric, we then estimate the gas volume number density of each fragment $ \rm n_{fr}$ from   $\rm \theta_{SMA}$ (in units of rad), and the averaged gas volume number density $ \rm n_{c}$ of the entire clump on larger scales from $\rm \theta_{ALTLASGAL}$ (in units of rad) 
as\footnote{ Since both the small-scale fragments and large-scale filaments show asymmetric structures, the simplification 2-D Gaussian fit cannot provide more precise estimation of the volume number density than the method  we use here.}
\begin{eqnarray}\label{irdc:volume_d}
 \rm n_{fr}=\frac{N_{fr}}{\theta_{SMA}\it D} ~~~~~~~~~~~~~~(cm^{-3})\\
  \rm n_{c}=\frac{N_{c}}{\theta_{ATLAS}\it D} ~~~~~~~~~~~~~~(cm^{-3})
    \end{eqnarray}

We assume that  the core-scale structures (size of $\sim$0.1\,pc) seen in the SMA-COMP configuration result from the fragmentation from the pc-scale clumps.
If fragmentation is governed by the thermal Jeans instabilities,  the thermal Jeans length $\rm \lambda_{th-J}$ and mass $\rm M_{th-J}$ of the entire clump can be estimated from the  averaged density and temperature in this clump,
 \begin{eqnarray}
\rm \lambda_{th-J}&=&\rm c_s(\frac{\pi}{G\mu m_Hn_{c}})^{1/2}   \nonumber\\
                             &=& \rm 0.067(\frac{T_{ATLAS}}{10\,K})^{1/2}(\frac{n_{c}}{10^5\,cm^{-3}})^{-1/2}\,pc    \label{irdc:jean_l}     \\
\rm M_{th-J}&=&\rm \frac{c_s^3}{6}(\frac{\pi^5}{G^3\mu m_Hn_{c}})^{1/2}  \nonumber\\
                    &=&\rm 0.912(\frac{T_{ATLAS}}{10\,K})^{3/2}(\frac{n_{c}}{10^5\,cm^{-3}})^{-1/2}\,M_{\odot}   \label{irdc:jean_m}
 ,\end{eqnarray}
 where G is the gravitational constant ($\rm 6.67\times10^{-8}~cm^3g^{-1}s^{-2}$) and $\rm c_s=[k_BT_{ATLAS}/(\mu m_H)]^{1/2}$ is the speed of sound at temperature $\rm T_{ATLAS}$. \\

\begin{table*}
\caption[Parameters of the fragments]{The parameters measured  from SMA observations at 1.1\,mm, including the coordinates,  specific intensity at continuum peak, total flux density, and projected source size.  }\label{irdc:fragI}

  \small
\centering

\scalebox{0.9}{
\begin{tabular}{c|cp{1.3cm}p{1.3cm}p{1.cm}p{0.8cm}p{1.6cm}p{0.7cm}| p{0.7cm} p{1.5cm} p{1.5cm}| p{1cm} p{1cm}}\hline\hline

Source           &    &R.A.                 &Dec.                          &$\rm I_\nu^{\it a}$     &$\rm S_\nu^{\it a}$  &$\rm \theta_{fr}^{\it a,b}$   &Con.$^{\it b}$   &$\rm T_{SMA}$          &$\rm N_{fr}^{\it c}$                    & $\rm n_{fr}^{\it c} $                               & $\rm \lambda_{fr}^{\it d} $    & $\rm M_{fr} $                \\

                 &&J[2000]     &J[2000]     &($\rm\frac{mJy}{beam}$)         &(mJy)                &($\rm \arcsec\times\arcsec$) & & $\rm (K)$      & $\rm 10^{23}(cm^{-2})$      & $\rm 10^5 (cm^{-3})$         & $\rm (pc)$     & $\rm (M_{\odot})$   \\
\hline
\hline

G28.34\,S              &    &$\rm 18^h42^m$     &$\rm -04^{\circ}04^{''}$     &          &            &                                         &          &15  &                &                                      &                       &\\
                  &A   &$\rm 46^s.426$     &$\rm 15^{'}.54$              &23.5      &69.8        &$\rm 5.2\arcsec\times3.6\arcsec$         &0.78      &    & 6.38           & 47.24                                              &\multirow{2}{*}{0.24}   & 53.45\\     
                  &B   &$\rm 46^s.904$     &$\rm 07^{'}.08$              &8.4       &81.4        &$\rm 9.2\arcsec\times6.6\arcsec$         &0.78      &    & 2.28                 & 16.88                                    &                        & 62.33\\
\hline
IRDC\,18530   &    & $\rm 18^h55^m$   &$\rm 02^{\circ}17^{''}$  &      &        &      &      &15   &       &       &          &\\
                     &A   &$\rm 29^s.841$     &$\rm 09^{'}.87$              &5.4         &8.4        &$\rm 3.6\arcsec\times2.5\arcsec$        &0.77   &         & 1.55 & 12.09 & & 6.16   \\ 
                     &B   &$\rm 29^s.983$     &$\rm 08^{'}.88$              &5.3         &9.6        &$\rm 3.8\arcsec\times2.7\arcsec$         &0.77  &         & 1.52 & 11.87 &\multirow{2}{*}{0.07} & 7.03 \\
                   &C   &$\rm 30^s.165$     &$\rm 07^{'}.81$              &4.0     &8.3     &$\rm 4.5\arcsec\times2.7\arcsec$     &0.78    &     & 1.15 & 8.96 & & 6.08 \\
                     &D   &$\rm 30^s.525$     &$\rm 11^{'}.69$              &3.7     &6.5     &$\rm 6.1\arcsec\times1.7\arcsec$     &0.78 &     & 1.06 & 8.29 & & 4.76  \\
 \hline

IRDC\,18306  &    & $\rm 18^h33^m$   &$\rm -08^{\circ}32^{''}$  &      &        &       &       &15   &       &       &          &\\
                     &A   &$\rm 31^s.609$     &$\rm 29^{'}.36$              & 13.8     &45.3    &$\rm 5.8\arcsec\times4.0\arcsec$     &0.86 & & 5.17           &57.39                   &\multirow{2}{*}{0.15}     & 21.24 \\      
                     &B   &$\rm 32^s.230$     &$\rm 29^{'}.58$              &9.2     &71.4    &$\rm 11.2\arcsec\times4.9\arcsec$    &0.86  &      & 3.44 & 38.26  & & 33.49 \\
\hline

IRDC\,18308  &    & $\rm 18^h33^m$   &$\rm -08^{\circ}$                     &       &        &     &      &15    &       &       &          &\\
                     &A   &$\rm 35^s.194$     &$\rm 35^{''}59^{'}.55$              &8.3     &42.6     &$\rm 8.9\arcsec\times4.0\arcsec$     &0.87 &     & 3.46 & 33.24 & & 29.84  \\      
                     &B   &$\rm 34^s.990$     &$\rm 36^{''}04^{'}.53$              &7.1     &78.0     &$\rm 15.5\arcsec\times5.0\arcsec$     &0.87 &     & 2.96 & 28.43                  &0.12       & 54.64  \\
                     &C   &$\rm 35^s.194$     &$\rm 36^{''}09^{'}.52$              &11.3     &49.8     &$\rm 6.9\arcsec\times4.9\arcsec$    &0.8 &     & 4.30 & 41.25                  &       & 34.89\\
\hline
 \hline
  \multicolumn{10}{l}{{\bf Note.} {\it a.} Properties are estimated from the maps with COMP configuration}\\
 \multicolumn{10}{l}{~~~~~~~~~~{\it b.}$\rm \theta_{fr}$ is from 2-D Gaussian fits}\\
 \multicolumn{10}{l}{~~~~~~~~~~{\it c.}  $\rm H_2$ column density  and the average volume number density at the continuum peak are estimated at 15 K. }\\
  \multicolumn{10}{l}{~~~~~~~~~~{\it d.} $\rm \lambda_{fr}$ is the mean projected separation of concentrations}\\
\end{tabular}
}

\end{table*}

\begin{table*}
\caption[Two hypotheses of fragmentation]{Parameters from 30\,m and APEX observations and estimates based on two hypotheses of fragmentation (thermal Jeans fragmentation and turbulent Jeans fragmentation). }\label{irdc:fragII}

\centering
\scalebox{0.78}{
\begin{tabular}{c|cc|cccc|ccccccc|cc}
\hline\hline

Source      &$\rm \Delta \upsilon \it ^a$      &$\rm \sigma_{obs}\it ^a$    &$\rm T_{30\,m}$      &$\rm \sigma_{th}$     &$\rm \sigma_{Nth}$     &$\rm n_{c}\it ^b$     &$\rm \lambda_{th-J}\it ^c$      &$\rm M_{th-J}\it ^c$   &$\rm \lambda_{Nth-J}\it ^d$      &$\rm M_{Nth-J}\it ^d$    &$\rm \lambda_{cl}\it ^{e,g}$      &$\rm M_{cl}\it ^{e,g}$     &$\rm (M/l)_{cl} \it ^{e,g}$  &$\rm (M/l)_{ATLAS} \it ^{f,g}$  &Con.$\it ^h$\\
     &$\rm (km~s^{-1})$     &$\rm (km~s^{-1})$   &(K)   &$\rm (km~s^{-1})$    &$\rm (km~s^{-1})$    & ($\rm 10^5~cm^{-3}$)       & (pc)    & $\rm (M_{\odot})$   & (pc)    & $\rm (M_{\odot})$   & (pc)    & $\rm (M_{\odot})$     & $\rm M_{\odot}pc^{-1}$   & $\rm M_{\odot}pc^{-1}$\\

\hline
\hline
&&     &12     &$\rm 0.14$     &$\rm 1.29\pm0.03$     &$\rm 1.07$     &$\rm 0.07$     &$\rm 1.16$     &$\rm 0.45$     &$\rm 291$     &$\rm 1.57$     &$\rm 1240$   &   &$\rm1690\pm481$   &\\
G28.34\,S    &$\rm 3.07\pm0.06$     &$\rm 1.30\pm0.03$     &15     &$\rm 0.15$     &$\rm 1.29\pm0.03$     &$\rm 0.72$     &$\rm 0.10$     &$\rm 1.97$     &$\rm 0.55$     &$\rm 354$     &$\rm 1.91$     &$\rm 1510$   &$\rm 788$    &$\rm1146\pm326$  &$\rm 0.63\pm0.04$\\
&&     &18     &$\rm 0.17$     &$\rm 1.29\pm0.03$     &$\rm 0.54$     &$\rm 0.12$     &$\rm 2.99$     &$\rm 0.63$     &$\rm 409$     &$\rm 2.21$     &$\rm 1740$   &&$\rm858\pm244$  &\\
\hline
&&     &12     &$\rm 0.14$     &$\rm 0.84\pm0.03$     &$\rm 0.79$     &$\rm 0.08$     &$\rm 1.34$     &$\rm 0.34$     &$\rm 94.60$     &--     &--  &--&--     &\\
18530     &$\rm 2.01\pm0.06$     &$\rm 0.85\pm0.03$     &15     &$\rm 0.15$     &$\rm 0.84\pm0.03$     &$\rm 0.54$     &$\rm 0.11$     &$\rm 2.28$     &$\rm 0.41$     &$\rm 115$     &--     &--   &--  &--     &$\rm 0.62\pm0.04$\\
&&     &18     &$\rm 0.17$     &$\rm 0.84\pm0.03$     &$\rm 0.40$     &$\rm 0.14$     &$\rm 3.47$     &$\rm 0.48$     &$\rm 133$     &--     &--   &--&--     &\\
\hline
&&     &12     &$\rm 0.14$     &$\rm 0.86\pm0.04$     &$\rm 0.85$     &$\rm 0.08$     &$\rm 1.30$     &$\rm 0.34$     &$\rm 97.10$     &--     &--    &--&--    &\\
18306     &$\rm 2.05\pm0.10$     &$\rm 0.87\pm0.04$     &15     &$\rm 0.15$     &$\rm 0.86\pm0.04$     &$\rm 0.57$     &$\rm 0.11$     &$\rm 2.21$     &$\rm 0.41$     &$\rm 118$     &--     &--   &--&--     &$\rm 0.64\pm0.07$\\
&&     &18     &$\rm 0.17$     &$\rm 0.85\pm0.04$     &$\rm 0.43$     &$\rm 0.14$     &$\rm 3.36$     &$\rm 0.47$     &$\rm 136$     &--     &--   &--&--     &\\
\hline
&&     &12     &$\rm 0.14$     &$\rm 0.88\pm0.04$     &$\rm 0.52$     &$\rm 0.10$     &$\rm 1.67$     &$\rm 0.44$     &$\rm 132$     &$\rm 1.54$     &$\rm 561$     &&$\rm391\pm34$  &\\
18308     &$\rm 2.09\pm0.09$     &$\rm 0.89\pm0.04$     &15     &$\rm 0.15$     &$\rm 0.87\pm0.04$     &$\rm 0.35$     &$\rm 0.14$     &$\rm 2.83$     &$\rm 0.53$     &$\rm 160$     &$\rm 1.87$     &$\rm 681$     &$\rm 365$   &$\rm265\pm23$  &$\rm 0.60\pm0.07$\\
&&     &18     &$\rm 0.17$     &$\rm 0.87\pm0.04$     &$\rm 0.26$     &$\rm 0.18$     &$\rm 4.30$     &$\rm 0.62$     &$\rm 185$     &$\rm 2.16$     &$\rm 787$    &&$\rm198\pm17$   &\\

\hline
 \hline
 \multicolumn{15}{l}{{\bf Note.} {\it a.} Measured from $\rm H^{13}CO^+~(1\rightarrow0)$  in a 40\arcsec$\times$40\arcsec~region (from [20\arcsec, 20\arcsec] to  [-20\arcsec, -20\arcsec] offset, corresponding to $\sim 1.5\times10^5$ AU) obtained  from 30\,m observations.  }\\
\multicolumn{15}{l}{~~~~~~~~~ {\it b.} $\rm n_{c}$ is measured from ATLASGAL dust emission at $870\,\mu$m.}\\
\multicolumn{15}{l}{~~~~~~~~~~{\it c.} Thermal Jeans length ($\rm  \lambda_{th-J}$) and mass ($\rm M_{th-J}$) are predicted from ATLASGAL dust emission  at $870\,\mu$m.}\\
\multicolumn{15}{l}{~~~~~~~~~~{\it d.} Turbulent Jeans length ($\rm  \lambda_{Nth-J}$) and mass ($\rm M_{Nth-J}$) are predicted  from 30\,m observations on $\rm H^{13}CO^+~(1\rightarrow0)$.}\\
\multicolumn{15}{l}{~~~~~~~~~~{\it e.} The critical mass and length of cylindrical fragmentation  are predicted  from 30\,m observations on the velocity dispersion of $\rm H^{13}CO^+~(1\rightarrow0)$, Eqs.~\ref{irdc:mcritical}--\ref{irdc:m_cyl}.}\\
\multicolumn{15}{l}{~~~~~~~~~~{\it f.} The linear mass densities at different dust temperatures (assuming $\rm T_{ATLAS}=T_{30\,m}$)  are predicted  from ATLASGAL dust emission at $870\,\mu$m.}\\
\multicolumn{15}{l}{~~~~~~~~~~{\it g.}  The values in IRDC\,18306 and  IRDC\,18308 are not given, because the fragments in each source do not   align well along the large-scale filamentary direction. 
}\\
\multicolumn{15}{l}{~~~~~~~~~~{\it h.} Concentration of the gas in the filamentary scale is estimated  from ATLASGAL dust emission at $870\,\mu$m (temperature independent).}\\
\end{tabular}
}

\end{table*}

We assume  $\rm T_{SMA}=15~K$ for the following reasons: (1) We did not detect the SMA-targeted lines with low upper energy levels ($\rm E_u/k\sim 25~K$); (2) Low temperatures have been estimated from other observations in  G28.34\,S, including the SABOCA 350\,$\mu$m survey (13\,K as upper limit, 11\arcsec resolution; \citealt{ragan13}),  SPIRE 500\,$\mu$m  observations (16\,K, 36\arcsec resolution), and  the rotation temperature of  $\rm NH_3$ derived from VLA+Effelsberg (13-15 K, 5\arcsec resolution, \citealp{wangy08}; private communication with K. Wang); (3) It is  unclear whether a protostar has already formed in these sources. Dust and gas may either have higher temperature in the envelope region (i.e.,  structures probed by ATLASGAL, SABOCA, and SPIRE 500\,$\mu$m) than at the SMA-probed center because of  UV-shielding, or the envelope can be colder than the central core if the protostar has  formed (i.e., in G28.34\,S). In either case, we give the estimates of the large-scale ATLASGAL clumps at  ad hoc temperatures  12\,K, 15\,K, and 18\,K.\\

As shown in Tables~\ref{irdc:fragI}--\ref{irdc:fragII}, the volume number densities in the SMA fragments  are at least ten times higher than the averaged volume number densities found in the ATLASGAL clumps. Moreover, the average projected separation  between the fragments (0.07--0.24 pc) is comparable to the predicted thermal Jeans length of the entire clump (0.07--0.18 pc) at a temperature of $\rm T_{ATLASGAL}>12\,K$ and $\rm T_{SMA}\sim15\,K$. However,  the fragment masses are  significantly higher (more than four times)   than the predicted thermal Jeans masses of the entire clumps, even without accounting for interferometric-filtering. In particular,  fragment masses are $\rm > 10\,M_{\odot}$ in our sample, except for IRDC\,18530. A similar discrepency has also been reported in fragmentation studies in several other IRDCs (e.g., \citealt{wangk14,beuther15} ). \\

{ Different hypotheses have been proposed for the ``non-thermal support'', as listed below. \\}

 \begin{itemize}
 \item {\bf Turbulent Jeans fragmentation:}\\
   Since thermal  Jeans fragmentation describes the fragmentation processes when  the internal pressure is dominated by the thermal motion, the above-mentioned discrepency
    indicates that non-thermal motions, such as turbulence may play an important role in fragmentation  (e.g.,\citealt{wangk11,wangk14}). \\

To investigate whether  non-thermal pressure provides extra kinetic energy against the { gravitational energy} of the fragments, we need to measure the velocity dispersion over the entire clump, which can be derived from the linewidth of the dense gas tracer(s).  The linewidths of all the species are on average $\rm 2\text{--}3~km\,s^{-1}$ (see{ Table~\ref{tab:lineprofile}}). $\rm H^{13}CO^{+}~(1\rightarrow0)$ has a high critical density ($\rm n_{crit}>10^4~cm^{-3}$; \citealt{bergin07,csengeri11,shirley15}), low $\rm E_u/k_B$ (4\,K), and no hyperfine multiplet. Its velocity dispersion is not affected by Galactic spiral arms
(e.g., the line profiles of CO, $\rm C^{18}O$, and $\rm C^{17}O$ along the line of sight are contributions from various Galactic arms; see \citealt{beuther07c}), and it has a symmetric Gaussian profile. Therefore, we use this line to measure the observed velocity dispersion in the line of sight $\rm \sigma_{obs}=\Delta \upsilon/\sqrt{(8\rm ln2)}$.  Here we assume that the averaged gas temperature in the entire clump is $\rm T_{30\,m}=T_{ATLAS}$. Comparing $\rm \sigma_{obs}$ with the thermally broadened velocity dispersion of  $\rm H^{13}CO^{+}$ $\rm \sigma_{th}=\sqrt{8\rm ln2k_BT_{30\,m}/(30\,m_H)}$, we estimate the non-thermally broadened velocity dispersion $\rm \sigma_{Nth}=\sqrt{\rm \sigma_{obs}^2-\sigma_{th}^2}$ (see Table~\ref{irdc:fragII}).  Within the gas temperature $\rm T_{30\,m}$ range of  12--18\,K, we find that $\rm \sigma_{obs}$  is  dominated by non-thermal motions throughout the entire  source clump (size of $\sim10^5$ AU). Therefore, assuming that turbulence dominates the non-thermal motion, we replace $\rm c_s$ with $\rm \sigma_{obs}$ in Eqs.~\ref{irdc:jean_l} and \ref{irdc:jean_m} and list the turbulent Jeans length $\rm \lambda_{Nth-J}$ and mass $\rm M_{Nth-J}$  in Table~\ref{irdc:fragII}:

 \begin{eqnarray}
\rm \lambda_{Nth-J}&=&\rm \sigma_{obs}(\frac{\pi}{G\mu m_Hn_{c}})^{1/2}  \nonumber\\
                             &=& \rm 0.35(\frac{\sigma_{obs}}{km\,s^{-1}})(\frac{n_{c}}{10^5\,cm^{-3}})^{-1/2}\,pc   \label{irdc:tur_l}     \\
\rm M_{Nth-J}&=&\rm \frac{\sigma_{obs}^3}{6}(\frac{\pi^5}{G^3\mu m_Hn_{c}})^{1/2}  \nonumber\\
                    &=&\rm 136.33(\frac{\sigma_{obs}}{km\,s^{-1}})^{3}(\frac{n_{c}}{10^5\,cm^{-3}})^{-1/2}\,M_{\odot}  \label{irdc:tur_m}
 .\end{eqnarray}
 
The above equations do not take magnetic fields into account. 
{
In such a regime, the non-thermal motions provide sufficient extra { support} in the cloud even at low temperatures, and may increase the fragmentation scale.
Both $\rm \lambda_{Nth-J}$ and $\rm M_{Nth-J}$ are higher than the values observed with SMA, and so this may be a viable method of support. Of course it should be noted that, if the turbulence lead to shocks occurring in the region, the increased density may lead instead to further fragmentation \citep{dobbs05}. Note that our data is insufficient to isolate multiple velocity components, given its spectral and spatial resolution. If these are present, the velocity dispersion, and therefore turbulent Jeans length/mass, would be overestimated.\\
}

The large-scale filamentary gas distribution at $870\,\mu$m and the small-scale SMA fragments in G28.34\,S and IRDC\,18308 are well aligned with roughly the same projected separation (Figure~\ref{irdc:fig:conti}). Similar feature has been reported in other HMSFRs, e.g.,  { 
G28.34\,S P1 \citep{zhang09,wangk11}, G11.11-0.12 \citep{wangk14}, and NGC\,7538\,S \citep{feng16c}. } Simulations (e.g., \citealp{chandrasekhar53,nagasawa87,bastien91,inutsuka92,fischera12}) also  predicted that isothermal, ``cylindrical" gas collapses into a chain of fragments, with equal spatial separation $\rm \lambda_{cl}$  along the filament under  self gravity, and then the fragments  grow due to { ``varicose" or ``sausage" fluid} instability \citep{jackson10}. 
Without  magnetic fields, the critical linear mass density $\rm (M/l)_{crit}$ represents the  linear mass over which self-gravity overcomes the internal thermal and non-thermal pressure. The critical separation $\rm \lambda_{cl}$ and critical mass of fragment $\rm M_{cl}$ in the large-scale clump  were originally given in \citet{chandrasekhar53,nagasawa87}. In the case that the thermal motions are not sufficient to be against gravitational energy, we replace the sound speed with  the velocity dispersion $\rm \sigma_{obs}$, and have \citep{fiege00,wangk14}:
\begin{eqnarray}
\rm (M/l)_{crit}&=&\rm 2\sigma_{obs}^2/G \nonumber\\
                         &=&\rm 465.02(\frac{\sigma_{obs}}{1\,km~s^{-1}})^2\,M_{\odot}pc^{-1}  \label{irdc:mcritical}\\
\rm \lambda_{cl}&=&\rm22\sigma_{obs}(4 \pi G\mu m_H n_c)^{-1/2} \nonumber\\
                          &=&\rm1.25(\frac{\sigma_{obs}}{1\,km~s^{-1}})(\frac{n_{c}}{10^5\,cm^{-3}})^{-1/2} \,pc\label{irdc:spac_cyl}\\
\rm M_{cl}&=&\rm(M/l)_{crit} \lambda_{cl} \nonumber\\
                  &=&\rm580.38(\frac{\sigma_{obs}}{1\,km~s^{-1}})^3(\frac{n_{c}}{10^5\,cm^{-3}})^{-1/2}  \,M_{\odot}\label{irdc:m_cyl} 
\end{eqnarray}

In Table~\ref{irdc:fragII}, we  list the critical mass and separation of each source estimated from cylindrical fragmentation. Non-thermal motions, such as  turbulence in this special case, can provide support against { gravitational collapse of the} mass up to $\rm \sim800\,M_{\odot}$. 
In particular, the  hypothesis  of cylindrical fragmentation predicts high linear mass densities  for both G28.34\,S ($\rm 788\,M_{\odot}pc^{-1}$) and IRDC\,18308 ($\rm 365\,M_{\odot}pc^{-1}$), which  are similar to that of IRDC G11.11-0.12 found in \citet{kainulainen13} and \citet{wangk14}.  
{
These predictions are upper limits, because possible unresolved velocity components may lead to an overestimation of the $\rm H^{13}CO^+$ velocity dispersion from the 30\,m observations (at the angular resolution of 30\arcsec).\\

To compare the predicted upper limit with the values from observations, we also measure the linear mass of the above ATLASGAL filamentary clumps $\rm(M/l)_{ATLAS}$ from their dust emission maps\footnote{ Since a 2-D Gaussian is not suitable for modeling a filamentary structure, we use the same algorithm that measures the total flux per unit projection size to measure the linear masses in rectangles, which are centered on the mapping center of 30 m observations and individually have lengths that are one, two, or three times  the local FWHM width, and then derive the mean. } at the angular resolution of 18\arcsec. We find that if G28.34\,S is warmer than 18 K ($\rm 855\pm244\,M_{\odot}pc^{-1}$) and IRDC 18308 is warmer than 15 K ($\rm 265\pm23\,M_{\odot}pc^{-1}$, Table~\ref{irdc:fragII}), the predicted upper limits are higher than the observed linear mass of the ATLASGAL sources. If magnetic field were taken into account, the predicted linear mass would be even higher (e.g., \citealp{kirk15}),  i.e.,  both clumps are likely stable against cylindrical fragmentation. \\
\color{black}}

\item {\bf Other possibilities:\\}
{
We do not find obvious velocity gradients in our sources, so the rotation energy is negligible compared to the gravitational energy\footnote{Even if the velocity distribution of our source is biased by our field of view, rotation energy is in general not comparable to the gravitational energy \citep{goodman93,bihr15}.}. Nevertheless, other factors in addition to non-thermal motions may also lead to a ``mass/separation discrepancy''  between the observations and thermal Jeans predications, where the projected separation between the fragments are comparable to the Jeans length, but the observed mass is higher than the Jeans mass. The possible factors are the following:
 \\
 }

(1) Magnetic fields may suppress fragmentation (e.g., \citealp{commercon11}). \\
(2) An inhomogeneous density profile {  in the core may alter its fragmentation properties.} For example, recent numerical simulations by \citet{girichidis11} show that less fragmentation occurs in steeper initial density profiles, and the final fragments are more massive than from a flat initial density profile (by a factor of up to 20 in the extreme case). \\
(3) The ``precise'' Jeans length is difficult to quantify. Since the sensitivity of the SMA observations does not detect lower mass fragments, the separation between the fragments is an upper limit of the Jeans length. { Conversely, the projection of a} 3-D distribution onto the the plane of the sky reduces the observed fragment separation, which makes the measured length a lower limit of  the Jeans length. It is difficult to assess quantitatively how these two factors affect the measured separation between dust peaks. These uncertainties make the Jeans length comparison problematic. \\
(4) As the density increases, the local Jeans mass decreases, cores can  fragment further. Therefore, based on the current data, we cannot rule out the possibility of unresolved internal sub-fragments (condensations).\\

\end{itemize}

In short, the mass of the initial fragments resolved from 1.1\,mm SMA observation are already high (on average $\rm >10\,M_{\odot}$). 
Thermal energy alone cannot  compete against the self-gravity of these fragments; however, the combination of thermal and non-thermal motions can  provide sufficient energy in opposing the further fragmentation. Moreover, a steep initial density profile of the entire clump and the magnetic field  may also increase the upper limit of the final fragment mass. With the current data, it is difficult to measure the precise Jeans length and resolve internal substructures in the fragments. Therefore, we cannot  differentiate between the above scenarios.\\


\subsection{Infall motion}\label{sec:infall}

\begin{figure}
\centering
\includegraphics[width=8cm]{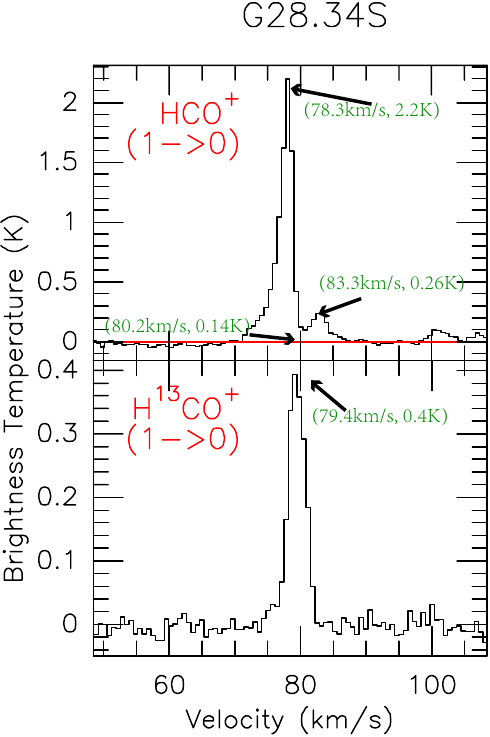}
\caption[]{Asymmetric line profile of $\rm HCO^+(1\rightarrow0)$ compared to $\rm H^{13}CO^+(1\rightarrow0)$ in G28.34\,S, indicating infall motion with a speed of $\rm \sim 0.84\,km~s^{-1}$.
\label{irdc:h13co+}}
\end{figure}

We find that the line profiles of  $\rm HCO^+~(1\rightarrow0)$, which is likely optically thick, shows an asymmetric non-Gaussian profile, especially in G28.34\,S. In contrast,  $\rm H^{13}CO^+~(1\rightarrow0)$, which is optically thin, shows a fairly symmetric profile  (Figure~\ref{irdc:h13co+}). Following \citet{mardones97}, we measure the normalized velocity difference between the two lines $\rm \delta V_{H^{13}CO^+}=(V_{HCO^+}-V_{H^{13}CO^+})/\Delta \upsilon_{H^{13}CO^+}$ in each source. Except for IRDC\,18308, 
$\rm  \delta V_{H^{13}CO^+}$  in other sources are significant\footnote{Following the method in  \citet{mardones97}, we  set the threshold of ``significance'' as 5 times  the uncertainty of $\rm  \delta V_{H^{13}CO^+}$.}   (Table~\ref{irdc:ATLAS}), indicative of violent dynamics. 
In particular, line profile of  $\rm HCO^+~(1\rightarrow0)$ in G28.34\,S shows double intensity peaks of blue- and redshifted gas, indicating a potential  infall motion, with an infall speed $\sim \rm 0.84\,km~s^{-1}$ (derived by using Eq. 9 of \citealt{myers96}).
Nevertheless,  we note that the peak intensity of the redshifted gas is only 0.1 as strong as that of  the blueshifted gas, so we cannot rule out the possibility that double intensity peaks comes from multiple velocity components. If this is the case, the high-velocity (redshifted) gas of  $\rm H^{13}CO^+~(1\rightarrow0)$ may be too weak for  the observation sensitivity.
\\

\subsection{Molecular column density and abundance estimates}\label{sec:col}
We did not detect emissions of the higher-J  molecular transitions  in the fragments from the SMA observations. However, the lower-J transitions of the dense gas tracers are detected in the extended structures in the 30\,m line survey. This indicates that our sources are in the cold, young phase. To probe the chemical properties in these  gas clumps, we quantify their excitation conditions and molecular abundances  in this section.

\subsubsection{Molecular excitation temperatures and column densities}\label{chemipro}
We calculated the  molecular column densities averaged from the central 40\arcsec$\times$40\arcsec~region of each source. Since the optical depth may lead to large uncertainties of certain lines, we use different methods to correct their opacities. 
\begin{itemize}
\item {\bf $\rm \bf N_2H^+$, $\rm \bf C_2H$, HCN, and $\rm \bf H^{13}CN$} have multiplets
in their $J=1\rightarrow0$ or $\rm N=1\rightarrow0$ ($\rm C_2H$)  transitions.  We use the  HFS method 
 to fit these transitions (Figure~\ref{irdc:hyperfit}) and list their excitation temperatures $\rm T_{ex}$ in Table~\ref{irdc:tex} (see Appendix \ref{appen:hfs} for details). \\

From the fits, we find that the linewidths of the strongest components $\rm  \Delta \upsilon$ of the above species are in the range of $\rm  1.7\text{--}3.2~km\,s^{-1}$. This is consistent with the linewidths of the other species (Table~\ref{tab:lineprofile}), indicating that non-thermal motions dominate the velocity dispersion. 
$\rm N_2H^+~(J=1\rightarrow0)$ and $\rm C_2H^+~(N=1\rightarrow0)$ are fitted very well, but the fits for $\rm HCN~(J=1\rightarrow0)$ and $\rm H^{13}CN~(J=1\rightarrow0)$ show strong deviations. Most of the $\rm T_{ex}$ we derived from the HFS fits are below 10\,K (except for HCN in G28.34 S), which could be due to multiple factors:\\

(1) In the cold gas environment with densities of $\rm 10^4\text{--}10^5 \,cm^{-3}$, species with high critical densities are subthermally excited (e.g.,  at 10 K, the critical density of $\rm N_2H^+~(J=1\rightarrow0)$ is $\rm \sim6\times10^4~cm^{-3}$, and that of $\rm HCN~(J=1\rightarrow0)$ is $\rm \sim5\times10^5~cm^{-3}$; \citealt{shirley15}), so their $\rm T_{ex}$ is lower than the typical kinetic temperatures of IRDCs (e.g., \citealt{sridharan05,vasyunin11,fontani12}). This feature is  common in both high-mass and low-mass dark clouds (e.g., \citealt{caselli02a,crapsi05,miettinen11,fontani12}).\\

(2) To increase the signal-to-noise ratio for the weak emission, we extract the average spectra from the 40\arcsec$\times$40\arcsec~region, but the gas is not homogeneous as assumed. For one thing,  $\rm H^{13}CN$ is deficient at the continuum peak of  IRDC\,18530 and IRDC\,18308 (Figure~\ref{irdc:molint}). In addition, $\rm HCN~(J=1\rightarrow0)$ has an anomalous line profile in IRDC\,18306  (the HFS component with the strongest relative intensity shows weaker emission than the others).  This ``anomalousness'' has also been noted by \citet{afonso98} and was explained as  the inconsistency intensity ratios between multiplets of HCN, i.e., individual hyperfine lines may be enhanced or suppressed in dark clouds because of non-LTE \citep{loughnane12}. Therefore, the low $\rm T_{ex}$ we derive here are the averaged excitation temperatures from the  imperfectly fitted spectrum (Figure~\ref{irdc:hyperfit}.III).\\

(3) The HFS method is applicable when the source is in ideal LTE (assumption  {\scriptsize\encircle{1}} -{\scriptsize\encircle{3}}  in Appendix \ref{appen:hfs}) and multiplets are resolved (assumption {\scriptsize\encircle{4}}).  However, in our observations HCN and $\rm H^{13}CN$ may be non-LTE, and seven multiplet lines of $\rm N_2H^+$ are blended into three Gaussian peaks  at a velocity resolution of $\rm\sim0.641~km\,s^{-1}$ (Figure~\ref{irdc:hyperfit}.I). Therefore, the HFS method only gives the rough level of the $\rm T_{ex}$. \\

(4) Another factor which would affect the $\rm T_{ex}$ estimation is the filling factor  {\it f}, which is estimated by assuming that the species emission extent is equal to the area within the contour of half maximum integrated intensity  (Eq. \ref{irdc:sourcesize}--\ref{irdc:hyper_tex}). In most cases, the emission extends over the 30\,m primary beam, so {$f\rm \sim1$}. For species whose equivalent distribution diameter is smaller than the primary beam could have a {$f\rm \ll1$} (e.g., $\rm H^{13}CN$ in IRDC\,18306 and IRDC\,18308 have {$f<0.5$}). Moreover, the asymmetric spatial distribution of emission would also bring uncertainty (i.e., the equivalent distribution diameter depends on the velocity range we integrate and the edge of the emission extent we define).
 On average, $\rm T_{ex}$ is underestimated for {$f\rm \sim1$} and  will increase by a factor of 1.2--1.5 for {$f\sim \rm0.2$}.\\

Following the ``CTex"  approach introduced by \citet{caselli02b}, 
where $\rm N_2H^+$ with $\rm T_{ex}\sim5\,K$, we assume that the transitions of each molecule can be described by an approximately  constant  $\rm T_{ex}$. Then we estimate the column densities of the above species\footnote{A more precise approach is to use the non-LTE radiative transfer code RADEX. However, 
it does not significantly improve our results, when we do not know the source structure. 
} with Eq.~\ref{irdc:hyper_NT}.\\

\onecolumn
 \begin{figure*}
\begin{center}
\begin{tabular}{ll}
I. $\rm N_2H^+~(J=1\rightarrow0)$\\
\includegraphics[width=8cm]{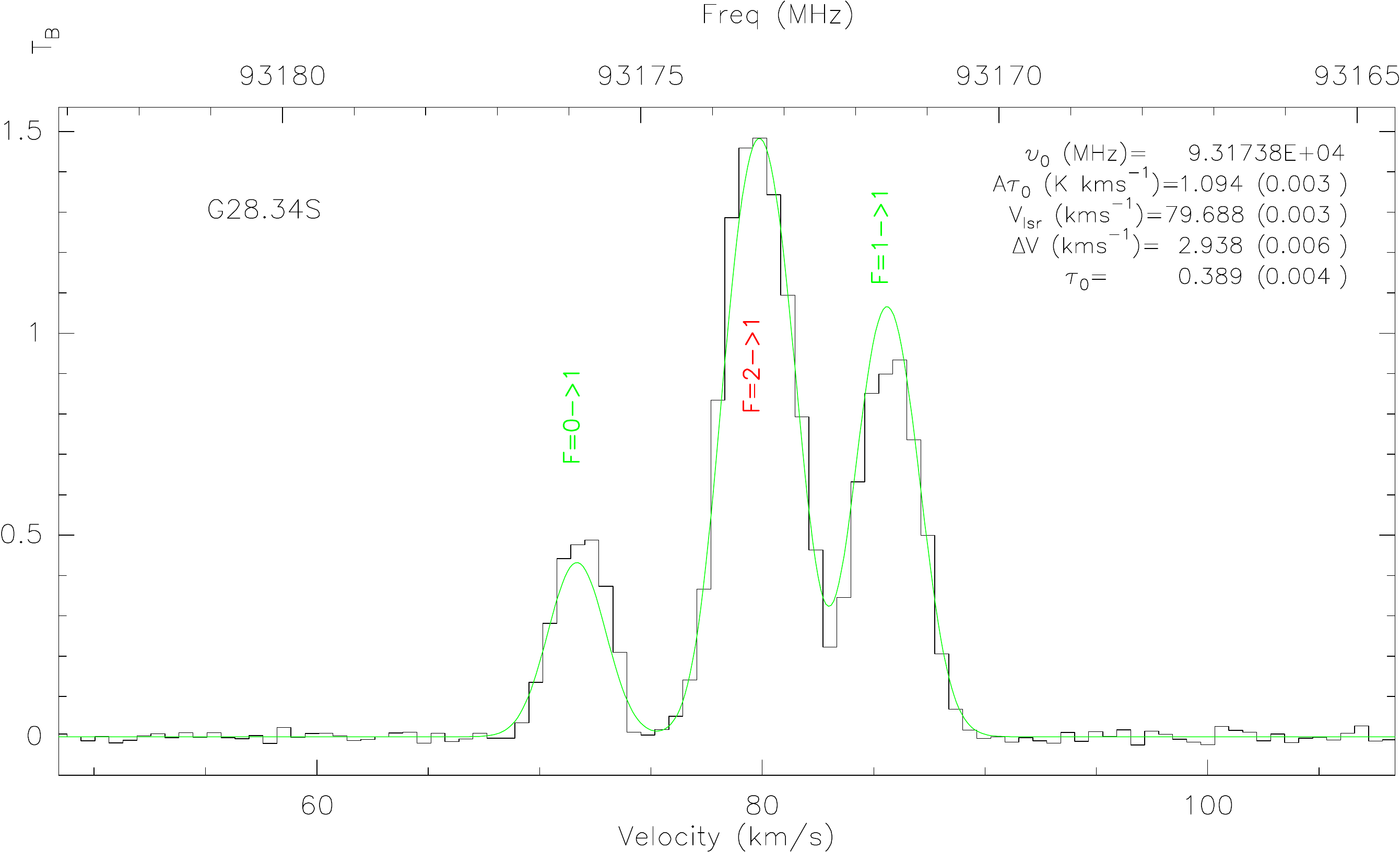}
&\includegraphics[width=8cm]{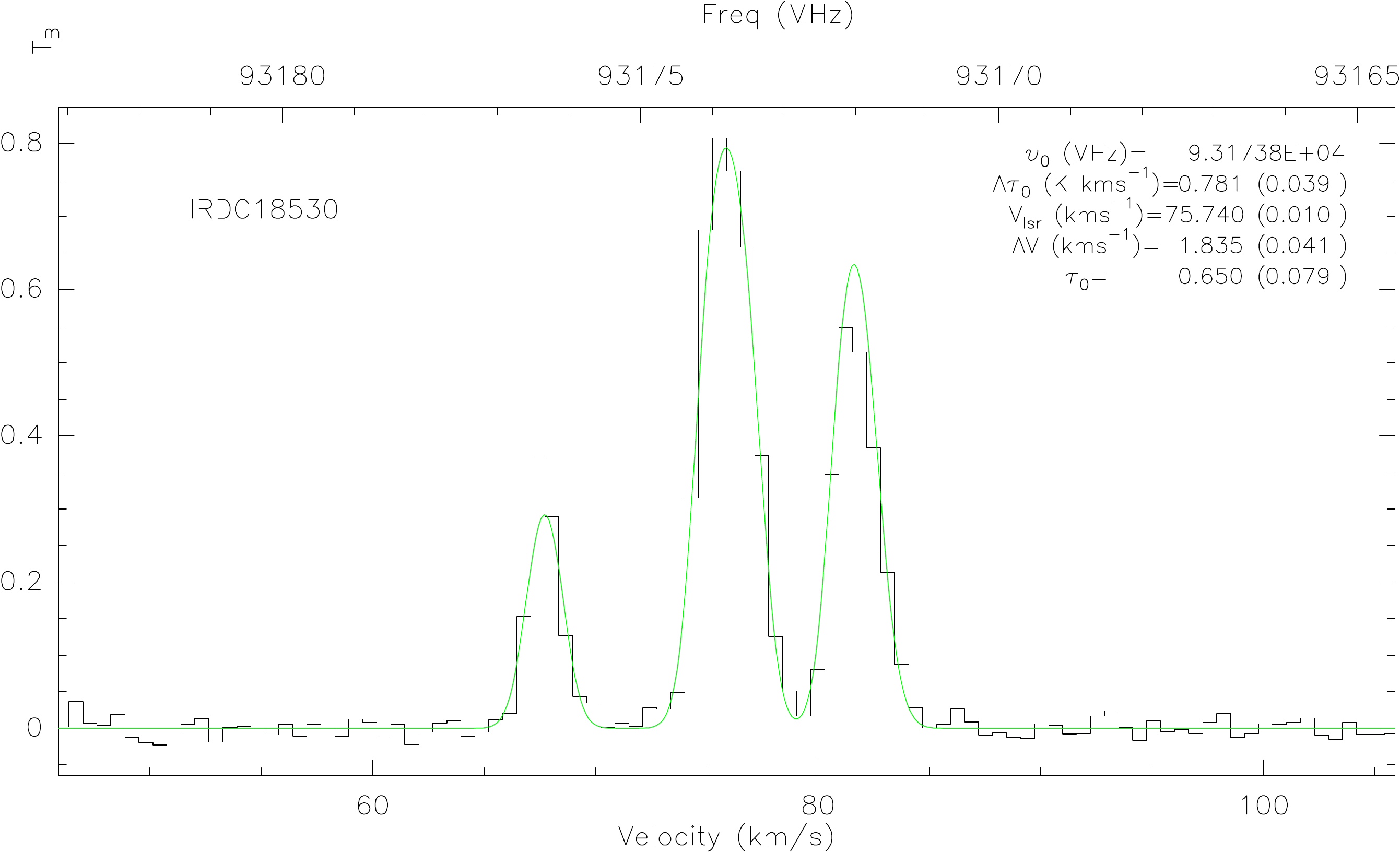}\\
\includegraphics[width=8cm]{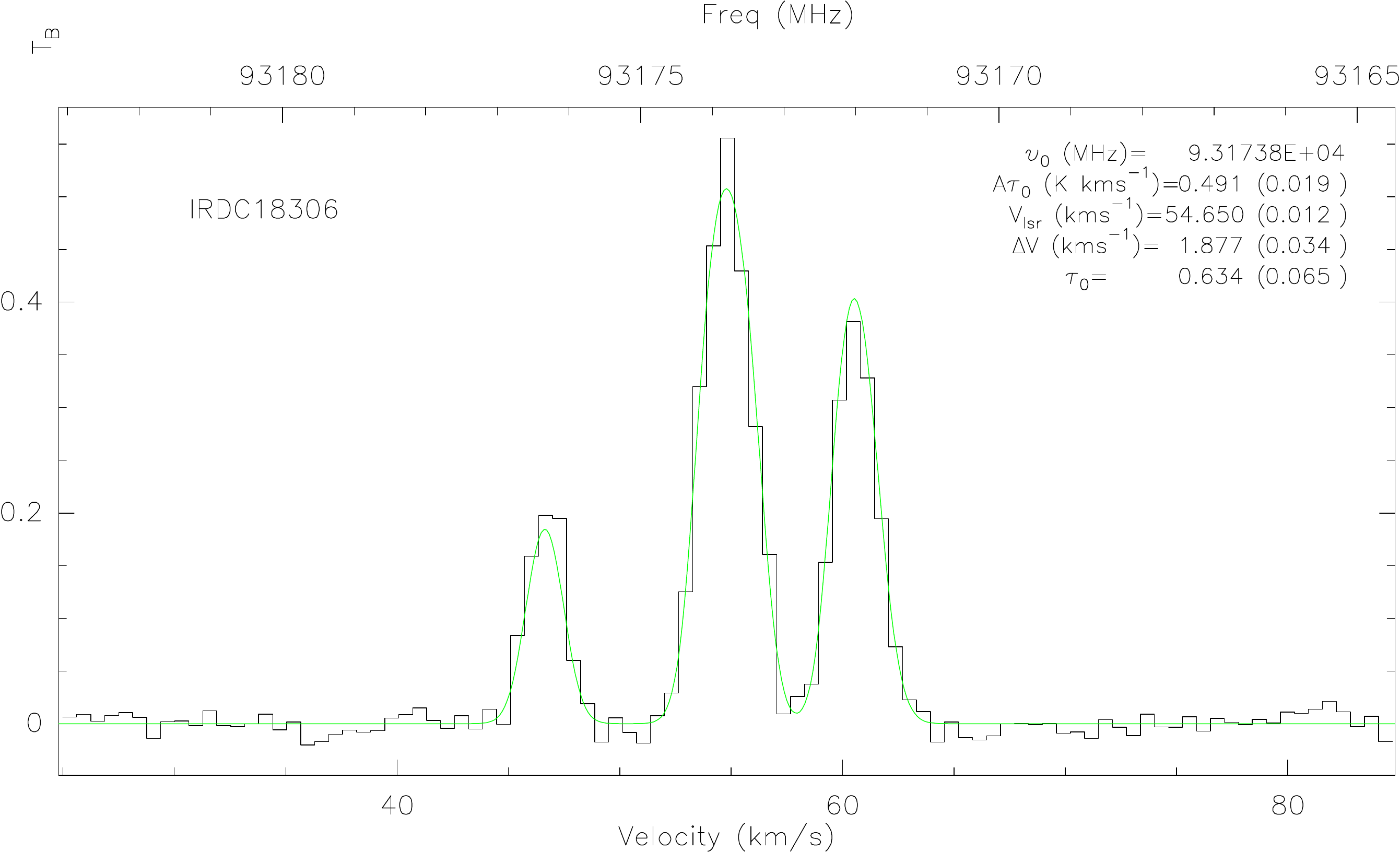}
&\includegraphics[width=8cm]{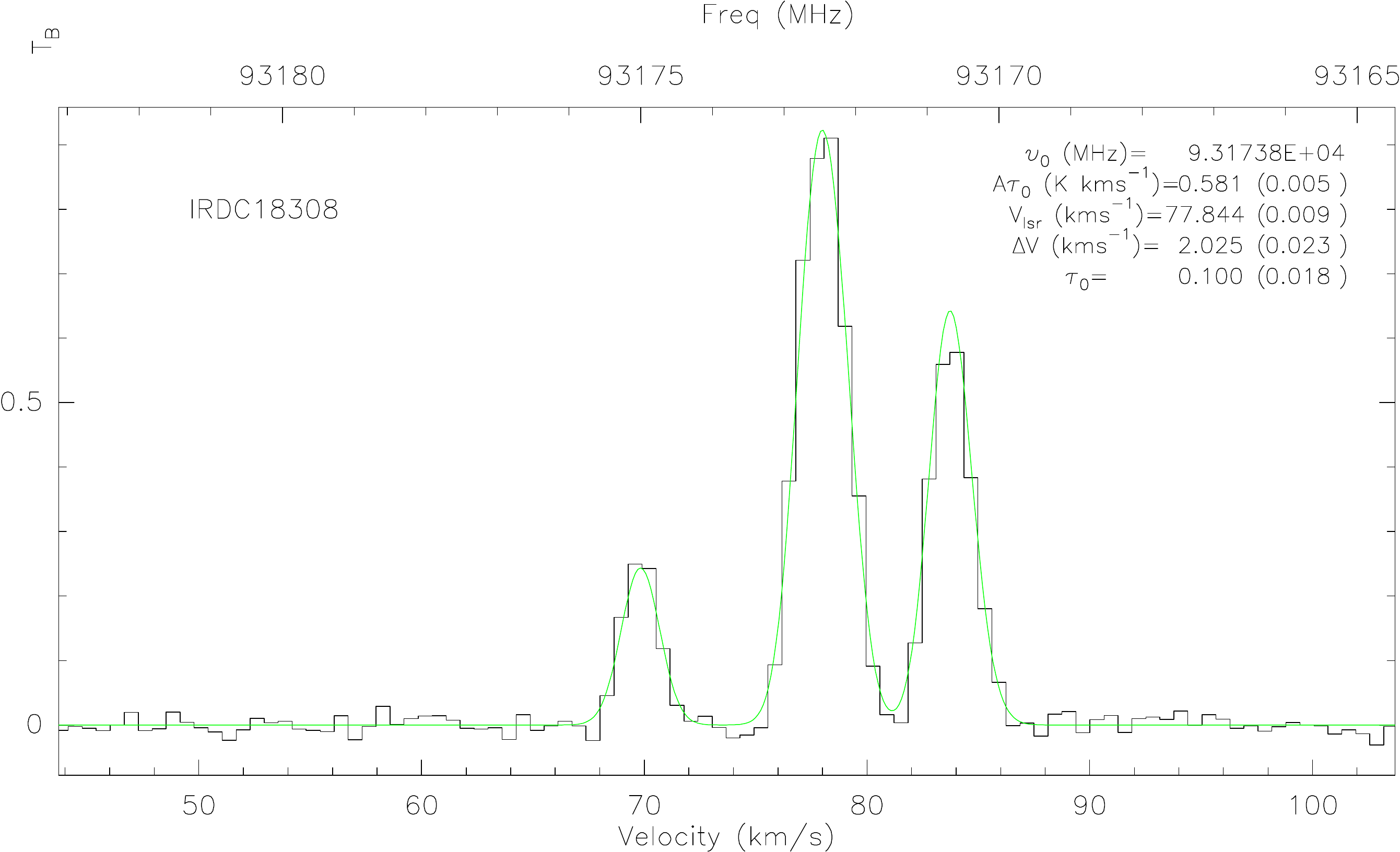}\\
\multicolumn{2}{l}{*. Hyperfine multiplet is fitted using seven components, which are not resolvable in our data, so we label  the three  strongest hyperfine lines. }
\end{tabular}

\begin{tabular}{ll}
II. $\rm C_2H~(N=1\rightarrow0)$\\
\includegraphics[width=8cm]{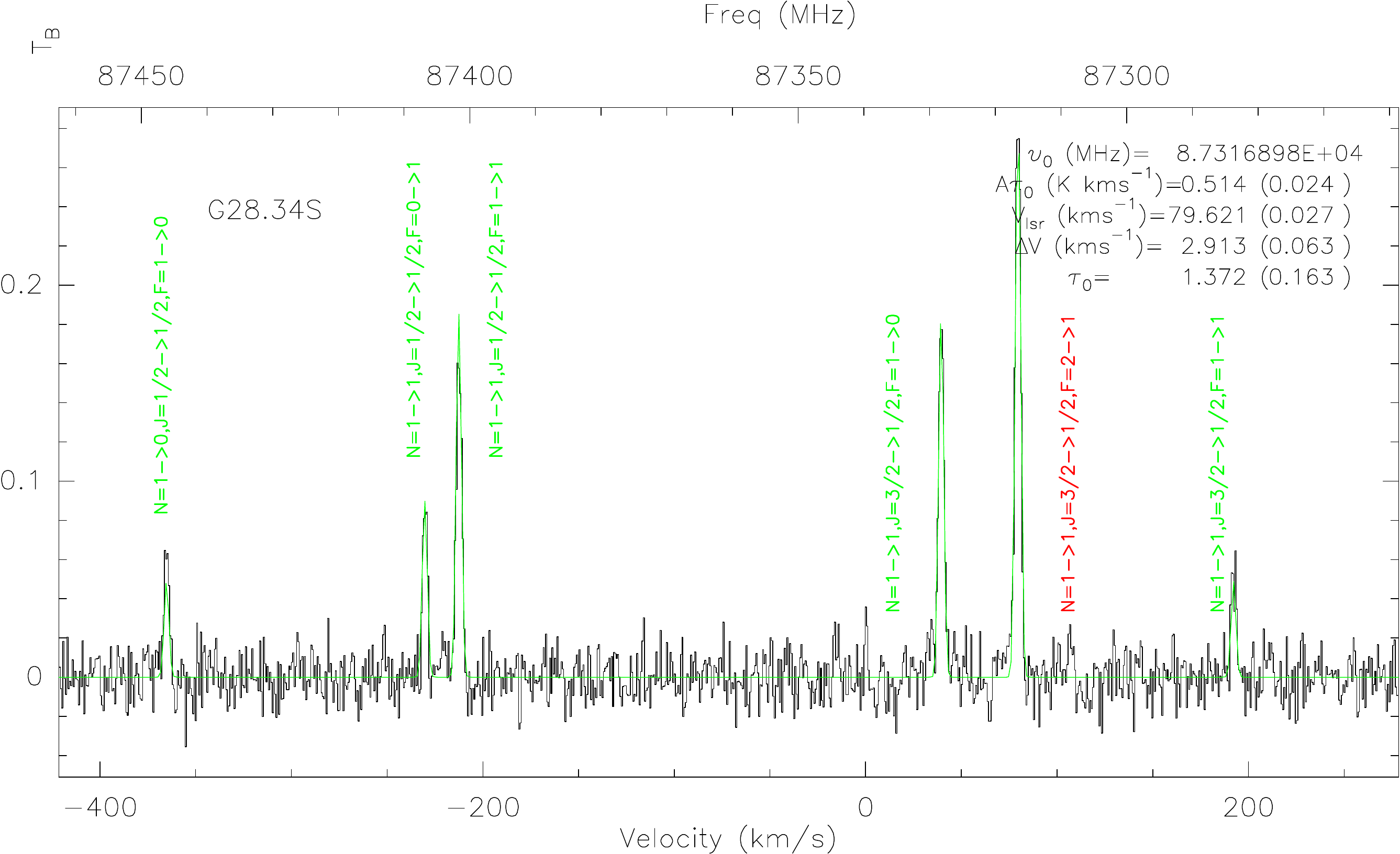}
&\includegraphics[width=8cm]{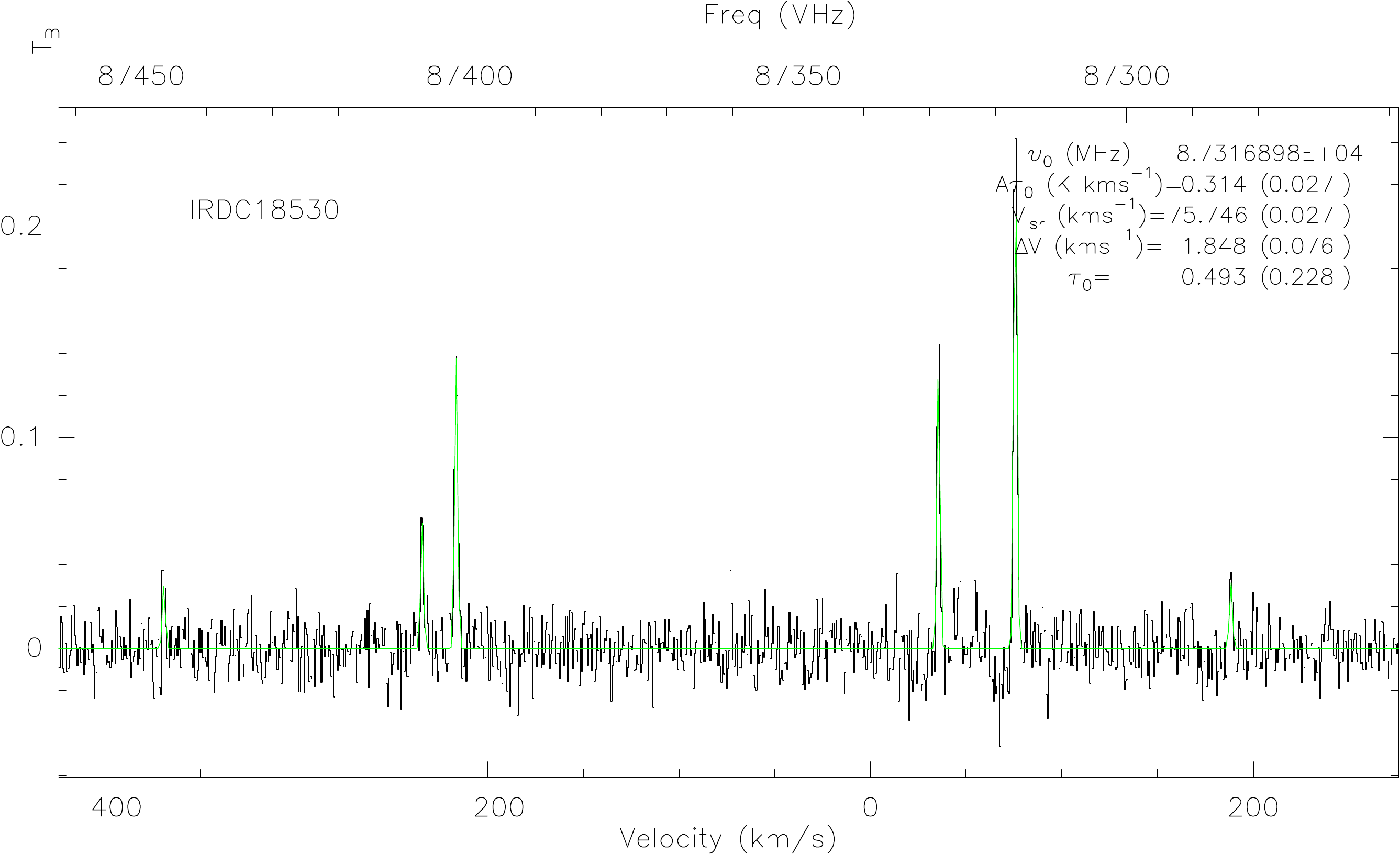}\\
\includegraphics[width=8cm]{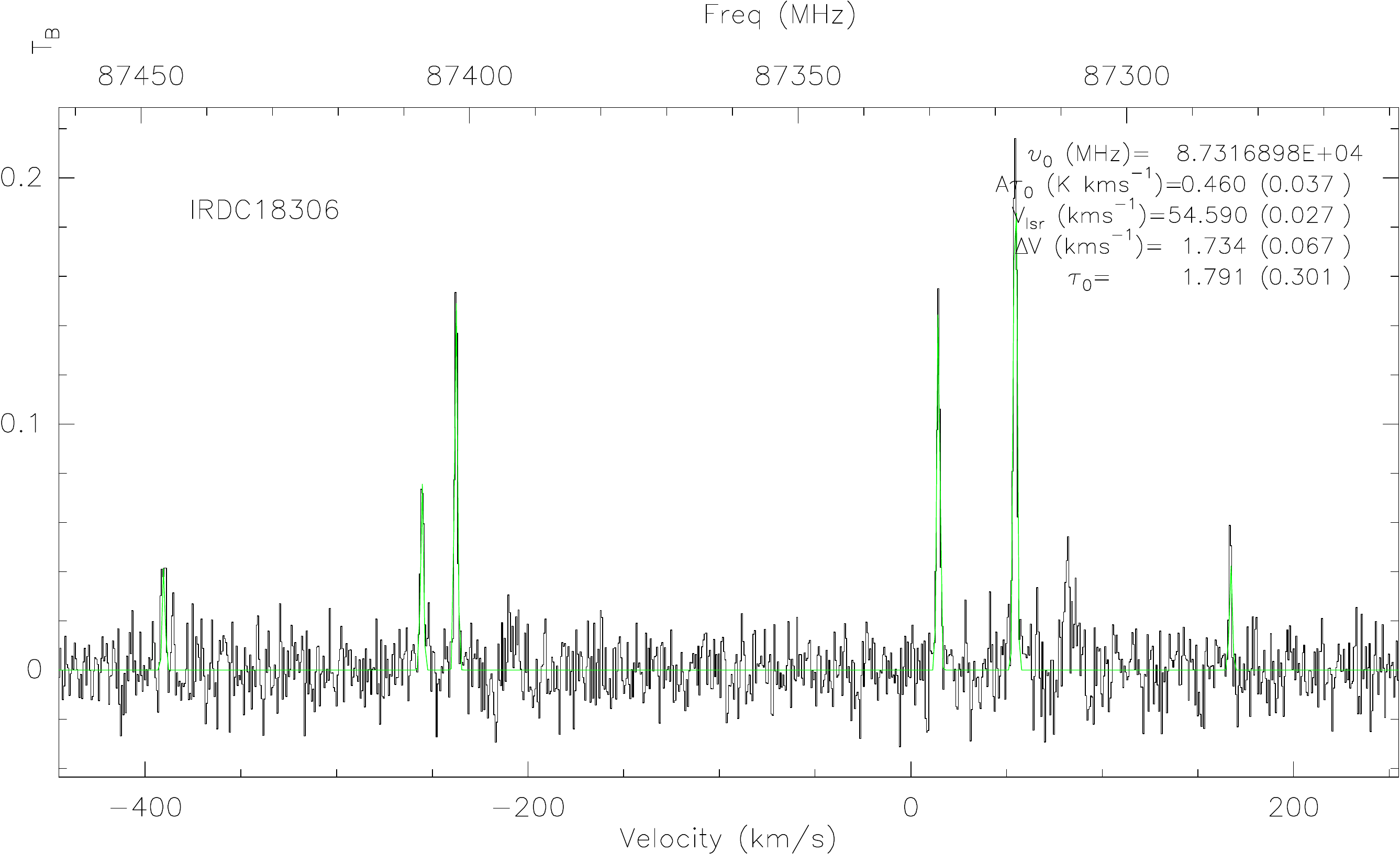}
&\includegraphics[width=8cm]{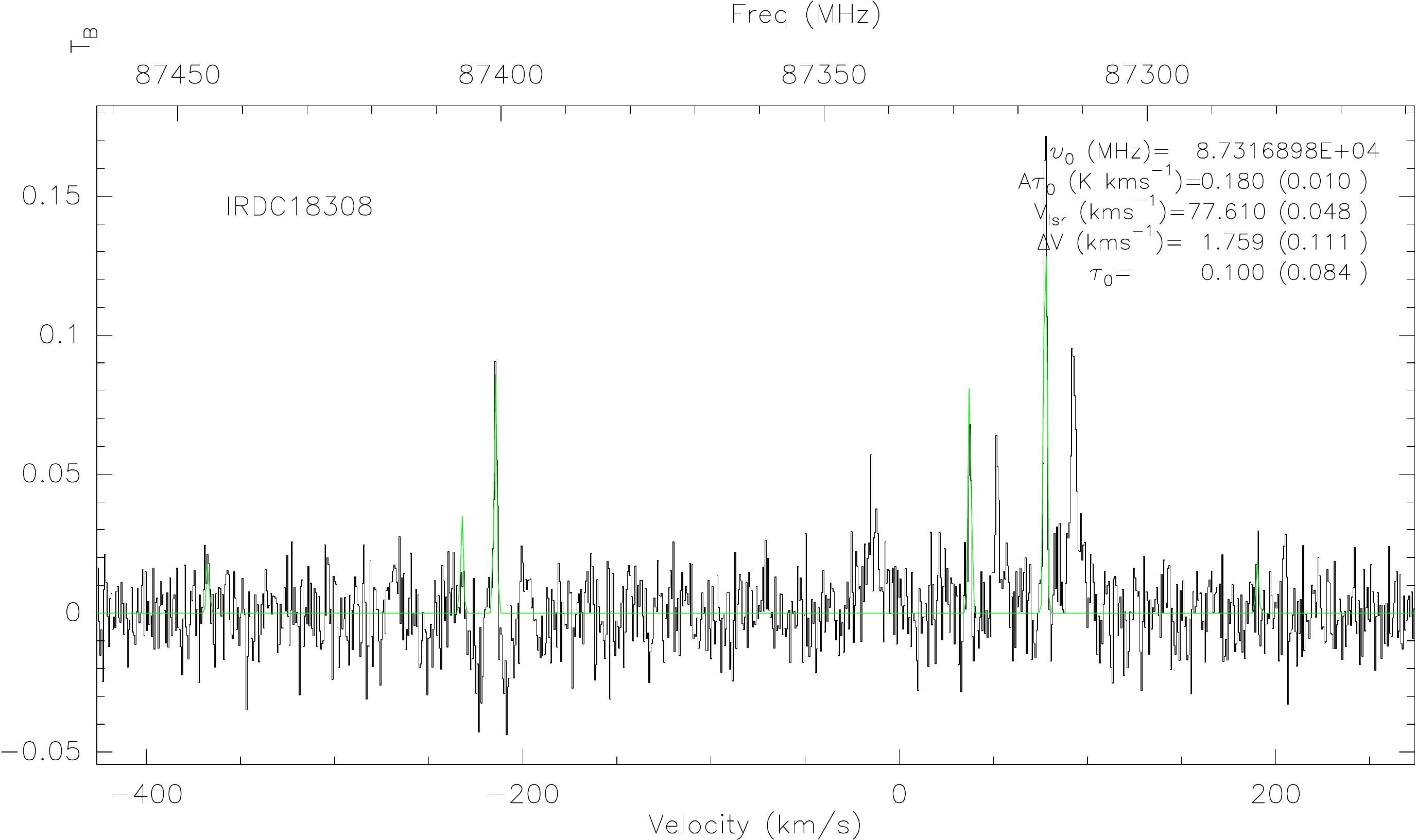}
\end{tabular}

\end{center}
\caption[hyperfine multiplet fits to the observed $\rm N_2H^+~(J=1\rightarrow0)$, $\rm C_2H~(N=1\rightarrow0)$, $\rm HCN~(J=1\rightarrow0)$, and $\rm H^{13}CN~(J=1\rightarrow0)$ from G28.34\,S, IRDC\,18530, IRDC\,18306 and IRDC 18308]{Hyperfine multiplet fits (green lines) to the observed $\rm N_2H^+~(J=1\rightarrow0)$, $\rm C_2H~(N=1\rightarrow0)$, $\rm HCN~(J=1\rightarrow0),$ and $\rm H^{13}CN~(J=1\rightarrow0)$ (black histograms) from each IRDC;  the fit parameters are in the right upper corner. The transitions of multiplets are labeled in the panels of G28.34\,S (with the strongest line in red).}\label{irdc:hyperfit}
\end{figure*}

\begin{figure*}
\ContinuedFloat
\begin{center}
\begin{tabular}{ll}
III. $\rm HCN~(J=1\rightarrow0)$\\
\includegraphics[width=8cm]{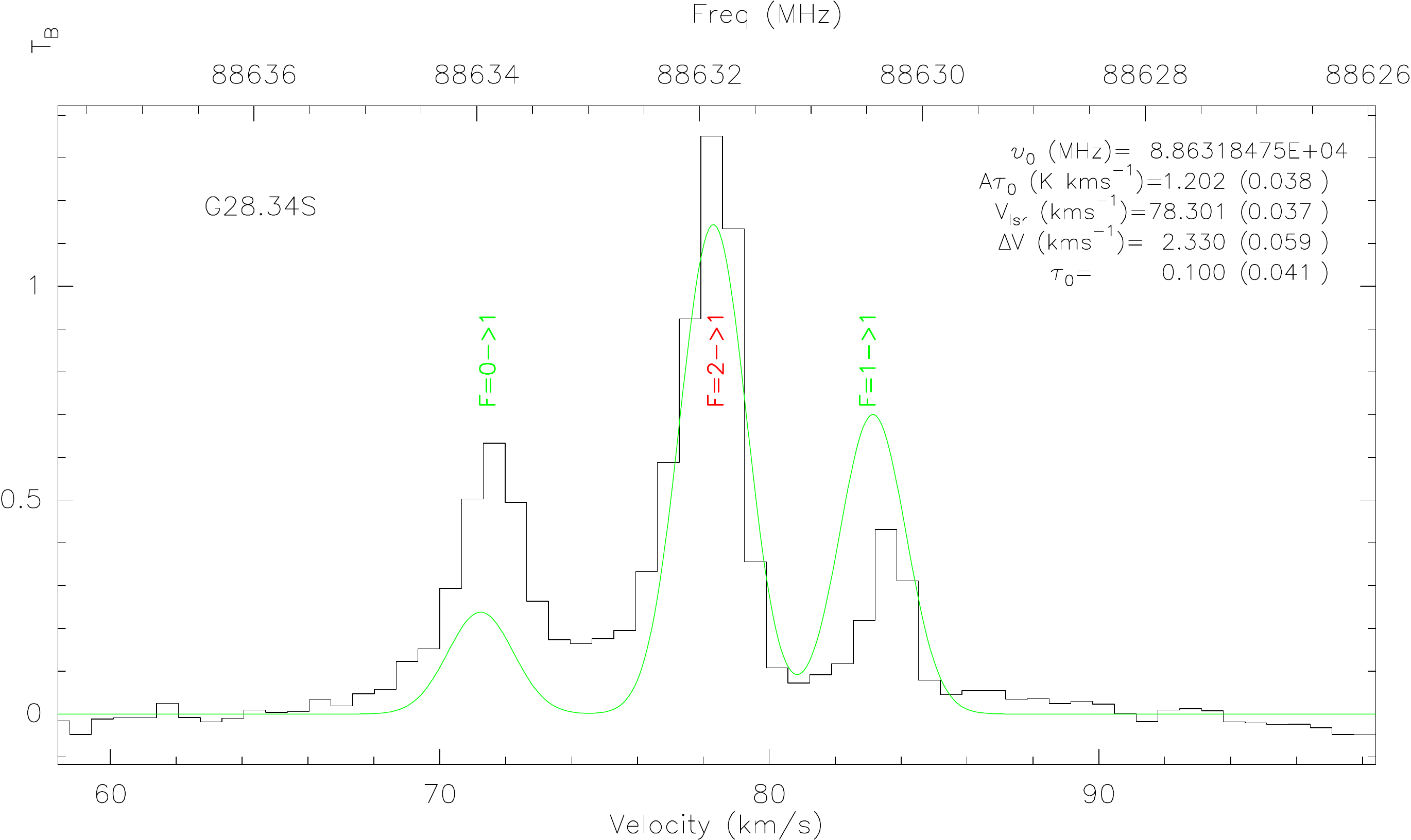}
&\includegraphics[width=8cm]{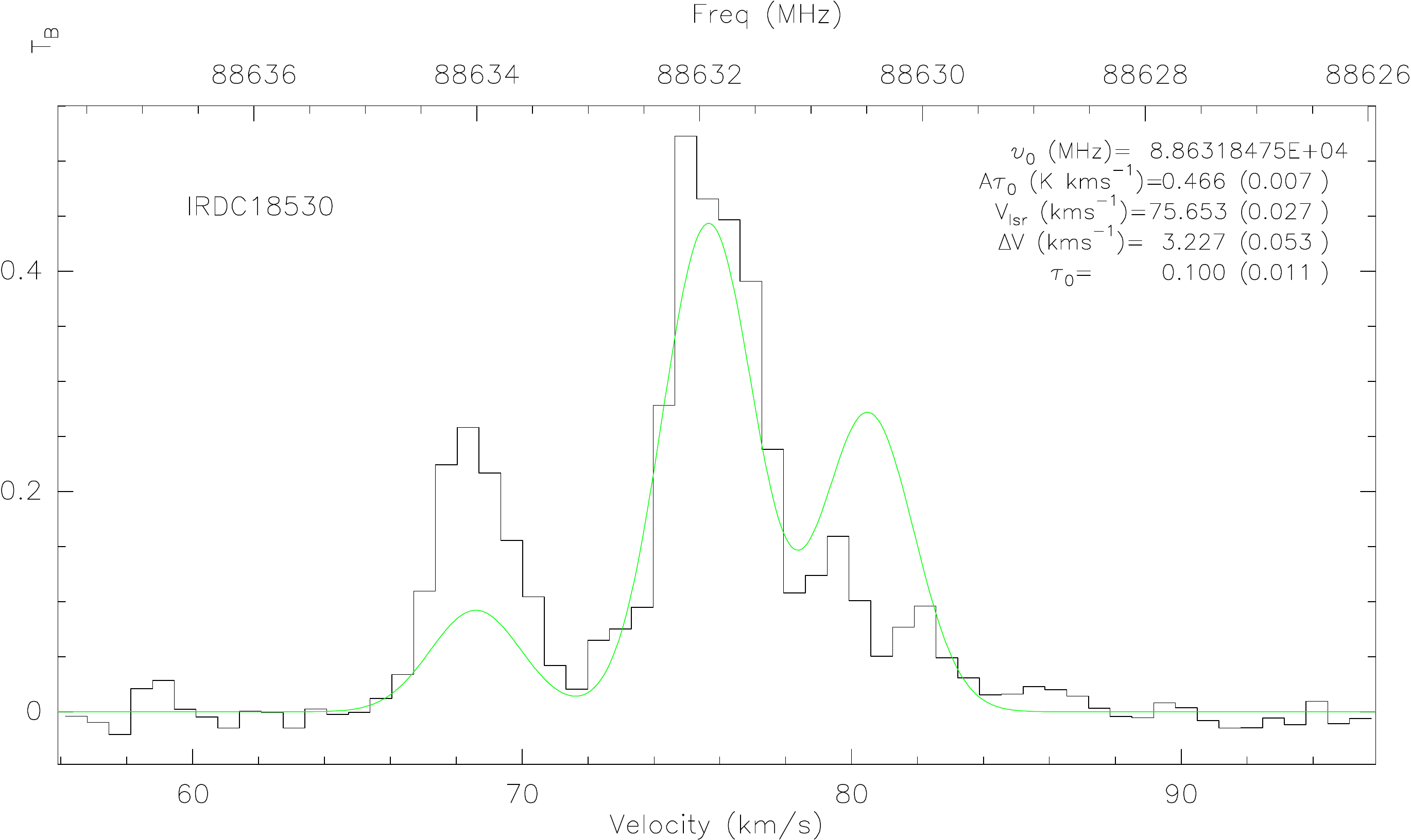}\\
\includegraphics[width=8cm]{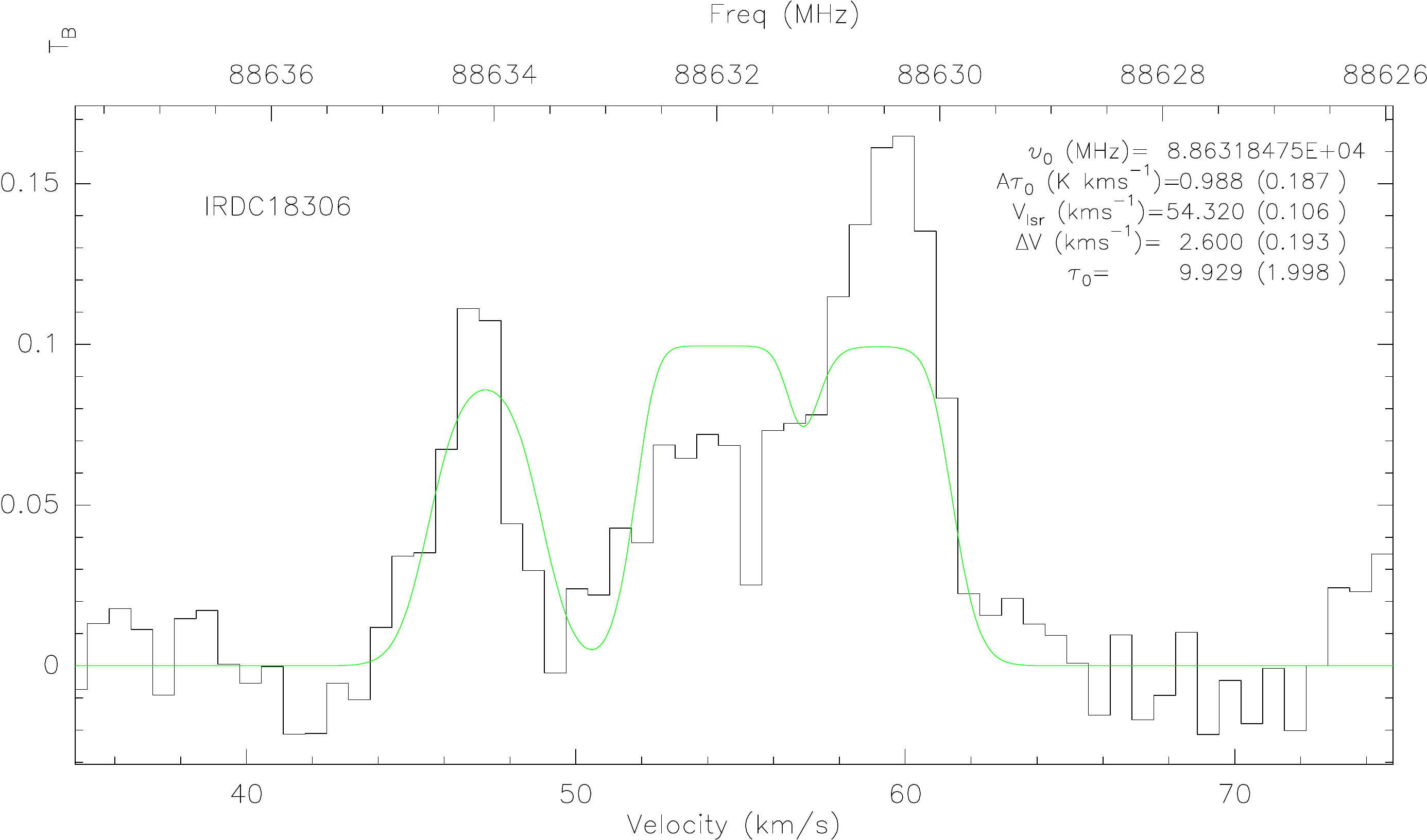}
&\includegraphics[width=8cm]{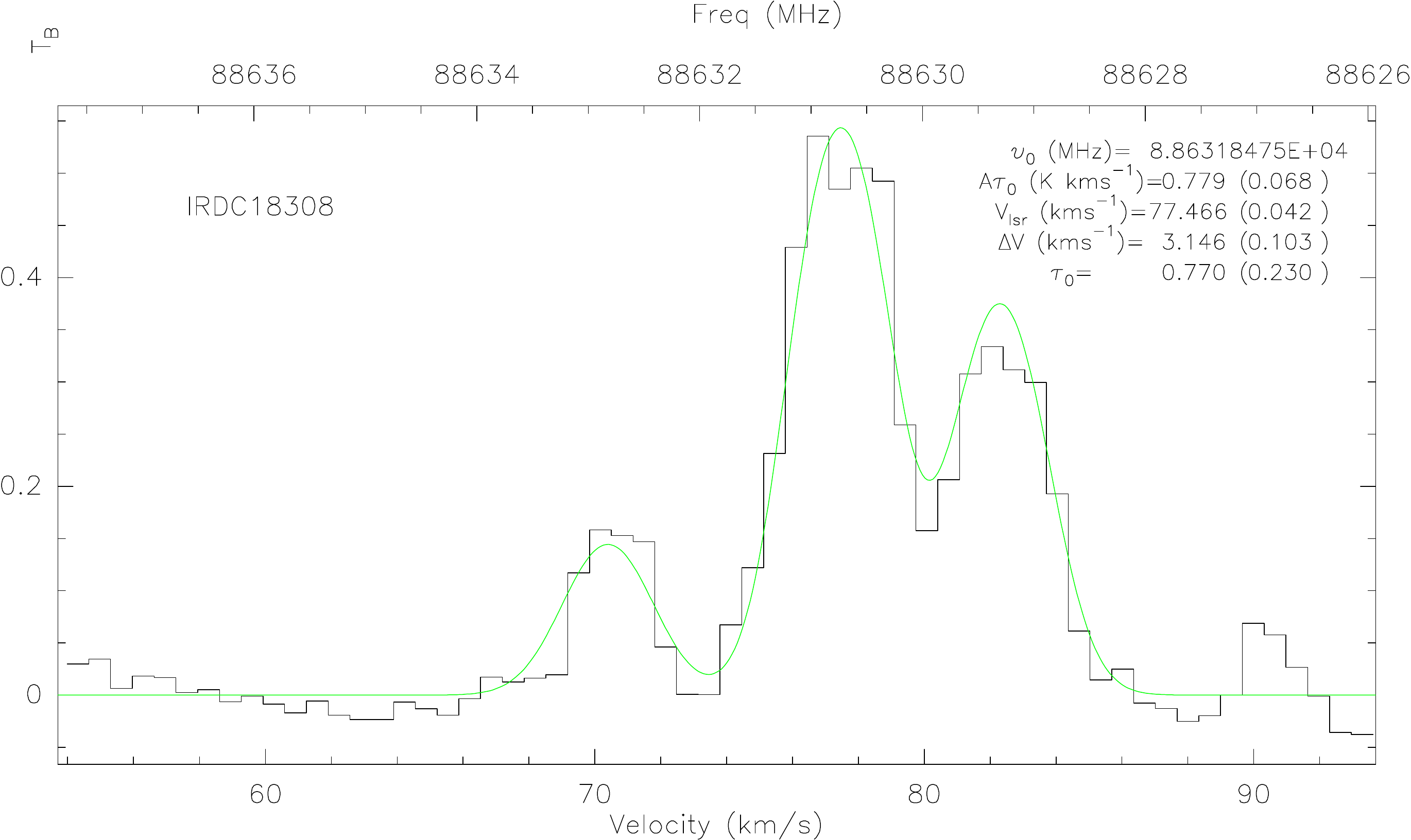}
\end{tabular}

\begin{tabular}{ll}
IV. $\rm H^{13}CN~(J=1\rightarrow0)$\\
\includegraphics[width=8cm]{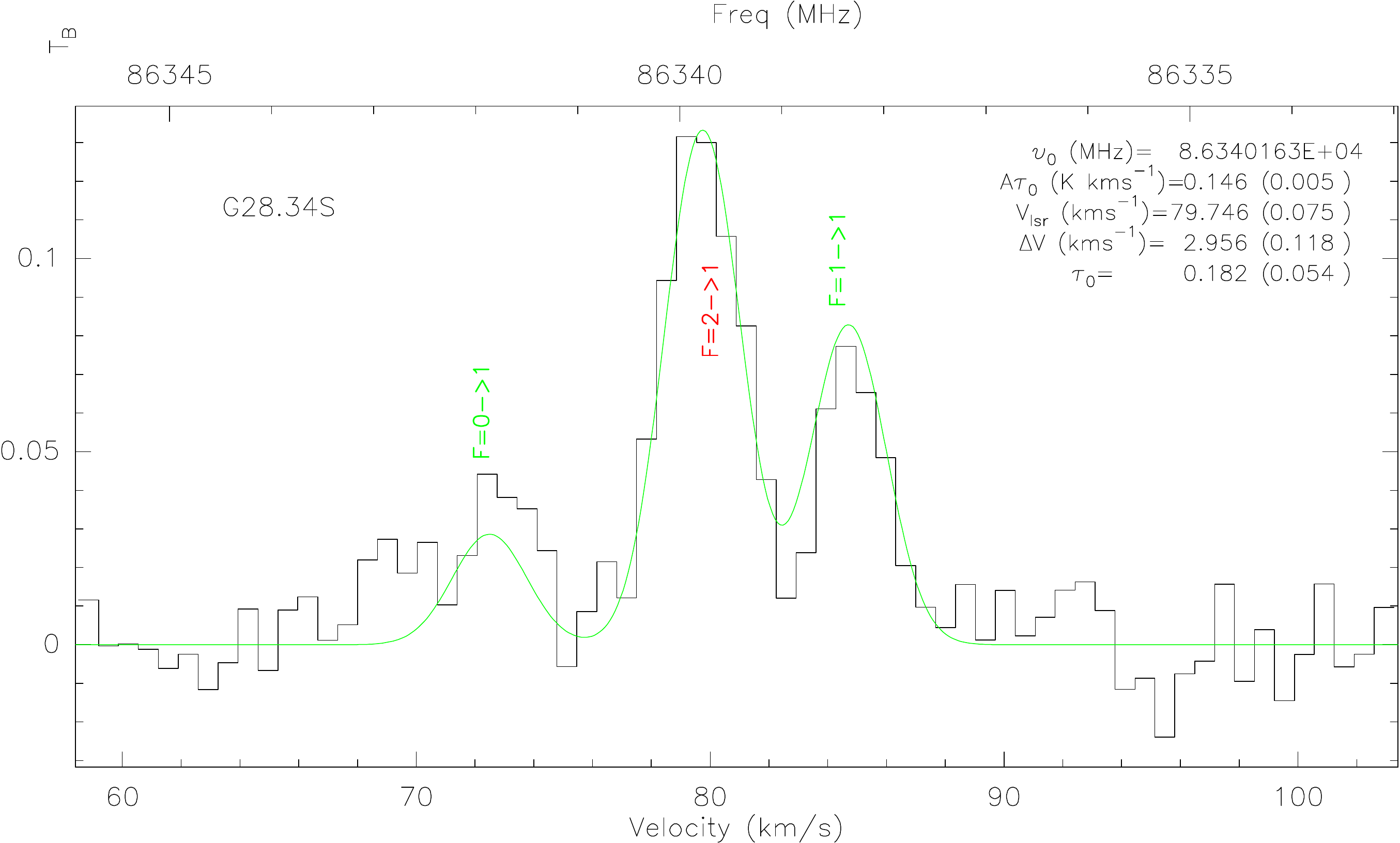}
&\includegraphics[width=8cm]{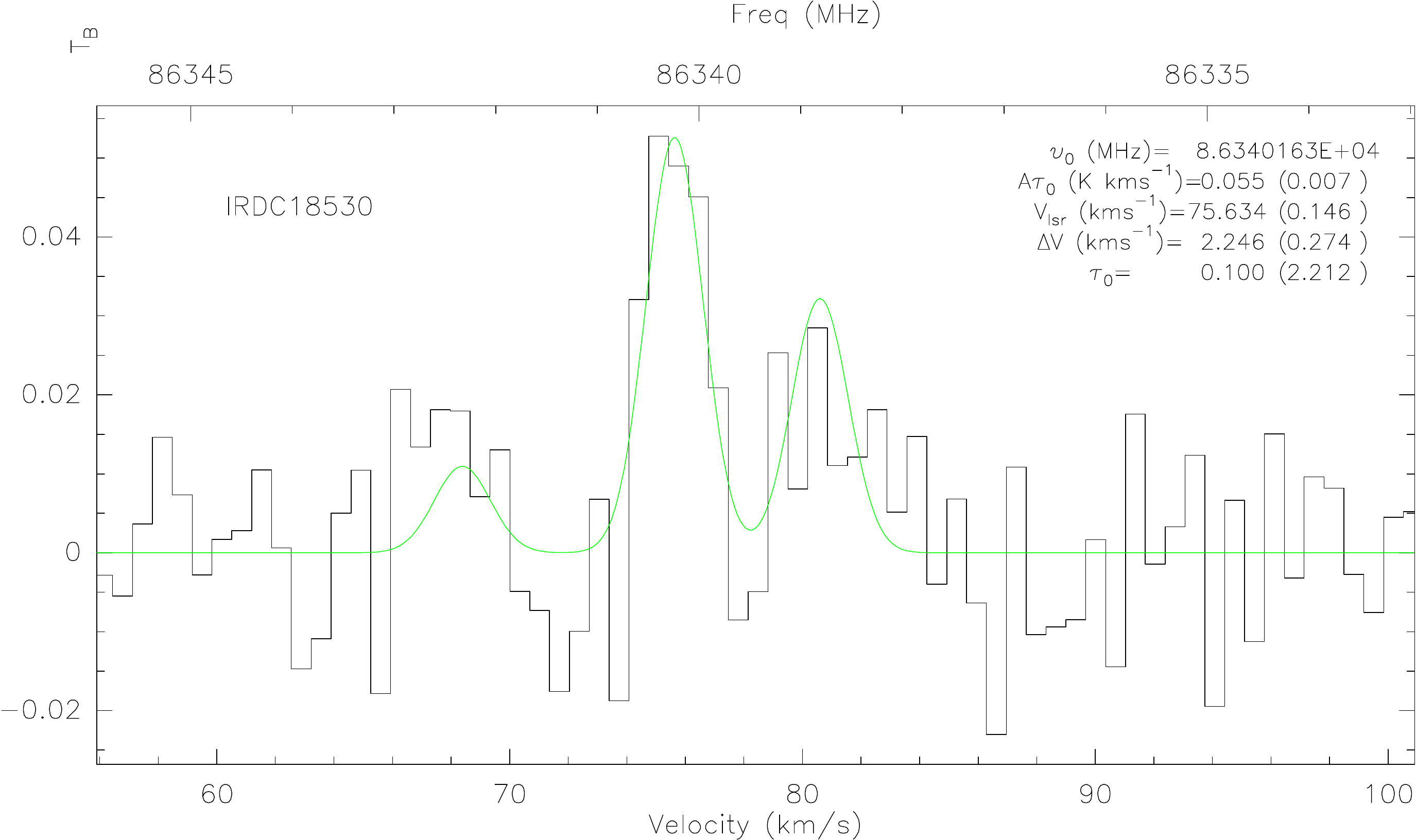}\\
\includegraphics[width=8cm]{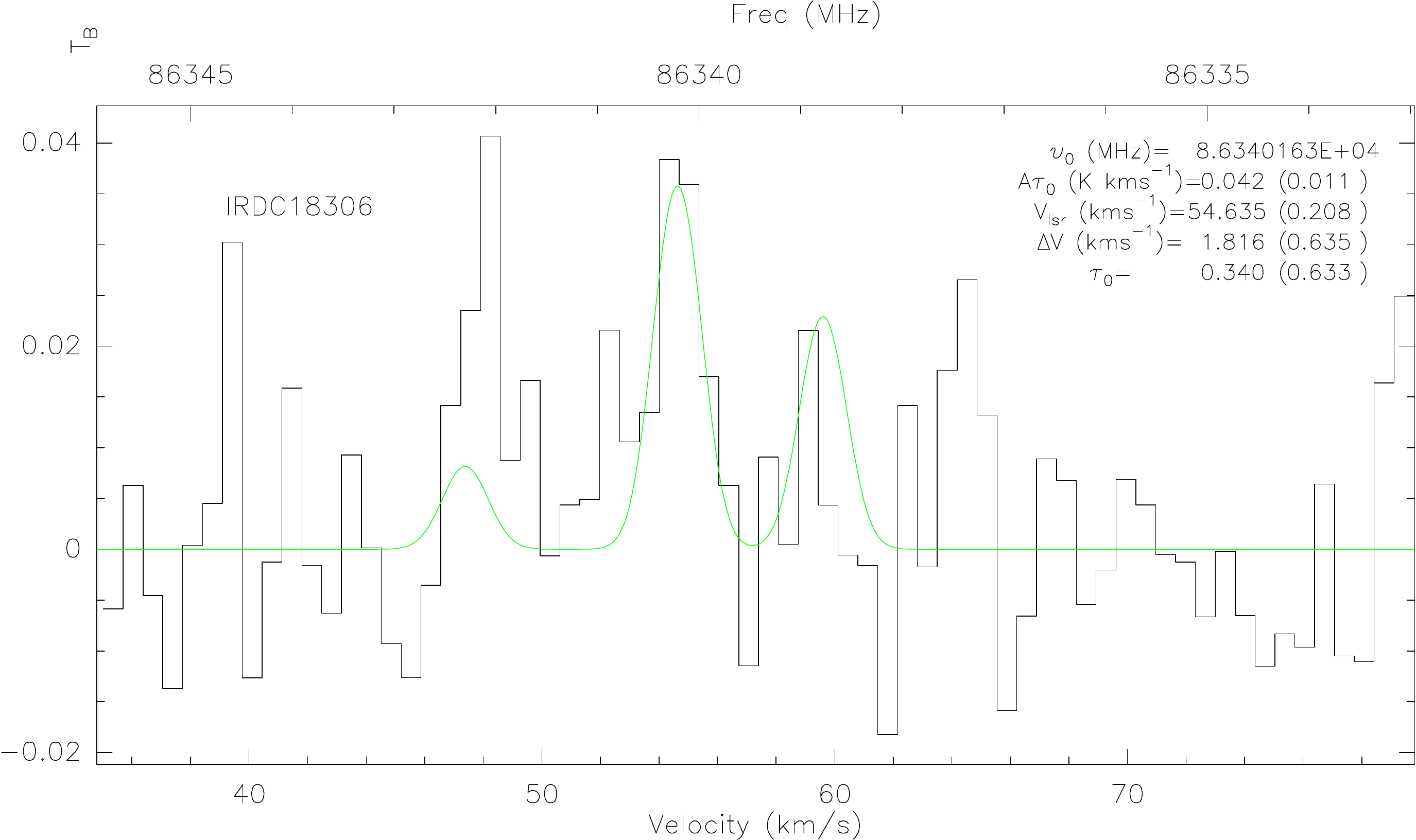}
&\includegraphics[width=8cm]{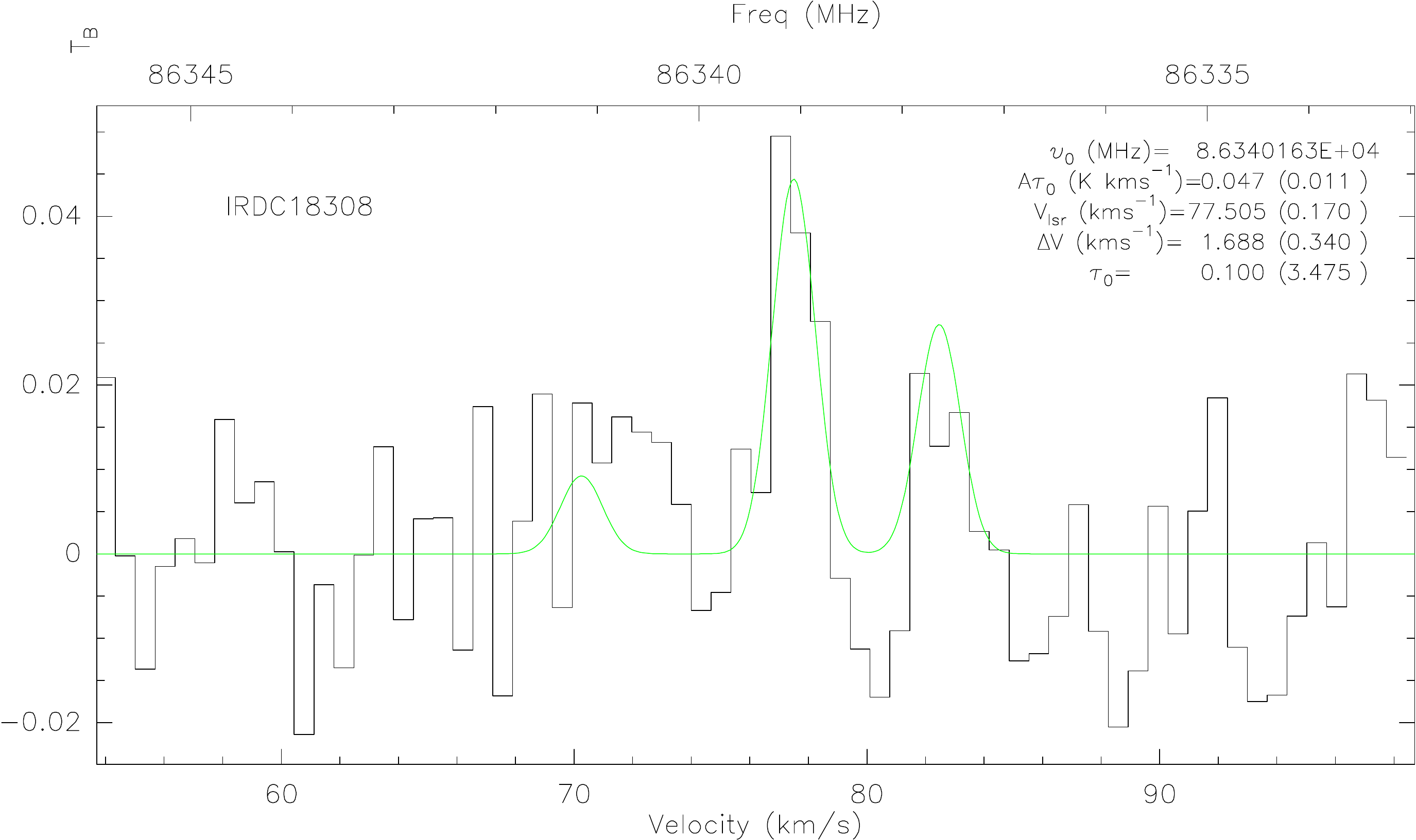}
\end{tabular}

\end{center}
\caption{(continued)}
\end{figure*} 

\twocolumn

\begin{table*}
\small
\begin{center}
\caption{$\rm T_{ex}$ derived from hyperfine fits to $\rm N_2H^+$, $\rm C_2H$, HCN, and $\rm H^{13}CN$ multiplets in four IRDCs}\label{irdc:tex}
\begin{tabular}{cp{2.5cm}p{2.5cm}p{2.5cm}p{2.5cm}}\hline\hline

Species                          & G28.34\,S      &IRDC\,18530$^a$                  &IRDC\,18306$^{a,b}$   &IRDC\,18308$^{a,b}$ \\
                                       & (K)      & (K)                             & (K)                & (K) \\
\hline
$\rm N_2H^+~(J=1\rightarrow0)$      &$\rm 5.81\pm0.04$    &$\rm 4.09\pm0.21$    &$\rm 3.62\pm0.11$    &$\rm 8.91\pm0.94$\\
$\rm C_2H~(N=1\rightarrow0)$       &$\rm 3.14\pm0.06$    &$\rm 3.26\pm0.17$    &$\rm 3.01\pm0.06$    &$\rm 4.68\pm0.98$\\
$\rm HCN~(J=1\rightarrow0)$     &$\rm 15.10\pm3.81$    &$\rm 7.70\pm0.55$    &$\rm 2.82\pm0.04^*$    &$\rm 3.87\pm0.34$\\
$\rm H^{13}CN~(J=1\rightarrow0)$     &$\rm 3.63\pm0.23$    &$\rm 3.35\pm0.62^*$    &$\rm 3.55\pm0.62^{*,\dag}$    &$\rm 5.17\pm2.42^{*,\dag}$\\

\hline\hline
\multicolumn{5}{l}{{\bf Note.} {\it a}.``*'' denotes the fits having large deviation from the observation.}\\
\multicolumn{5}{l}{~~~~~~~~  {\it b}.``$\dag$'' denotes species with filling factor $f\rm <0.5$.}\\
\end{tabular}
\end{center}
\end{table*}

\item {\bf Molecules without hyperfine multiplet}\\
CO  isotopologues are important gas tracers because of their  low dipole moments and thermalization near their critical densities (at 10--20 K, $\rm \sim10^3~cm^{-3}$ for $\rm ^{13}CO~(2\rightarrow1)$, $\rm \sim10^4~cm^{-3}$ for $\rm C^{18}O~(2\rightarrow1)$ and  $\rm C^{17}O~(2\rightarrow1)$). Therefore, they are in LTE in our dense gas clumps ($\rm n=10^4\text{--}10^5~cm^{-3}$), and their $\rm T_{ex}$ can be taken as the $\rm T_{kin}$ of the gas.\\

Lower-J transitions of the CO isotopologues (e.g., $\rm ^{13}CO~(2\rightarrow1)$)  may be optically thick, so a precise estimation of its $\rm T_{ex}$ needs opacity correction.  Assuming that fractionation (isotopic exchange reaction) of $\rm ^{12}C\Leftrightarrow {^{13}C}$ and of  $\rm {^{17}O} \Leftrightarrow  {^{16}O}\Leftrightarrow {^{18}O}$ are stable, 
we correct the optical depth $\tau$  and estimate $\rm T_{ex}$ of  CO isotopologue ($1\rightarrow0$) lines  (listed in Table~\ref{irdc:cotex}, see Appendix \ref{appen:cotemp} for details).\\

\begin{table*}
\small
\begin{center}
\caption[Optical depth $\tau$ and $\rm T_{ex}$ derived for CO ($1\rightarrow0$) isotopologues on four IRDCs]{Optical depth $\tau$ and $\rm T_{ex}$ derived for CO ($1\rightarrow0$) isotopologues, according to Eq.~\ref{eq:tau} and \ref{irdc:abun13co}-\ref{irdc:ratioab}.}\label{irdc:cotex}
\begin{tabular}{c|cc|cc|cc|cc}\hline\hline
Line   &\multicolumn{2}{c|}{G28.34\,S}   &\multicolumn{2}{c|}{IRDC\,18530}   &\multicolumn{2}{c|}{IRDC\,18306}    &\multicolumn{2}{c}{IRDC\,18308}  \\
\hline
   &$\tau$   &\footnotesize  $\rm T_{ex}~{(K)}$   &$\tau$   &\footnotesize  $\rm T_{ex}~{(K)}$   &$\tau$   &\footnotesize  $\rm T_{ex}~{(K)}$   &$\tau$   &\footnotesize  $\rm T_{ex}~{(K)}$  \\
$\rm C^{18}O \,(2\rightarrow1)^{\it a}$  &0.25   &14   &--   &--   &0.65   &8   &--   &--\\
$\rm ^{13}CO \,(2\rightarrow1)^{\it a}$   &2.6   &11   &--   &--   &5.8   &8   &0.9   &13\\
$\rm ^{13}CO \,(2\rightarrow1)^{\it b}$   &2.8   &11   &2.9   &9   &6.2   &7   &2.2   &11\\

\hline\hline
\multicolumn{9}{l}{{\bf Note.} {\it a}. Derived from  $\rm  C^{17}O \,(2\rightarrow1)$, which is assumed to be optically thin.}\\
\multicolumn{9}{l}{~~~~~~~~~~{\it b}. Derived from $\rm  C^{18}O\, (2\rightarrow1)$, which is assumed to be optically thin.}
\end{tabular}
\end{center}
\end{table*}

For G28.34\,S and IRDC\,18306, $\rm C^{18}O~(2\rightarrow1)$  is optically thin according to our optical depth measurements from  $\rm C^{17}O~(2\rightarrow1)$ (Table~\ref{irdc:cotex}). Since we did not observe $\rm C^{17}O~(1\rightarrow0)$ in IRDC\,18530, and the intensity of this line is only $\sim3\sigma$ rms  in IRDC\,18308, we take the optically thin $\rm C^{18}O~(2\rightarrow1)$ as a standard to estimate the other chemical parameters. \\

The values of  
$\rm T_{ex}$ derived from CO isotopologues are consistent in G28.34S (11--14\, K), indicating that CO is in LTE. This temperature range is also  consistent with the values derived from  VLA+Effelsberg  $\rm NH_3$ observations, the SABOCA 350 $\mu$m survey, and  SPIRE 500\,$\mu$m observations { (mentioned in Section \ref{fragmentation})}.  Most of the other lines  without hyperfine multiplets have low critical density (except for $\rm SiO~(2\rightarrow1)$), so  we assume they are coupled with dust and thermalized at $\rm T_{gas}= T_{dust}= T_{kin}\sim T_{ex,CO}$. The column densities of these species can be derived according to Appendix \ref{appen:colother}. In particular, the column densities of species whose rare isotopologue lines we detected ($\rm  HCO^+$, HNC, and $\rm ^{13}CO$)  are derived after opacity corrections.

\end{itemize}
Molecular column densities estimated using the above approaches are listed in Tables~\ref{irdc:column-HFS}--\ref{irdc:column-other}.

\subsubsection{ $\rm H_2$ column density estimates}
It is possible to estimate $\rm H_2$ column density at  clump-scale   from Eq.~ \ref{irdc:atlasgas}.
  At the temperature of $\rm T_{ATLAS}=T_{ex,CO}\sim10\,K$, average $\rm H_2$ column density measured from the ATLASGAL  in the $\rm 40\arcsec\times 40\arcsec$ region is on the order of $\rm 10^{23} \,cm^{-2}$ ($\rm 1g \,cm^{-2}$, Table~\ref{irdc:ATLAS}), which is consistent with  the threshold for high-mass star-forming clouds proposed by \citet{krumholz08}.

\begin{table*}
\small
\caption{CO depletion, HNC/HCN ratio, lower limit of ionization, and velocity asymmetric degree of $\rm H^{13}CO^+$ in four IRDCs }\label{irdc:ATLAS}
\begin{center}
\scalebox{0.85}{
\begin{tabular}{c|ccc|ccc|ccc|ccc}\hline\hline

Source            &{\it l}           & {\it b}     & $\rm {\it D}_{GC}^{\it a}$           & $\rm I_{0.87mm}$  &$\rm T_g$    &$\rm N_{H_2}$                      &\multicolumn{3}{c|}{$\mathscr{D}\rm_{CO}^{\it b}$}  &$\rm X(\frac{HNC}{HCN})^{\it c}$          &$x(e)^e$     &$\delta V_{\rm H^{13}CO^+}^{\it f}$\\
                      &$(^o)$     &$(^o)$   & (kpc)                             & (Jy/beam)         &(K)         &$\rm 10^{22}~(cm^{-2})$                 &$\rm C^{18}O$ &$\rm ^{13}CO$ &$\rm C^{17}O$ & &($\rm 10^{-7}$)    &\\ 
\hline
G28.34\,S           &28.324   &0.067     & 4.6         &$0.77\pm0.10$     &12                &$13.23\pm0.17$                  &$15.22\pm2.19$  &$\rm 13.71\pm6.51$   &$14.54\pm2.55$   &$\rm 1.56\pm0.77$         &1.10    &$\rm -0.51\pm0.01$\\
IRDC\,18530      &35.431   &0.137        & 4.5         &$0.56\pm0.07$     & 9                 &$17.12\pm0.21$                  &$36.47\pm5.24$  &$\rm 19.54\pm6.83$    &$--$  &$\rm 0.64\pm0.22$          &3.66    &$\rm -0.19\pm0.002$\\
IRDC\,18306     &23.296   &0.056         & 3.6          &$0.48\pm0.12$    & 8             &$19.10\pm0.48$                 &$49.81\pm13.46$          &$\rm 21.52\pm6.51$ &$41.20\pm11.87$ &$--^d$        &6.89    &$\rm 0.30\pm0.02$\\
IRDC\,18308      &23.251   &0.017        & 4.4         &$0.36\pm0.07$     & 12          &$6.12\pm1.20$                  &$19.45\pm4.31$            &$\rm 12.16\pm4.69$  &$24.04\pm7.03$ &$--^d$         &0.51   &$\rm 0.0009\pm0.005$\\
\hline\hline
\multicolumn{12}{l}{{\bf Note.} {\it a.} Assuming the galactocentric distance of the sun is 8 kpc.}\\
\multicolumn{12}{l}{~~~~~~~~~~{\it b.}  Estimated from $\rm C^{18}O$ and optical depth corrected $\rm ^{13}CO$ with Eqs.~\ref{irdc:abunco}-\ref{irdc:abunc18o}.}\\
\multicolumn{12}{l}{~~~~~~~~~~{\it c.}  Measured with the optical depth corrected HNC  and hyperfine fitted HCN at the same gas temperature $\rm T_{ex,CO}$.}\\
\multicolumn{12}{l}{~~~~~~~~~~{\it d.} The value is not given because of large uncertainties in HCN abundance, see Table~\ref{irdc:column-HFS}.}\\
\multicolumn{12}{l}{~~~~~~~~~~{\it e.} Lower limit is measured from Eq.~\ref{irdc:ionizationeq}, with hyperfine fitted $\rm N_2H^+$, the optical depth corrected $\rm ^{13}CO$, 
and $\rm H^{13}CO^+$ at the same gas temperature $\rm T_{ex,CO}$.} \\
\multicolumn{12}{l}{~~~~~~~~~~{\it f.} $\rm \delta V_{H^{13}CO^+}=(V_{HCO^+}-V_{H^{13}CO^+})/\Delta \upsilon_{H^{13}CO^+}$, \citet{mardones97}.} \\
\end{tabular}
}
\end{center}
\end{table*}

\subsubsection{ Molecular abundance with respect to $\rm H_2$}\label{irdc:abunest}
APEX has a comparable beam size (18\arcsec) to the IRAM 30\,m at 1\,mm (12\arcsec) and 3\,mm  (30\arcsec) .   Therefore, assuming that the averaged continuum specific intensity at 18\arcsec\ angular resolution (from APEX) is comparable to that at 12\arcsec--30\arcsec resolution (from the 30\,m), the average molecular abundances  with respect to $\rm H_2$ at the clump scale (within $\rm 40\arcsec\times 40\arcsec$ region)  
can be estimated
(Tables~\ref{irdc:column-HFS}--\ref{irdc:column-other}).
 One needs to keep in mind that the difference in the beam size  
 will lead to the overestimation of the 1\,mm molecular abundance and underestimation of the 3\,mm molecular abundance by a factor of  2--3 because the $\rm H_2$ column density  (Eq. \ref{irdc:atlasgas}) is underestimated at 1\,mm and overestimated at 3\,mm. \\

The column densities of $\rm N_2H^+$, $\rm C_2H$, HCN, and $\rm H^{13}CN$  are estimated with  respect to $\rm T_{ex}$ estimated from their HFS fittings. Since $\rm  T_{ex}$  is different from that of the dust temperature used to derive the $\rm H_2$ column density,  it introduces a large uncertainty for the column density estimates.
Previous molecular spectral line observations towards several IRDCs have shown the typical range of $\rm T_{kin}=10\text{--}20~K$  (e.g., \citealt{carey98, teyssier02,sridharan05,pillai06,sakai08,devine11,ragan11,zhang11}). In addition, spectral energy distributions of clumps within ``quiescent" IRDCs indicate a range of dust temperature $\rm T_{dust}=10\text{--}30~K$ (e.g., \citealt{henning10,rathborne10}). Therefore, 
we also estimate the column densities and abundances of $\rm N_2H^+$, $\rm C_2H$, HCN, and $\rm H^{13}CN$ at the same $\rm T_{kin}\sim T_{ex,CO}$ as the other species (see Table~\ref{irdc:column-HFS}). Comparison  between the values at these two sets of temperatures indicates that the molecular abundances are sensitive to the temperatures in the 5--15\,K regime, and 5\,K  difference in temperature brings the  abundance  uncertainty  to the order of 1--2 magnitudes. In the following analysis, we use the  abundances at $\rm \sim T_{ex,CO}$ to estimate the other chemical parameters.\\

\subsection{Chemical similarity and diversity of our sample compared to other HMSCs}\label{irdc:hcn}

 On the clump-core scale ($\sim$0.8 pc), most strong emission lines detected by the 30\,m observations are from N-bearing species, which show  anti-correlation spatial distribution with CO isotopologues in G28.34\,S and IRDC\,18308. Emissions from shock tracers (e.g., SiO in G28.34\,S and HNCO), infall tracers  (e.g., $\rm HCO^+$ and  HCN), and the detection of carbon ring (e.g.,  c-$\rm C_3H_2$) indicate that these cold, young clouds are not completely ``quiescent". \\

Although the above features are qualitatively similar to the low-mass prestellar objects (LMPO) (typical scale of $\sim$0.02 pc, \citealp{ohishi92,jorgensen04,tafalla06}), quantitative comparison between HMSCs and LMPOs requires observations on the same spatial scale.
We compare the molecular abundances in our sources with  previous spectral line studies in other HMSCs 
(e.g., \citealt{vasyunin09,gerner14}). {On the one hand}, all species in our sources have, on average, lower abundances than the above-mentioned studies, probably owing to our lower\footnote{The molecular abundance estimated in our study is, on average, one magnitude lower than the IRDC large sample study by \citet{gerner14}. However, if we assume a higher $\rm T_{kin}\sim15~K$ (as in Gerner et al. 2014), the molecular abundances are comparable.}   $\rm T_{kin}$ and the possibly overestimated $\rm H_2$ column density (Section \ref{irdc:abunest}). Therefore, a more precise estimation of the molecular abundance requires a more reliable temperature measurement. { On the other hand}, species have higher abundances in G28.34\,S than the others in our sample, again indicating that  it is a chemically more evolved source. \\

The typical dense gas tracers HCN and HNC are a pair of isomers whose abundance ratio  is strongly temperature dependent. It has been reported that the HCN/HNC ratio decreases with  increasing temperature, ranging from $\le1$ in the low-mass prestellar cores (e.g., \citealt{ohishi92,jorgensen04,tafalla06,sarrasin10}) to the order of unity in dark clouds (e.g.,  \citealt{hirota98,vasyunin11}), then above 10 in high-mass protostellar objects (e.g., \citealt{blake87,helmich97}), and reaching several tens in warm cores (e.g., Orion GMC, \citealt{goldsmith86,schilke92}). In our sources, the HCN/HNC ratio is almost unity in G28.34\,S and IRDC 18530 (Table~\ref{irdc:ATLAS}), which is consistent with the value in dark clouds. If we compare the abundance ratio in our study to the statistical criteria value given by \citet{jin15}, IRDC 18530 is likely a ``quiescent" IRDC, and G28.34S is more ``active".\\

\subsection{CO depletion}\label{irdc:codepletion}
The sublimation temperature of CO is around 20\,K \citep{caselli99,fontani06,aikawa08}, so most CO is frozen out (depleted) onto the grains  in the center of cold cloud. In contrast, N-bearing species are more resilient to depletion (e.g., \citealt{bergin02,caselli02b,jorgensen04}). Moreover, since CO destroys $\rm N_2H^+$ and decreases the deuterium fraction, the anti-correlated spatial distribution between  $\rm N_2H^+$/$\rm NH_2D$ and CO in G28.34\,S and IRDC\,18308 indicates strong depletion of CO (Figure~\ref{irdc:molint}).\\

The  depletion factor is defined as the ratio between the expected abundance of a species $\rm X_\alpha^E$ and the observed  abundance of that species $\rm X_\alpha^O$ (Lacy et al., 1994),
$\mathscr{D}_\alpha=\rm X_\alpha^E/X_\alpha^O$. 
We measure the CO depletion factor  averaged in the $\rm 40\arcsec\times40\arcsec$ region from $\rm J=2\rightarrow1$ lines of $\rm C^{17}O$, $\rm C^{18}O$, and $\rm ^{13}CO$ (with optical depth correction; see Table~\ref{irdc:ATLAS}). Depletion factors derived from $\rm C^{17}O$ and $\rm C^{18}O$ are  consistent in each IRDC (with a mean of 30). However, depletion factors derived from $\rm ^{13}CO$ are lower than those derived from  its rare isotopologues by a factor of 2 in IRDC\,18530 and IRDC\,18306,  indicating the canonical ratio between isotopes may have large uncertainties in both sources. One possibility is fractionation of $\rm ^{12}C\Leftrightarrow {^{13}C}$ and of $\rm {^{17}O} \Leftrightarrow  {^{16}O}\Leftrightarrow {^{18}O}$ in the low temperature environment ($\rm<10$~K in both sources). Nevertheless, CO seems to be heavily depleted in our sample, with a depletion factor in the range of 14--50 at $<15$\,K and $\rm n\sim10^5~cm^{-3}$, which is consistent with the range given by \citet{fontani12} from a large sample study.  In particular, a depletion of $\sim14$ in G28.34S is consistent with that in the nearby source G28.34\,S P1  \citep{zhang09}. Compared with G28.34\, S and  IRDC\,18308, IRDC\,18530 and IRDC\, 18306 have higher CO depletion factors on scales of 0.8\,pc, but they do not show significant anti-correlation distribution between CO and $\rm N_2H^+/NH_2D$ at the spatial resolution of 0.2--0.6\,pc. Therefore, high depletion of CO may happen on smaller scales ($\rm <0.2$\,pc);   confirming this possibility requires higher resolution maps.\\

 \subsection{Ionization}\label{irdc:iosec}
 The ionization fraction {\it x(e)} in the early stage of star formation is an important indicator of the cosmic ray impact in the dense, dark clouds. Based on the assumption that gas  is quasi-neutral,  {\it x(e)} can be simply estimated by summing up the abundance of all the molecular ions (e.g., \citealt{caselli02b,miettinen09}). Therefore, the ionization fraction in our sources is dominated by the abundance of $\rm N_2H^+$, and on average, at a lower limit of  $\rm10^{-8}$.\\
 
In addition, based on the assumption of chemical equilibrium, {\it x(e)} in dense molecular clouds can
also be derived from ion-neutral reaction. Using the same method as described in \citet{miettinen11}  and the chemical network in their Table 8, we only consider the reactions between $\rm N_2H^+$, $\rm HCO^+$, and CO (the abundance is derived from $\rm C^{18}O$ without any fractional reactions). The lower limit of {\it x(e)} can be derived as (see also \citealt{qi03})
\begin{eqnarray} 
&\rm HCO^++e \overset{\it \beta_{11}}{\rightarrow} CO+H,   \label{irdc:hco2co}\\
&\rm N_2H^++CO\overset{\it k_{11}}{\rightarrow} HCO^++N_2  \label{irdc:co2hco},\\
&x(e)\rm \ge\frac{\it k_{15}\rm X(N_2H^+)X(CO)}{\beta_{11}X(HCO^+)}          \label{irdc:ionizationeq}
,\end{eqnarray}
where the rate coefficients $\it \beta_{11}=\rm 2.4\times10^{-7}(\frac{T_{kin}}{300~K})^{-0.69}$ and $\it k_{15}\rm=8.8\times10^{-10}\rm cm^3s^{-1}$ are adopted from UMIST database \citep{woodall07}.\\

From Table~\ref{irdc:ATLAS}, we found that the lower limit of ionization rate is $\rm\sim10^{-7}$. This method makes some assumptions, e.g., CO depletion and  fractional reaction are not considered, as discussed in \citet{caselli98,lintott06,miettinen11}. Nevertheless, these  more precise, model-based estimations are consistent with those reported in HMSCs (e.g.,  \citealt{williams98,bergin99,chenhr10,miettinen11}) and the modeling result in \citet{gerner14}.  Moreover, the lower limit of  ionization rate in our large-scale  ($\sim$0.8 pc) HMSC envelopes is higher than the model-predicted values in the center of the low-mass prestellar objects 
(e.g., $\rm>10^{-9}$,  \citealt{caselli02b}). This is reasonable because the cosmic ray ionization rates in the envelope of HMSCs are higher than in the inner region of their low-mass counterparts. It is also possible that these HMSCs are in a much more active environment than low-mass prestellar cores, and hence they are affected by much more radiation of any kind. \\

In addition, as mentioned in Section \ref{distribution}, $\rm H^{13}CO^+$ is ``deficient" in regions where $\rm ^{13}CO$  is abundant in IRDC\,18530. This can be explained by Eqs. \ref{irdc:hco2co}--\ref{irdc:co2hco}, which show that  the rate coefficient $\it \beta_{11}$ is enhanced in this source.\\

\section{Conclusions}\label{irdc:conclusion}
To investigate the initial conditions of high-mass star formation, we carried out several observations with the SMA and IRAM 30\,m. In this paper, we  study the dynamic and chemical properties of four IRDCs (G28.34S, IRDC\,18530, IRDC\,18306, and IRDC\,18308), which exhibit strong (sub)mm continuum emission toward 70 $\mu$m extinction features.  At different spatial resolutions and wavelengths, this small sample study of IRDCs illustrates that the cold ($\rm T<15$ K), dense ($\rm \sim10^5~cm^{-3}$) high-mass star-forming regions are not completely  quiescent. 

Our conclusions are as follows: \\
\begin{enumerate}
\item At an angular resolution of 2\arcsec~($\sim 10^4$ AU at an average distance of 4 kpc), the 1.1\,mm SMA observations resolve each source into several fragments.  The average separation  between the fragments is 0.07--0.24 pc, comparable to the Jeans length. However,
the mass of each fragment is on average $\rm >10\,M_{\odot}$, exceeding the thermal Jeans mass of the entire clump by a factor of up to 30,  indicating that thermal pressure alone  does not dominate the fragmentation process.\\

\item In the 8\,GHz SMA observational band at 1.1\,mm, higher-J transitions (including dense gas tracers such as $\rm H^{13}CO^+~(2\rightarrow1)$)  are not detected, implying that  our sources are in the cold and young evolutionary status when lines with $\rm E_u/k_B>20~K$ may not be excited. However, our line survey at 1\,mm/3\,mm shows a number of emissions from lower-J transitions. In particular, a $>5\sigma$ rms detection of  SiO $\rm (2\rightarrow1)$ with broad line wings in G28.34\,S may be the signature of a potential  protostellar object, and the asymmetric $\rm HCO^{+}$ line profile in this source  traces  potential infall motion at  a speed of $\rm \sim 0.84~km\,s^{-1}$.  \\

\item Linewidths of all the 30\,m detected lines are on average $\rm 2\text{--}3~km\,s^{-1}$. Using the $\rm H^{13}CO^{+}~(1\rightarrow0)$ line, we calculate the velocity dispersion components and the critical mass based on a non-thermal fragmentation scenario. {Non-thermal motions can provide sufficient extra support to increase the scale of the fragmentation, making it possible to support fragments up to  several hundreds of $\rm M_{\odot}$ against gravity. Alternative explanations may include magnetic fields and steep core density profiles.} The mass of the well-aligned fragments in filament G28.34\,S and IRDC\,18308 is also in rough agreement with the scenario of ``cylindrical fragmentation''. \\

\item $\rm N_2H^+~(J=1\rightarrow0)$, $\rm C_2H~(N=1\rightarrow0)$, $\rm HCN~(J=1\rightarrow0)$, and $\rm H^{13}CN~(J=1\rightarrow0)$ have multiplets in our 30\,m data. Using a hyperfine multiplet fit program, we estimate their excitation temperature (on average 5--8 K) and conclude that they are subthermally excited. { The excitation temperatures of the CO isotopologues suggest that the sources are in the very early cold phase ($\sim 10$ K).}\\

\item Our large-scale  high-mass clumps exhibit   chemical features that are qualitatively similar to low-mass prestellar cores on small scales, including a large number of N-bearing species and the detection of carbon rings and molecular ions. {Moreover, G28.34\,S and IRDC\,18308 show anti-correlated spatial distributions between $\rm N_2H^+$/$\rm NH_2D$ and CO on a scale of 0.5\,pc.} We calculate the molecular abundances,  HCN/HNC ratio, CO depletion factor, and ionization fraction and find that G28.34\,S is slightly more chemically evolved than the other sources in our sample.\\

\item Large-scale line surveys measure the chemical features of the high-mass clumps hosting HMSCs, but are not sufficient to probe the chemical complexity of the fragments themselves. The above observations reveal non-thermal motions, which dominate the fragmentation process from the dark cloud to the high-mass cores. However, without line detection, we do not know the dynamics within these cores.  { Further studies are expected to uncover the chemical and dynamic features within the densest fragments of IRDCs.}

\end{enumerate}

\begin{acknowledgements}
We would like to thank the SMA staff and IRAM 30\,m staff for
their helpful support during the reduction of the SMA data and  the performance of the 
IRAM 30~m observations in service mode. We thank  J. Pineda, Y. Wang, Z. Y. Zhang, and K. Wang  for helpful discussions.  
 We thank the anonymous reviewer for the constructive comments. \\
This research made use of NASA's Astrophysics Data System.\\
The Submillimeter Array is a joint project between the Smithsonian Astrophysical Observatory and the Academia Sinica Institute of Astronomy and Astrophysics and is funded by the Smithsonian Institution and the Academia Sinica.
 \end{acknowledgements}

\bibliographystyle{aa}
\bibliography{starless_accepted_language_proof.bbl}

\begin{thebibliography}{143}
\expandafter\ifx\csname natexlab\endcsname\relax\def\natexlab#1{#1}\fi

\bibitem[{{Afonso} {et~al.}(1998){Afonso}, {Yun}, \& {Clemens}}]{afonso98}
{Afonso}, J.~M., {Yun}, J.~L., \& {Clemens}, D.~P. 1998, \aj, 115, 1111

\bibitem[{{Aikawa} {et~al.}(2008){Aikawa}, {Wakelam}, {Garrod}, \&
  {Herbst}}]{aikawa08}
{Aikawa}, Y., {Wakelam}, V., {Garrod}, R.~T., \& {Herbst}, E. 2008, \apj, 674,
  984

\bibitem[{{Bastien} {et~al.}(1991){Bastien}, {Arcoragi}, {Benz}, {Bonnell}, \&
  {Martel}}]{bastien91}
{Bastien}, P., {Arcoragi}, J.-P., {Benz}, W., {Bonnell}, I., \& {Martel}, H.
  1991, \apj, 378, 255

\bibitem[{{Bergin}(2003)}]{bergin03}
{Bergin}, E.~A. 2003, in SFChem 2002: Chemistry as a Diagnostic of Star
  Formation, ed. C.~L. {Curry} \& M.~{Fich}, 63

\bibitem[{{Bergin} {et~al.}(2002){Bergin}, {Alves}, {Huard}, \&
  {Lada}}]{bergin02}
{Bergin}, E.~A., {Alves}, J., {Huard}, T., \& {Lada}, C.~J. 2002, \apjl, 570,
  L101

\bibitem[{{Bergin} {et~al.}(1994){Bergin}, {Goldsmith}, {Snell}, \&
  {Ungerechts}}]{bergin94}
{Bergin}, E.~A., {Goldsmith}, P.~F., {Snell}, R.~L., \& {Ungerechts}, H. 1994,
  \apj, 431, 674

\bibitem[{{Bergin} {et~al.}(1999){Bergin}, {Plume}, {Williams}, \&
  {Myers}}]{bergin99}
{Bergin}, E.~A., {Plume}, R., {Williams}, J.~P., \& {Myers}, P.~C. 1999, \apj,
  512, 724

\bibitem[{{Bergin} {et~al.}(1996){Bergin}, {Snell}, \& {Goldsmith}}]{bergin96}
{Bergin}, E.~A., {Snell}, R.~L., \& {Goldsmith}, P.~F. 1996, \apj, 460, 343

\bibitem[{{Bergin} \& {Tafalla}(2007)}]{bergin07}
{Bergin}, E.~A. \& {Tafalla}, M. 2007, \araa, 45, 339

\bibitem[{{Beuther} \& {Henning}(2009)}]{beuther09b}
{Beuther}, H. \& {Henning}, T. 2009, \aap, 503, 859

\bibitem[{{Beuther} {et~al.}(2015){Beuther}, {Henning}, {Linz}, {Feng},
  {Ragan}, {Smith}, {Bihr}, {Sakai}, \& {Kuiper}}]{beuther15}
{Beuther}, H., {Henning}, T., {Linz}, H., {et~al.} 2015, \aap, 581, A119

\bibitem[{{Beuther} {et~al.}(2010){Beuther}, {Henning}, {Linz}, {Krause},
  {Nielbock}, \& {Steinacker}}]{beuther10b}
{Beuther}, H., {Henning}, T., {Linz}, H., {et~al.} 2010, \aap, 518, L78

\bibitem[{{Beuther} {et~al.}(2011){Beuther}, {Kainulainen}, {Henning}, {Plume},
  \& {Heitsch}}]{beuther11}
{Beuther}, H., {Kainulainen}, J., {Henning}, T., {Plume}, R., \& {Heitsch}, F.
  2011, \aap, 533, A17

\bibitem[{{Beuther} {et~al.}(2013){Beuther}, {Linz}, {Tackenberg}, {Henning},
  {Krause}, {Ragan}, {Nielbock}, {Launhardt}, {Bihr}, {Schmiedeke}, {Smith}, \&
  {Sakai}}]{beuther13b}
{Beuther}, H., {Linz}, H., {Tackenberg}, J., {et~al.} 2013, \aap, 553, A115

\bibitem[{{Beuther} {et~al.}(2002){Beuther}, {Schilke}, {Menten}, {Motte},
  {Sridharan}, \& {Wyrowski}}]{beuther02}
{Beuther}, H., {Schilke}, P., {Menten}, K.~M., {et~al.} 2002, \apj, 566, 945

\bibitem[{{Beuther} {et~al.}(2008){Beuther}, {Semenov}, {Henning}, \&
  {Linz}}]{beuther08a}
{Beuther}, H., {Semenov}, D., {Henning}, T., \& {Linz}, H. 2008, \apjl, 675,
  L33

\bibitem[{{Beuther} \& {Sridharan}(2007)}]{beuther07c}
{Beuther}, H. \& {Sridharan}, T.~K. 2007, \apj, 668, 348

\bibitem[{{Beuther} {et~al.}(2012){Beuther}, {Tackenberg}, {Linz}, {Henning},
  {Krause}, {Ragan}, {Nielbock}, {Launhardt}, {Schmiedeke}, {Schuller},
  {Carlhoff}, {Nguyen-Luong}, \& {Sakai}}]{beuther12b}
{Beuther}, H., {Tackenberg}, J., {Linz}, H., {et~al.} 2012, \aap, 538, A11

\bibitem[{{Beuther} {et~al.}(2009){Beuther}, {Zhang}, {Bergin}, \&
  {Sridharan}}]{beuther09}
{Beuther}, H., {Zhang}, Q., {Bergin}, E.~A., \& {Sridharan}, T.~K. 2009, \aj,
  137, 406

\bibitem[{{Bihr} {et~al.}(2015){Bihr}, {Beuther}, {Linz}, {Ragan}, {Hennemann},
  {Tackenberg}, {Smith}, {Krause}, \& {Henning}}]{bihr15}
{Bihr}, S., {Beuther}, H., {Linz}, H., {et~al.} 2015, \aap, 579, A51

\bibitem[{{Blake} {et~al.}(1987){Blake}, {Sutton}, {Masson}, \&
  {Phillips}}]{blake87}
{Blake}, G.~A., {Sutton}, E.~C., {Masson}, C.~R., \& {Phillips}, T.~G. 1987,
  \apj, 315, 621

\bibitem[{{Bontemps} {et~al.}(2010){Bontemps}, {Motte}, {Csengeri}, \&
  {Schneider}}]{bontemps10}
{Bontemps}, S., {Motte}, F., {Csengeri}, T., \& {Schneider}, N. 2010, \aap,
  524, A18

\bibitem[{{Bronfman} {et~al.}(1996){Bronfman}, {Nyman}, \& {May}}]{bronfman96}
{Bronfman}, L., {Nyman}, L.-A., \& {May}, J. 1996, \aaps, 115, 81

\bibitem[{{Butler} \& {Tan}(2009)}]{butler09}
{Butler}, M.~J. \& {Tan}, J.~C. 2009, \apj, 696, 484

\bibitem[{{Carey} {et~al.}(1998){Carey}, {Clark}, {Egan}, {Price}, {Shipman},
  \& {Kuchar}}]{carey98}
{Carey}, S.~J., {Clark}, F.~O., {Egan}, M.~P., {et~al.} 1998, \apj, 508, 721

\bibitem[{{Caselli} {et~al.}(2002{\natexlab{a}}){Caselli}, {Benson}, {Myers},
  \& {Tafalla}}]{caselli02c}
{Caselli}, P., {Benson}, P.~J., {Myers}, P.~C., \& {Tafalla}, M.
  2002{\natexlab{a}}, \apj, 572, 238

\bibitem[{{Caselli} {et~al.}(1999){Caselli}, {Walmsley}, {Tafalla}, {Dore}, \&
  {Myers}}]{caselli99}
{Caselli}, P., {Walmsley}, C.~M., {Tafalla}, M., {Dore}, L., \& {Myers}, P.~C.
  1999, \apjl, 523, L165

\bibitem[{{Caselli} {et~al.}(1998){Caselli}, {Walmsley}, {Terzieva}, \&
  {Herbst}}]{caselli98}
{Caselli}, P., {Walmsley}, C.~M., {Terzieva}, R., \& {Herbst}, E. 1998, \apj,
  499, 234

\bibitem[{{Caselli} {et~al.}(2002{\natexlab{b}}){Caselli}, {Walmsley},
  {Zucconi}, {Tafalla}, {Dore}, \& {Myers}}]{caselli02a}
{Caselli}, P., {Walmsley}, C.~M., {Zucconi}, A., {et~al.} 2002{\natexlab{b}},
  \apj, 565, 331

\bibitem[{{Caselli} {et~al.}(2002{\natexlab{c}}){Caselli}, {Walmsley},
  {Zucconi}, {Tafalla}, {Dore}, \& {Myers}}]{caselli02b}
{Caselli}, P., {Walmsley}, C.~M., {Zucconi}, A., {et~al.} 2002{\natexlab{c}},
  \apj, 565, 344

\bibitem[{{Chambers} {et~al.}(2009){Chambers}, {Jackson}, {Rathborne}, \&
  {Simon}}]{chambers09}
{Chambers}, E.~T., {Jackson}, J.~M., {Rathborne}, J.~M., \& {Simon}, R. 2009,
  \apjs, 181, 360

\bibitem[{{Chandrasekhar} \& {Fermi}(1953)}]{chandrasekhar53}
{Chandrasekhar}, S. \& {Fermi}, E. 1953, \apj, 118, 116

\bibitem[{{Chen} {et~al.}(2010){Chen}, {Liu}, {Su}, \& {Zhang}}]{chenhr10}
{Chen}, H.-R., {Liu}, S.-Y., {Su}, Y.-N., \& {Zhang}, Q. 2010, \apjl, 713, L50

\bibitem[{{Chin} {et~al.}(1996){Chin}, {Henkel}, {Whiteoak}, {Langer}, \&
  {Churchwell}}]{chin96}
{Chin}, Y.-N., {Henkel}, C., {Whiteoak}, J.~B., {Langer}, N., \& {Churchwell},
  E.~B. 1996, \aap, 305, 960

\bibitem[{{Chira} {et~al.}(2013){Chira}, {Beuther}, {Linz}, {Schuller},
  {Walmsley}, {Menten}, \& {Bronfman}}]{chira13}
{Chira}, R.-A., {Beuther}, H., {Linz}, H., {et~al.} 2013, \aap, 552, A40

\bibitem[{{Codella} {et~al.}(2001){Codella}, {Bachiller}, {Nisini}, {Saraceno},
  \& {Testi}}]{codella01}
{Codella}, C., {Bachiller}, R., {Nisini}, B., {Saraceno}, P., \& {Testi}, L.
  2001, \aap, 376, 271

\bibitem[{{Commer{\c c}on} {et~al.}(2011){Commer{\c c}on}, {Hennebelle}, \&
  {Henning}}]{commercon11}
{Commer{\c c}on}, B., {Hennebelle}, P., \& {Henning}, T. 2011, \apjl, 742, L9

\bibitem[{{Crapsi} {et~al.}(2005){Crapsi}, {Caselli}, {Walmsley}, {Myers},
  {Tafalla}, {Lee}, \& {Bourke}}]{crapsi05}
{Crapsi}, A., {Caselli}, P., {Walmsley}, C.~M., {et~al.} 2005, \apj, 619, 379

\bibitem[{{Csengeri} {et~al.}(2011){Csengeri}, {Bontemps}, {Schneider},
  {Motte}, \& {Dib}}]{csengeri11}
{Csengeri}, T., {Bontemps}, S., {Schneider}, N., {Motte}, F., \& {Dib}, S.
  2011, \aap, 527, A135

\bibitem[{{Devine} {et~al.}(2011){Devine}, {Chandler}, {Brogan}, {Churchwell},
  {Indebetouw}, {Shirley}, \& {Borg}}]{devine11}
{Devine}, K.~E., {Chandler}, C.~J., {Brogan}, C., {et~al.} 2011, \apj, 733, 44

\bibitem[{{Dobbs} {et~al.}(2005){Dobbs}, {Bonnell}, \& {Clark}}]{dobbs05}
{Dobbs}, C.~L., {Bonnell}, I.~A., \& {Clark}, P.~C. 2005, \mnras, 360, 2

\bibitem[{{Draine}(2011)}]{draine11}
{Draine}, B.~T. 2011, {Physics of the Interstellar and Intergalactic Medium}

\bibitem[{{Feng} {et~al.}(2016){Feng}, {Beuther}, {Semenov}, {Henning}, {Linz},
  {Mills}, \& {Teague}}]{feng16c}
{Feng}, S., {Beuther}, H., {Semenov}, D., {et~al.} 2016, ArXiv e-prints

\bibitem[{{Fiege} \& {Pudritz}(2000)}]{fiege00}
{Fiege}, J.~D. \& {Pudritz}, R.~E. 2000, \mnras, 311, 85

\bibitem[{{Fischera} \& {Martin}(2012)}]{fischera12}
{Fischera}, J. \& {Martin}, P.~G. 2012, \aap, 542, A77

\bibitem[{{Fontani} {et~al.}(2006){Fontani}, {Caselli}, {Crapsi}, {Cesaroni},
  {Molinari}, {Testi}, \& {Brand}}]{fontani06}
{Fontani}, F., {Caselli}, P., {Crapsi}, A., {et~al.} 2006, \aap, 460, 709

\bibitem[{{Fontani} {et~al.}(2012){Fontani}, {Giannetti}, {Beltr{\'a}n},
  {Dodson}, {Rioja}, {Brand}, {Caselli}, \& {Cesaroni}}]{fontani12}
{Fontani}, F., {Giannetti}, A., {Beltr{\'a}n}, M.~T., {et~al.} 2012, \mnras,
  423, 2342

\bibitem[{{Fontani} {et~al.}(2014){Fontani}, {Sakai}, {Furuya}, {Sakai},
  {Aikawa}, \& {Yamamoto}}]{fontani14}
{Fontani}, F., {Sakai}, T., {Furuya}, K., {et~al.} 2014, \mnras, 440, 448

\bibitem[{{Frerking} {et~al.}(1982){Frerking}, {Langer}, \&
  {Wilson}}]{frerking82}
{Frerking}, M.~A., {Langer}, W.~D., \& {Wilson}, R.~W. 1982, \apj, 262, 590

\bibitem[{{Fuller} {et~al.}(2005){Fuller}, {Williams}, \&
  {Sridharan}}]{fuller05}
{Fuller}, G.~A., {Williams}, S.~J., \& {Sridharan}, T.~K. 2005, \aap, 442, 949

\bibitem[{{Garrod} {et~al.}(2008){Garrod}, {Weaver}, \& {Herbst}}]{garrod08}
{Garrod}, R.~T., {Weaver}, S.~L.~W., \& {Herbst}, E. 2008, \apj, 682, 283

\bibitem[{{Gerner} {et~al.}(2014){Gerner}, {Beuther}, {Semenov}, {Linz},
  {Vasyunina}, {Bihr}, {Shirley}, \& {Henning}}]{gerner14}
{Gerner}, T., {Beuther}, H., {Semenov}, D., {et~al.} 2014, \aap, 563, A97

\bibitem[{{Giannetti} {et~al.}(2014){Giannetti}, {Wyrowski}, {Brand},
  {Csengeri}, {Fontani}, {Walmsley}, {Nguyen Luong}, {Beuther}, {Schuller},
  {G{\"u}sten}, \& {Menten}}]{giannetti14}
{Giannetti}, A., {Wyrowski}, F., {Brand}, J., {et~al.} 2014, \aap, 570, A65

\bibitem[{{Girichidis} {et~al.}(2011){Girichidis}, {Federrath}, {Banerjee}, \&
  {Klessen}}]{girichidis11}
{Girichidis}, P., {Federrath}, C., {Banerjee}, R., \& {Klessen}, R.~S. 2011,
  \mnras, 413, 2741

\bibitem[{{Goldsmith} {et~al.}(1986){Goldsmith}, {Irvine}, {Hjalmarson}, \&
  {Ellder}}]{goldsmith86}
{Goldsmith}, P.~F., {Irvine}, W.~M., {Hjalmarson}, A., \& {Ellder}, J. 1986,
  \apj, 310, 383

\bibitem[{{Goodman} {et~al.}(1993){Goodman}, {Benson}, {Fuller}, \&
  {Myers}}]{goodman93}
{Goodman}, A.~A., {Benson}, P.~J., {Fuller}, G.~A., \& {Myers}, P.~C. 1993,
  \apj, 406, 528

\bibitem[{{Helmich} \& {van Dishoeck}(1997)}]{helmich97}
{Helmich}, F.~P. \& {van Dishoeck}, E.~F. 1997, \aaps, 124, 205

\bibitem[{{Henning} {et~al.}(2010){Henning}, {Linz}, {Krause}, {Ragan},
  {Beuther}, {Launhardt}, {Nielbock}, \& {Vasyunina}}]{henning10}
{Henning}, T., {Linz}, H., {Krause}, O., {et~al.} 2010, \aap, 518, L95

\bibitem[{{Hildebrand}(1983)}]{hildebrand83}
{Hildebrand}, R.~H. 1983, \qjras, 24, 267

\bibitem[{{Hirota} {et~al.}(1998){Hirota}, {Yamamoto}, {Mikami}, \&
  {Ohishi}}]{hirota98}
{Hirota}, T., {Yamamoto}, S., {Mikami}, H., \& {Ohishi}, M. 1998, \apj, 503,
  717

\bibitem[{{Ho} {et~al.}(2004){Ho}, {Moran}, \& {Lo}}]{ho04}
{Ho}, P.~T.~P., {Moran}, J.~M., \& {Lo}, K.~Y. 2004, \apjl, 616, L1

\bibitem[{{Hofner} {et~al.}(2001){Hofner}, {Wiesemeyer}, \&
  {Henning}}]{hofner01}
{Hofner}, P., {Wiesemeyer}, H., \& {Henning}, T. 2001, \apj, 549, 425

\bibitem[{{Horn} {et~al.}(2004){Horn}, {M{\o}llendal}, {Sekiguchi}, {Uggerud},
  {Roberts}, {Herbst}, {Viggiano}, \& {Fridgen}}]{horn04}
{Horn}, A., {M{\o}llendal}, H., {Sekiguchi}, O., {et~al.} 2004, \apj, 611, 605

\bibitem[{{Inutsuka} \& {Miyama}(1992)}]{inutsuka92}
{Inutsuka}, S.-I. \& {Miyama}, S.~M. 1992, \apj, 388, 392

\bibitem[{{Jackson} {et~al.}(2010){Jackson}, {Finn}, {Chambers}, {Rathborne},
  \& {Simon}}]{jackson10}
{Jackson}, J.~M., {Finn}, S.~C., {Chambers}, E.~T., {Rathborne}, J.~M., \&
  {Simon}, R. 2010, \apjl, 719, L185

\bibitem[{{Jim{\'e}nez-Serra} {et~al.}(2010){Jim{\'e}nez-Serra}, {Caselli},
  {Tan}, {Hernandez}, {Fontani}, {Butler}, \& {van Loo}}]{jimenez10}
{Jim{\'e}nez-Serra}, I., {Caselli}, P., {Tan}, J.~C., {et~al.} 2010, \mnras,
  406, 187

\bibitem[{{Jim{\'e}nez-Serra} {et~al.}(2009){Jim{\'e}nez-Serra},
  {Mart{\'{\i}}n-Pintado}, {Caselli}, {Viti}, \&
  {Rodr{\'{\i}}guez-Franco}}]{jimenez09}
{Jim{\'e}nez-Serra}, I., {Mart{\'{\i}}n-Pintado}, J., {Caselli}, P., {Viti},
  S., \& {Rodr{\'{\i}}guez-Franco}, A. 2009, \apj, 695, 149

\bibitem[{{Jin} {et~al.}(2015){Jin}, {Lee}, \& {Kim}}]{jin15}
{Jin}, M., {Lee}, J.-E., \& {Kim}, K.-T. 2015, \apjs, 219, 2

\bibitem[{{Jones} {et~al.}(2008){Jones}, {Burton}, {Cunningham}, {Menten},
  {Schilke}, {Belloche}, {Leurini}, {Ott}, \& {Walsh}}]{jones08}
{Jones}, P.~A., {Burton}, M.~G., {Cunningham}, M.~R., {et~al.} 2008, \mnras,
  386, 117

\bibitem[{{J{\o}rgensen} {et~al.}(2004){J{\o}rgensen}, {Sch{\"o}ier}, \& {van
  Dishoeck}}]{jorgensen04}
{J{\o}rgensen}, J.~K., {Sch{\"o}ier}, F.~L., \& {van Dishoeck}, E.~F. 2004,
  \aap, 416, 603

\bibitem[{{Kainulainen} {et~al.}(2013){Kainulainen}, {Ragan}, {Henning}, \&
  {Stutz}}]{kainulainen13}
{Kainulainen}, J., {Ragan}, S.~E., {Henning}, T., \& {Stutz}, A. 2013, \aap,
  557, A120

\bibitem[{{Kirk} {et~al.}(2006){Kirk}, {Johnstone}, \& {Di Francesco}}]{kirk06}
{Kirk}, H., {Johnstone}, D., \& {Di Francesco}, J. 2006, \apj, 646, 1009

\bibitem[{{Kirk} {et~al.}(2015){Kirk}, {Klassen}, {Pudritz}, \&
  {Pillsworth}}]{kirk15}
{Kirk}, H., {Klassen}, M., {Pudritz}, R., \& {Pillsworth}, S. 2015, \apj, 802,
  75

\bibitem[{{Krumholz} \& {McKee}(2008)}]{krumholz08}
{Krumholz}, M.~R. \& {McKee}, C.~F. 2008, \nat, 451, 1082

\bibitem[{{Lee} {et~al.}(2001){Lee}, {Myers}, \& {Tafalla}}]{lee01}
{Lee}, C.~W., {Myers}, P.~C., \& {Tafalla}, M. 2001, \apjs, 136, 703

\bibitem[{{Lintott} \& {Rawlings}(2006)}]{lintott06}
{Lintott}, C.~J. \& {Rawlings}, J.~M.~C. 2006, \aap, 448, 425

\bibitem[{{Linz} {et~al.}(2010){Linz}, {Krause}, {Beuther}, {Henning}, {Klein},
  {Nielbock}, {Stecklum}, {Steinacker}, \& {Stutz}}]{linz10}
{Linz}, H., {Krause}, O., {Beuther}, H., {et~al.} 2010, \aap, 518, L123

\bibitem[{{Loughnane} {et~al.}(2012){Loughnane}, {Redman}, {Thompson}, {Lo},
  {O'Dwyer}, \& {Cunningham}}]{loughnane12}
{Loughnane}, R.~M., {Redman}, M.~P., {Thompson}, M.~A., {et~al.} 2012, \mnras,
  420, 1367

\bibitem[{{Lovas}(2004)}]{lovas04}
{Lovas}, F.~J. 2004, Phys. Chem. Ref., 33, 117

\bibitem[{{Mardones} {et~al.}(1997){Mardones}, {Myers}, {Tafalla}, {Wilner},
  {Bachiller}, \& {Garay}}]{mardones97}
{Mardones}, D., {Myers}, P.~C., {Tafalla}, M., {et~al.} 1997, \apj, 489, 719

\bibitem[{{Maret} {et~al.}(2011){Maret}, {Hily-Blant}, {Pety}, {Bardeau}, \&
  {Reynier}}]{maret11}
{Maret}, S., {Hily-Blant}, P., {Pety}, J., {Bardeau}, S., \& {Reynier}, E.
  2011, \aap, 526, A47

\bibitem[{{Matthews} \& {Irvine}(1985)}]{matthews85}
{Matthews}, H.~E. \& {Irvine}, W.~M. 1985, \apjl, 298, L61

\bibitem[{{Meier} \& {Turner}(2005)}]{meier05}
{Meier}, D.~S. \& {Turner}, J.~L. 2005, \apj, 618, 259

\bibitem[{{Miettinen} {et~al.}(2009){Miettinen}, {Harju}, {Haikala},
  {Kainulainen}, \& {Johansson}}]{miettinen09}
{Miettinen}, O., {Harju}, J., {Haikala}, L.~K., {Kainulainen}, J., \&
  {Johansson}, L.~E.~B. 2009, \aap, 500, 845

\bibitem[{{Miettinen} {et~al.}(2011){Miettinen}, {Hennemann}, \&
  {Linz}}]{miettinen11}
{Miettinen}, O., {Hennemann}, M., \& {Linz}, H. 2011, \aap, 534, A134

\bibitem[{{Motte} {et~al.}(2007){Motte}, {Bontemps}, {Schilke}, {Schneider},
  {Menten}, \& {Brogui{\`e}re}}]{motte07}
{Motte}, F., {Bontemps}, S., {Schilke}, P., {et~al.} 2007, \aap, 476, 1243

\bibitem[{{M{\"u}ller} {et~al.}(2005){M{\"u}ller}, {Schl{\"o}der}, {Stutzki},
  \& {Winnewisser}}]{muller05}
{M{\"u}ller}, H.~S.~P., {Schl{\"o}der}, F., {Stutzki}, J., \& {Winnewisser}, G.
  2005, Journal of Molecular Structure, 742, 215

\bibitem[{{Myers} {et~al.}(1983){Myers}, {Linke}, \& {Benson}}]{myers83}
{Myers}, P.~C., {Linke}, R.~A., \& {Benson}, P.~J. 1983, \apj, 264, 517

\bibitem[{{Myers} {et~al.}(1996){Myers}, {Mardones}, {Tafalla}, {Williams}, \&
  {Wilner}}]{myers96}
{Myers}, P.~C., {Mardones}, D., {Tafalla}, M., {Williams}, J.~P., \& {Wilner},
  D.~J. 1996, \apjl, 465, L133

\bibitem[{{Nagasawa}(1987)}]{nagasawa87}
{Nagasawa}, M. 1987, Progress of Theoretical Physics, 77, 635

\bibitem[{{Nguyen-Lu'o'ng} {et~al.}(2013){Nguyen-Lu'o'ng}, {Motte}, {Carlhoff},
  {Louvet}, {Lesaffre}, {Schilke}, {Hill}, {Hennemann}, {Gusdorf}, {Didelon},
  {Schneider}, {Bontemps}, {Duarte-Cabral}, {Menten}, {Martin}, {Wyrowski},
  {Bendo}, {Roussel}, {Bernard}, {Bronfman}, {Henning}, {Kramer}, \&
  {Heitsch}}]{nguyen13}
{Nguyen-Lu'o'ng}, Q., {Motte}, F., {Carlhoff}, P., {et~al.} 2013, \apj, 775, 88

\bibitem[{{Ohishi} {et~al.}(1992){Ohishi}, {Irvine}, \& {Kaifu}}]{ohishi92}
{Ohishi}, M., {Irvine}, W.~M., \& {Kaifu}, N. 1992, in IAU Symposium, Vol. 150,
  Astrochemistry of Cosmic Phenomena, ed. P.~D. {Singh}, 171

\bibitem[{{Ossenkopf} \& {Henning}(1994)}]{ossenkopf94}
{Ossenkopf}, V. \& {Henning}, T. 1994, \aap, 291, 943

\bibitem[{{Palau} {et~al.}(2014){Palau}, {Rizzo}, {Girart}, \&
  {Henkel}}]{palau14}
{Palau}, A., {Rizzo}, J.~R., {Girart}, J.~M., \& {Henkel}, C. 2014, \apjl, 784,
  L21

\bibitem[{{Palumbo} {et~al.}(1997){Palumbo}, {Geballe}, \&
  {Tielens}}]{palumbo97}
{Palumbo}, M.~E., {Geballe}, T.~R., \& {Tielens}, A.~G.~G.~M. 1997, \apj, 479,
  839

\bibitem[{{Pickett} {et~al.}(1998){Pickett}, {Poynter}, {Cohen}, {Delitsky},
  {Pearson}, \& {M{\"u}ller}}]{pickett98}
{Pickett}, H.~M., {Poynter}, R.~L., {Cohen}, E.~A., {et~al.} 1998, \jqsrt, 60,
  883

\bibitem[{{Pillai} {et~al.}(2011){Pillai}, {Kauffmann}, {Wyrowski}, {Hatchell},
  {Gibb}, \& {Thompson}}]{pillai11}
{Pillai}, T., {Kauffmann}, J., {Wyrowski}, F., {et~al.} 2011, \aap, 530, A118

\bibitem[{{Pillai} {et~al.}(2006){Pillai}, {Wyrowski}, {Carey}, \&
  {Menten}}]{pillai06}
{Pillai}, T., {Wyrowski}, F., {Carey}, S.~J., \& {Menten}, K.~M. 2006, \aap,
  450, 569

\bibitem[{{Pineda} {et~al.}(2008){Pineda}, {Caselli}, \& {Goodman}}]{pineda08}
{Pineda}, J.~E., {Caselli}, P., \& {Goodman}, A.~A. 2008, \apj, 679, 481

\bibitem[{{Pirogov} {et~al.}(2003){Pirogov}, {Zinchenko}, {Caselli},
  {Johansson}, \& {Myers}}]{pirogov03}
{Pirogov}, L., {Zinchenko}, I., {Caselli}, P., {Johansson}, L.~E.~B., \&
  {Myers}, P.~C. 2003, \aap, 405, 639

\bibitem[{{Qi} {et~al.}(2003){Qi}, {Kessler}, {Koerner}, {Sargent}, \&
  {Blake}}]{qi03}
{Qi}, C., {Kessler}, J.~E., {Koerner}, D.~W., {Sargent}, A.~I., \& {Blake},
  G.~A. 2003, \apj, 597, 986

\bibitem[{{Ragan} {et~al.}(2012{\natexlab{a}}){Ragan}, {Henning}, {Krause},
  {Pitann}, {Beuther}, {Linz}, {Tackenberg}, {Balog}, {Hennemann}, {Launhardt},
  {Lippok}, {Nielbock}, {Schmiedeke}, {Schuller}, {Steinacker}, {Stutz}, \&
  {Vasyunina}}]{ragan12}
{Ragan}, S., {Henning}, T., {Krause}, O., {et~al.} 2012{\natexlab{a}}, \aap,
  547, A49

\bibitem[{{Ragan} {et~al.}(2009){Ragan}, {Bergin}, \& {Gutermuth}}]{ragan09}
{Ragan}, S.~E., {Bergin}, E.~A., \& {Gutermuth}, R.~A. 2009, \apj, 698, 324

\bibitem[{{Ragan} {et~al.}(2006){Ragan}, {Bergin}, {Plume}, {Gibson}, {Wilner},
  {O'Brien}, \& {Hails}}]{ragan06}
{Ragan}, S.~E., {Bergin}, E.~A., {Plume}, R., {et~al.} 2006, \apjs, 166, 567

\bibitem[{{Ragan} {et~al.}(2011){Ragan}, {Bergin}, \& {Wilner}}]{ragan11}
{Ragan}, S.~E., {Bergin}, E.~A., \& {Wilner}, D. 2011, \apj, 736, 163

\bibitem[{{Ragan} {et~al.}(2012{\natexlab{b}}){Ragan}, {Heitsch}, {Bergin}, \&
  {Wilner}}]{ragan12b}
{Ragan}, S.~E., {Heitsch}, F., {Bergin}, E.~A., \& {Wilner}, D.
  2012{\natexlab{b}}, \apj, 746, 174

\bibitem[{{Ragan} {et~al.}(2013){Ragan}, {Henning}, \& {Beuther}}]{ragan13}
{Ragan}, S.~E., {Henning}, T., \& {Beuther}, H. 2013, \aap, 559, A79

\bibitem[{{Rathborne} {et~al.}(2010){Rathborne}, {Jackson}, {Chambers},
  {Stojimirovic}, {Simon}, {Shipman}, \& {Frieswijk}}]{rathborne10}
{Rathborne}, J.~M., {Jackson}, J.~M., {Chambers}, E.~T., {et~al.} 2010, \apj,
  715, 310

\bibitem[{{Rathborne} {et~al.}(2006){Rathborne}, {Jackson}, \&
  {Simon}}]{rathborne06}
{Rathborne}, J.~M., {Jackson}, J.~M., \& {Simon}, R. 2006, \apj, 641, 389

\bibitem[{{Rawlings} {et~al.}(2004){Rawlings}, {Redman}, {Keto}, \&
  {Williams}}]{rawlings04}
{Rawlings}, J.~M.~C., {Redman}, M.~P., {Keto}, E., \& {Williams}, D.~A. 2004,
  \mnras, 351, 1054

\bibitem[{{Rawlings} {et~al.}(2000){Rawlings}, {Taylor}, \&
  {Williams}}]{rawlings00}
{Rawlings}, J.~M.~C., {Taylor}, S.~D., \& {Williams}, D.~A. 2000, \mnras, 313,
  461

\bibitem[{{Russeil} {et~al.}(2010){Russeil}, {Zavagno}, {Motte}, {Schneider},
  {Bontemps}, \& {Walsh}}]{russeil10}
{Russeil}, D., {Zavagno}, A., {Motte}, F., {et~al.} 2010, \aap, 515, A55

\bibitem[{{Sakai} {et~al.}(2008){Sakai}, {Sakai}, {Kamegai}, {Hirota},
  {Yamaguchi}, {Shiba}, \& {Yamamoto}}]{sakai08}
{Sakai}, T., {Sakai}, N., {Kamegai}, K., {et~al.} 2008, \apj, 678, 1049

\bibitem[{{Sanhueza} {et~al.}(2012){Sanhueza}, {Jackson}, {Foster}, {Garay},
  {Silva}, \& {Finn}}]{sanhueza12}
{Sanhueza}, P., {Jackson}, J.~M., {Foster}, J.~B., {et~al.} 2012, \apj, 756, 60

\bibitem[{{Sarrasin} {et~al.}(2010){Sarrasin}, {Abdallah}, {Wernli}, {Faure},
  {Cernicharo}, \& {Lique}}]{sarrasin10}
{Sarrasin}, E., {Abdallah}, D.~B., {Wernli}, M., {et~al.} 2010, \mnras, 404,
  518

\bibitem[{{Sault} {et~al.}(1995){Sault}, {Teuben}, \& {Wright}}]{sault95}
{Sault}, R.~J., {Teuben}, P.~J., \& {Wright}, M.~C.~H. 1995, in Astronomical
  Society of the Pacific Conference Series, Vol.~77, Astronomical Data Analysis
  Software and Systems IV, ed. R.~A. {Shaw}, H.~E. {Payne}, \& J.~J.~E.
  {Hayes}, 433

\bibitem[{{Schilke} {et~al.}(1992){Schilke}, {Walmsley}, {Pineau Des Forets},
  {Roueff}, {Flower}, \& {Guilloteau}}]{schilke92}
{Schilke}, P., {Walmsley}, C.~M., {Pineau Des Forets}, G., {et~al.} 1992, \aap,
  256, 595

\bibitem[{{Schuller} {et~al.}(2009){Schuller}, {Menten}, {Contreras},
  {Wyrowski}, {Schilke}, {Bronfman}, {Henning}, {Walmsley}, {Beuther},
  {Bontemps}, {Cesaroni}, {Deharveng}, {Garay}, {Herpin}, {Lefloch}, {Linz},
  {Mardones}, {Minier}, {Molinari}, {Motte}, {Nyman}, {Reveret}, {Risacher},
  {Russeil}, {Schneider}, {Testi}, {Troost}, {Vasyunina}, {Wienen}, {Zavagno},
  {Kovacs}, {Kreysa}, {Siringo}, \& {Wei{\ss}}}]{schuller09}
{Schuller}, F., {Menten}, K.~M., {Contreras}, Y., {et~al.} 2009, \aap, 504, 415

\bibitem[{{Shirley}(2015)}]{shirley15}
{Shirley}, Y.~L. 2015, \pasp, 127, 299

\bibitem[{{Sohn} {et~al.}(2007){Sohn}, {Lee}, {Park}, {Lee}, {Myers}, \&
  {Lee}}]{sohn07}
{Sohn}, J., {Lee}, C.~W., {Park}, Y.-S., {et~al.} 2007, \apj, 664, 928

\bibitem[{{Sridharan} {et~al.}(2005){Sridharan}, {Beuther}, {Saito},
  {Wyrowski}, \& {Schilke}}]{sridharan05}
{Sridharan}, T.~K., {Beuther}, H., {Saito}, M., {Wyrowski}, F., \& {Schilke},
  P. 2005, \apjl, 634, L57

\bibitem[{{Tackenberg} {et~al.}(2012){Tackenberg}, {Beuther}, {Henning},
  {Schuller}, {Wienen}, {Motte}, {Wyrowski}, {Bontemps}, {Bronfman}, {Menten},
  {Testi}, \& {Lefloch}}]{tackenberg12}
{Tackenberg}, J., {Beuther}, H., {Henning}, T., {et~al.} 2012, \aap, 540, A113

\bibitem[{{Tafalla} {et~al.}(1998){Tafalla}, {Mardones}, {Myers}, {Caselli},
  {Bachiller}, \& {Benson}}]{tafalla98}
{Tafalla}, M., {Mardones}, D., {Myers}, P.~C., {et~al.} 1998, \apj, 504, 900

\bibitem[{{Tafalla} {et~al.}(2002){Tafalla}, {Myers}, {Caselli}, {Walmsley}, \&
  {Comito}}]{tafalla02}
{Tafalla}, M., {Myers}, P.~C., {Caselli}, P., {Walmsley}, C.~M., \& {Comito},
  C. 2002, \apj, 569, 815

\bibitem[{{Tafalla} {et~al.}(2006){Tafalla}, {Santiago-Garc{\'{\i}}a}, {Myers},
  {Caselli}, {Walmsley}, \& {Crapsi}}]{tafalla06}
{Tafalla}, M., {Santiago-Garc{\'{\i}}a}, J., {Myers}, P.~C., {et~al.} 2006,
  \aap, 455, 577

\bibitem[{{Teyssier} {et~al.}(2002){Teyssier}, {Hennebelle}, \&
  {P{\'e}rault}}]{teyssier02}
{Teyssier}, D., {Hennebelle}, P., \& {P{\'e}rault}, M. 2002, \aap, 382, 624

\bibitem[{{Vasyunina} {et~al.}(2009){Vasyunina}, {Linz}, {Henning}, {Stecklum},
  {Klose}, \& {Nyman}}]{vasyunin09}
{Vasyunina}, T., {Linz}, H., {Henning}, T., {et~al.} 2009, \aap, 499, 149

\bibitem[{{Vasyunina} {et~al.}(2011){Vasyunina}, {Linz}, {Henning},
  {Zinchenko}, {Beuther}, \& {Voronkov}}]{vasyunin11}
{Vasyunina}, T., {Linz}, H., {Henning}, T., {et~al.} 2011, \aap, 527, A88

\bibitem[{{Walawender} {et~al.}(2005){Walawender}, {Bally}, {Kirk}, \&
  {Johnstone}}]{walawender05}
{Walawender}, J., {Bally}, J., {Kirk}, H., \& {Johnstone}, D. 2005, \aj, 130,
  1795

\bibitem[{{Wang} {et~al.}(2015){Wang}, {Testi}, {Ginsburg}, {Walmsley},
  {Molinari}, \& {Schisano}}]{wangk15}
{Wang}, K., {Testi}, L., {Ginsburg}, A., {et~al.} 2015, \mnras, 450, 4043

\bibitem[{{Wang} {et~al.}(2014){Wang}, {Zhang}, {Testi}, {Tak}, {Wu}, {Zhang},
  {Pillai}, {Wyrowski}, {Carey}, {Ragan}, \& {Henning}}]{wangk14}
{Wang}, K., {Zhang}, Q., {Testi}, L., {et~al.} 2014, \mnras, 439, 3275

\bibitem[{{Wang} {et~al.}(2011){Wang}, {Zhang}, {Wu}, \& {Zhang}}]{wangk11}
{Wang}, K., {Zhang}, Q., {Wu}, Y., \& {Zhang}, H. 2011, \apj, 735, 64

\bibitem[{{Wang} {et~al.}(2008){Wang}, {Zhang}, {Pillai}, {Wyrowski}, \&
  {Wu}}]{wangy08}
{Wang}, Y., {Zhang}, Q., {Pillai}, T., {Wyrowski}, F., \& {Wu}, Y. 2008, \apjl,
  672, L33

\bibitem[{{Wang} {et~al.}(2006){Wang}, {Zhang}, {Rathborne}, {Jackson}, \&
  {Wu}}]{wangy06}
{Wang}, Y., {Zhang}, Q., {Rathborne}, J.~M., {Jackson}, J., \& {Wu}, Y. 2006,
  \apjl, 651, L125

\bibitem[{{Wienen} {et~al.}(2012){Wienen}, {Wyrowski}, {Schuller}, {Menten},
  {Walmsley}, {Bronfman}, \& {Motte}}]{wienen12}
{Wienen}, M., {Wyrowski}, F., {Schuller}, F., {et~al.} 2012, \aap, 544, A146

\bibitem[{{Williams} {et~al.}(1998){Williams}, {Bergin}, {Caselli}, {Myers}, \&
  {Plume}}]{williams98}
{Williams}, J.~P., {Bergin}, E.~A., {Caselli}, P., {Myers}, P.~C., \& {Plume},
  R. 1998, \apj, 503, 689

\bibitem[{{Wilson} \& {Rood}(1994)}]{wilson94}
{Wilson}, T.~L. \& {Rood}, R. 1994, \araa, 32, 191

\bibitem[{{Woodall} {et~al.}(2007){Woodall}, {Ag{\'u}ndez}, {Markwick-Kemper},
  \& {Millar}}]{woodall07}
{Woodall}, J., {Ag{\'u}ndez}, M., {Markwick-Kemper}, A.~J., \& {Millar}, T.~J.
  2007, \aap, 466, 1197

\bibitem[{{Wu} {et~al.}(2010){Wu}, {Evans}, {Shirley}, \& {Knez}}]{wu10}
{Wu}, J., {Evans}, II, N.~J., {Shirley}, Y.~L., \& {Knez}, C. 2010, \apjs, 188,
  313

\bibitem[{{Zhang} \& {Wang}(2011)}]{zhang11}
{Zhang}, Q. \& {Wang}, K. 2011, \apj, 733, 26

\bibitem[{{Zhang} {et~al.}(2015){Zhang}, {Wang}, {Lu}, \&
  {Jim{\'e}nez-Serra}}]{zhang15}
{Zhang}, Q., {Wang}, K., {Lu}, X., \& {Jim{\'e}nez-Serra}, I. 2015, \apj, 804,
  141

\bibitem[{{Zhang} {et~al.}(2009){Zhang}, {Wang}, {Pillai}, \&
  {Rathborne}}]{zhang09}
{Zhang}, Q., {Wang}, Y., {Pillai}, T., \& {Rathborne}, J. 2009, \apj, 696, 268

\bibitem[{{Zinchenko} {et~al.}(2000){Zinchenko}, {Henkel}, \&
  {Mao}}]{zinchenko00}
{Zinchenko}, I., {Henkel}, C., \& {Mao}, R.~Q. 2000, \aap, 361, 1079

\end{thebibliography}

\newpage
\appendix
\section{Molecular column density based on the HFS fit}\label{appen:hfs}
The HFS method is based on the CLASS software package. In the fit model, the relative intensities with respect to the reference line (we set it as the strongest one) are converted from the $ \ell(\rm 300 K)$ value in the Splatalogue database.\\

Four assumptions are made in the HFS fit for one species: \\
{\scriptsize\encircle{1}} All the components of the multiplet have the same excitation temperature. \\
{\scriptsize\encircle{2}} The lines and their optical depths have Gaussian profiles. \\
{\scriptsize\encircle{3}} All the multiplets have the same linewidth. \\
{\scriptsize\encircle{4}} The multiplet components do not overlap.\\

One important parameter is the beam filling factor {\it f}. Although the distributions of most species are much more extended than the 30\,m primary beam, the emission extents of some species (e.g., $\rm H^{13}CN$ in IRDC\,18306 and IRDC\,18308) are smaller than the beam. To estimate  the filling factor (on the first order), we define an equivalent angular diameter of the emission extent $\theta_{mol}$ as \citep{wu10}
\begin{equation}
\rm \theta_{mol}=2(\frac{\mathscr {A}_{1/2}}{\pi}-\frac{{\theta_{30\,m}}^2}{4})^{1/2}~~~~~~~~~~~(rad)\label{irdc:sourcesize}
,\end{equation}
where $\rm \mathscr {A}_{1/2}$ is molecular extent  within the contour of half maximum integrated intensity and $\rm \theta_{30\,m}$ is the primary beam of 30\,m at the molecular rest frequency. \\

 With the beam filling factor $f=\rm \theta_{mol}^2/(\theta_{mol}^2+\theta_{30\,m}^2)$, background radiation temperature $\rm T_{bg}$ (assumed to be 2.7\,K), and the intensity in units of temperature $J_\nu(\rm T)=(h\nu/k_B)/[exp(h\nu/k_BT)-1]$, this method defines an effective intensity $A$ as
\begin{equation}
A=f\rm [{\it J_\nu}(T_{ex})-{\it J_\nu}(T_{bg})]~~~~~~~~~~~(K)\label{irdc:A}
.\end{equation}

The output parameters are $\rm V_{lsr}$, the optical depth $\tau_0$, the linewidth $\rm \Delta \upsilon$, and the value of effective intensity affected by the line optical depth $A\tau_0$ of the reference line. Therefore, with the light speed $c$,  the Planck constant $h$, and a constant factor $\rm \eta=2.7964 \times10^{-16}$ derived from \citet{pickett98},
the excitation temperature of the multiplet $\rm T_{ex}$ and the molecular column density $\rm N_{T,\alpha}$  at the frequency $\nu$ (in Hz) are given as
\begin{equation}\label{irdc:hyper_tex}
\rm T_{ex}=
\frac{h\nu/k_B}{ln[\frac{h\nu/k_B}{{\it A/f+J_\nu}(T_{bg})}+1]}~~~~~~~~~~~(K) 
\end{equation}
\begin{equation}\label{irdc:hyper_NT}
\rm N_{T,\alpha}=\frac{8 \pi \nu}{  \ell(T_{ex})\eta c^3} \tau_0 \Delta \upsilon~~~~~~~~~~~(cm^{-2})
\end{equation}

\section{Opacities and excitation temperatures of CO isotopologue lines}\label{appen:cotemp}
$\rm ^{13}CO~(2\rightarrow1)$, $\rm C^{18}O~(2\rightarrow1)$, and $\rm C^{17}O~(2\rightarrow1)$ are all detected in our data. To estimate their excitation temperatures, we need to know their optical depths. Assuming that fractionation (isotopic exchange reaction) of $\rm ^{12}C\Leftrightarrow {^{13}C}$ and of  $\rm {^{17}O} \Leftrightarrow  {^{16}O}\Leftrightarrow {^{18}O}$ are stable,
we can estimate the optical depth at the line center of a given transition $\rm \tau_{\alpha,0}$ by measuring the ratio between the main beam brightness temperature of the main line $\rm T_{mb,~\alpha,0}$ and its  isotopologue $\rm T_{mb,~\beta,0}$,   \citep{myers83}: \\

  \begin{equation}\label{eq:tau}
\rm \frac{1- exp(-\tau_{\alpha,0}/\Re_{\alpha})}{1-exp(-\tau_{\alpha,0})}\approx \frac{T_{mb, ~\beta,0}}{T_{mb, ~\alpha,0}}\\
  .\end{equation}
Here $\Re_{\alpha}$ is the intrinsic abundance of the main isotope (e.g., $\rm ^{12}C$) compared to its rare  isotope (e.g., $\rm ^{13}C$) in the ISM  (e.g., \citealt{wilson94,chin96}). \\

Since the line profiles of CO isotopologues in our sample  are contributions from various Galactic arms \citep{beuther07c},  we use a multi-Gaussian profile to fit their $\rm T_{mb}$. 
Then, assuming that $\rm C^{17}O~(2\rightarrow1)$ is optically thin and that all the lines are in LTE, we use Eq.~\ref{eq:tau} and  the canonical ratio between isotopes (the following Eq.~\ref{irdc:abun13co}-\ref{irdc:abunc17o}, \citealt{frerking82,wilson94,giannetti14}) to correct the optical depth of $\rm ^{13}CO~(2\rightarrow1)$ and $\rm C^{18}O~(2\rightarrow1)$. 
\begin{eqnarray}
&\rm X_{CO}^E=9.5\times10^{-5} exp(1.105-0.13{\it D}_{GC})  \label{irdc:abunco}\\
&\rm \Re_{^{12}C/^{13}C}^E=7.5{\it D}_{GC}+7.6  \label{irdc:abun13co}\\
&\rm  \Re_{^{16}O/^{18}O}^E=58.8{\it D}_{GC}+37.1   \label{irdc:abunc18o}\\
&\rm \Re_{^{16}O/^{17}O}^E=3.52\times(58.8{\it D}_{GC}+37.1)  \label{irdc:abunc17o}  
\end{eqnarray}

where $\rm {\it D}_{GC}$ (kpc) is the galactic center distance, and  $\rm X_{\alpha}^E$ is the  canonical (undepleted) abundance of isotopologue $\alpha$ with respect to $\rm H_2$.\\

Subsequently, using Eq.~\ref{irdc:ratioab}, we estimate $\rm T_{ex}$ of each CO isotopologue (see Table~\ref{irdc:cotex}):
\begin{equation}
\rm T_{ex}=
\frac{h\nu_\alpha/k_B}{ln[\frac{h\nu_\alpha/k_B}{{T_{mb,\alpha,0}/(1-e^{-\tau_\alpha})+J_\alpha}(T_{bg})}+1]}~~~~~~~~~~~(K) \label{irdc:ratioab}
\end{equation}

\section{Molecular column densities for species without hyperfine multiplets}\label{appen:colother}
 Consistent $\rm T_{ex}$ of CO  isotopologues (Table \ref{irdc:cotex}) indicates they are in LTE. Since most of the other lines without hyperfine multiples   have low critical density (except for $\rm SiO~(2\rightarrow1)$),  we assume they are coupled with dust and thermalized at $\rm T_{gas}= T_{dust}= T_{kin}\sim T_{ex,CO}$. The column densities of these species can be derived as 

 \begin{equation}
 \rm N_{T,\alpha}=
\rm \frac{k_B}{h c\it f} \frac{(e^\frac{h\nu}{k_BT_{ex,CO}}-1)}{\ell(T_{ex,CO})}T_{ex,CO} \int \tau_\upsilon  d\upsilon  ~~~~~~~~~~\rm(cm^{-2})
.\end{equation}
Assuming that  the observed emission homogeneously  distributes in the primary beam of 30\,m,
the integration of measured  main beam brightness temperature within the  velocity range $\rm \int T_{mb}(\upsilon)  d\upsilon$  can be  substituted for the last term in the above equation: 
\begin{equation}\label{eq:correc}
\rm T_{ex,CO} \int \tau_\upsilon  d\upsilon \cong \frac{\tau_{\alpha,0}}{1-e^{-\tau_{\alpha,0}}}\int T_{mb}(\upsilon)  d\upsilon~~~~~~~~~~~~(K~km\,s^{-1})\\ 
.\end{equation}

\noindent (1) For species with neither a hyperfine multiplet nor a rare isotopologue, we assume they are optically thin, therefore, $\frac{\tau_{\alpha,0}}{1-e^{-\tau_{\alpha,0}}}\sim1$.

\noindent (2)  $\rm \bf HCO^+$, HNC, and $\rm \bf ^{13}CO$:
 $\rm ^{13}CO~(2\rightarrow1)$ is optically thick in at least two of our sources when measured  with   $\rm C^{17}O~(2\rightarrow1)$, and  the other lines  may have the same issue. As we have the rare isotopologues of $\rm HCO^+~(1\rightarrow0)$ and  $\rm HNC~(1\rightarrow0)$ in our data, we use Eq.~\ref{eq:tau} to obtain their optical depths and correct their column densities by assuming that both  $\rm H^{13}CO^+~(1\rightarrow0)$ and $\rm HN^{13}C~(1\rightarrow0)$ are optically thin.\\

In sources where the emission of species  $\le3\sigma$ rms, we give their column density upper limit by assuming that {\it f} is unity.
\setcounter{table}{0}
\renewcommand{\thetable}{A\arabic{table}}

\begin{table*}
\small
\caption[Identified lines from 30\,m dataset]{Identified lines from the 30\,m line survey in G28.34\,S;  the lines detected only in the ``extra band" are in blue.  Lines  are imaged in Figure~\ref{irdc:molint}. }\label{irdc:tab:line}
\begin{center}
\begin{tabular}{llllllll}\hline\hline

Freq. & Mol. &Transition  &$  \ell(\rm 300 K)$  &$\rm S\mu^2$       &$\rm E_l$   &$\rm E_u/k$  &\\
(MHz) &        &                  &                 &($\rm D^2$)         & (cm)           &  (K)              &      \\    
\hline
{\color{blue}85139.103}         &$\rm OCS$               &$\rm J=7\rightarrow6$             &-3.79540                    &3.58037                        &8.51990        &16.34417                      &\\
{\color{blue}85338.894}     &c-$\rm C_3H_2$          &$\rm 2_{1,2}\rightarrow1_{0,1}$   &-3.74410                     &48.14839                       &1.63320        &6.44539                                   &\\


{\color{blue}85455.667}         &$\rm CH_3C_2H$      &$\rm 5_1\rightarrow4_1$               &-4.81870                    &1.79654                        &10.72420       &19.53084                                       &\\
{\color{blue}85457.300}     &$\rm CH_3C_2H$          &$\rm 5_0\rightarrow4_0$               &-4.79050                    &1.87136                        &5.70120        &12.30399                                       &\\
85926.278     &$\rm NH_2D$           &$\rm 1_{1,1,0s}\rightarrow1_{0,1,0a}$     &-3.45800       &28.59712               &11.50630       &20.67869                                 &\\
86338.733     &$\rm H^{13}CN$          &$\rm J=1\rightarrow0, F=1\rightarrow1$  &-3.02460      &8.91106                  &0.00000 &4.14358        &\\
86340.163     &$\rm H^{13}CN$          &$\rm J=1\rightarrow0, F=2\rightarrow1$  &-2.80270      &14.85302             &0.00000     &4.14365        &\\
86342.251     &$\rm H^{13}CN$          &$\rm J=1\rightarrow0, F=0\rightarrow1$  &-3.50170      &2.97026                  &0.00000 &4.14375        &\\
86754.288     &$\rm H^{13}CO^+$    &$\rm J=1\rightarrow0$                               &-2.28080                       &15.21089                       &0.00000        &4.16353                                        &\\
86846.960     &$\rm SiO$                 &$\rm J=2\rightarrow1$                         &-2.48320                       &19.19714                       &1.44850        &6.25203                                        &\\
87090.850       &$\rm HN^{13}C$     &$\rm J=1\rightarrow0$                              &-2.59520                       &7.28184                        &0.00000        &4.17968                                        &\\
87284.105       &$\rm C_2H$                 &$\rm N=1\rightarrow0, J=3/2\rightarrow1/2, F=1\rightarrow1$        &-5.18050       &0.10046                        &0.00150        &4.19111                                        &\\
87316.898       &$\rm C_2H$                     &$\rm N=1\rightarrow0, J=3/2\rightarrow1/2, F=2\rightarrow1$        &-4.18850       &0.98557                        &0.00150        &4.19268                                        &\\
87328.585       &$\rm C_2H$                     &$\rm N=1\rightarrow0, J=3/2\rightarrow1/2, F=1\rightarrow0$        &-4.49110       &0.49087                        &0.00000        &4.19109                                        &\\
87401.989       &$\rm C_2H$                     &$\rm N=1\rightarrow0, J=1/2\rightarrow1/2, F=1\rightarrow1$        &-4.49040       &0.49084                        &0.00150        &4.19677                                        &\\
87407.165       &$\rm C_2H$                     &$\rm N=1\rightarrow0, J=1/2\rightarrow1/2, F=0\rightarrow1$        &-4.88650       &0.19715                        &0.00150        &4.19702                                        &\\
87446.470       &$\rm C_2H$                     &$\rm N=1\rightarrow0, J=1/2\rightarrow1/2, F=1\rightarrow0$        &-5.17890       &0.10046                        &0.00000        &4.19674                                        &\\
87925.237               &$\rm HNCO$                     &$\rm 4_{0,4}\rightarrow3_{0,3}$            &-3.73180    &9.98656                        &4.39940        &10.54940                                       &\\
88630.416               &$\rm HCN$                  &$\rm J=1\rightarrow0,F=1\rightarrow1$              &-2.99110                       &8.91247                        &0.00000        &4.25356                                        &\\
88631.848               &$\rm HCN$                      &$\rm J=1\rightarrow0,F=2\rightarrow1$          &-2.76930                       &14.85197                       &0.00000        &4.25363                                        &\\
88633.936               &$\rm HCN$                      &$\rm J=1\rightarrow0,F=0\rightarrow1$          &-3.46820                       &2.97073                        &0.00000        &4.25373                                        &\\
89188.525     &$\rm HCO^+$              &$\rm J=1\rightarrow0$                          &-2.26080                       &15.21022                       &0.00000        &4.28035                                        &\\
90663.568     &$\rm HNC$                        &$\rm J=1\rightarrow0$                          &-2.52180                       &9.30176                        &0.00000        &4.35114                                        &\\
90979.023     &$\rm HC_3N$              &$\rm J=10\rightarrow9$                         &-2.28480                       &139.25442                      &13.65650       &24.01482                                       &\\
92494.308     &$\rm ^{13}CS$            &$\rm J=2\rightarrow1$                          &-2.80010                       &15.33521                       &1.54270        &6.65859                                 &\\
93171.880           &$\rm N_2H^{+*}$            &$\rm J=1\rightarrow0, F1=1\rightarrow1$                &-2.78440                       &37.25038                       &0.00000        &4.47152                        &\\
93173.700       &$\rm N_2H^{+*}$        &$\rm J=1\rightarrow0, F1=2\rightarrow1$                &-2.56250                       &62.08887                       &0.00000        &4.47161                            &\\
93176.130           &$\rm N_2H^{+*}$            &$\rm J=1\rightarrow0, F1=0\rightarrow1$                &-3.26140                       &12.41913                       &0.00000        &4.47172                            &\\
218222.192    &$\rm H_2CO$              &$\rm 3_{0,3}\rightarrow2_{0,2}$         &-2.76900                       &16.30796                       &7.28640        &20.95640                                       &\\
219560.354      &$\rm C^{18}O$          &$\rm J=2\rightarrow1$                          &-4.17940                       &0.02440                        &3.66190        &15.80580                                       &\\
220398.684      &$\rm ^{13}CO$          &$\rm J=2\rightarrow1$                          &-4.17490                       &0.04869                        &3.67590        &15.86618                                       &\\
{\color{blue}224714.385}        &$\rm C^{17}O$          &$\rm J=2\rightarrow1$                          &-4.15090                       &0.02432                        &3.74790        &16.17689                                       &\\

\hline \hline
\multicolumn{7}{l}{$\rm^*.~ N_2H^+$ hyperfine multiplets are blended, we only list the three  strongest hyperfine lines.}
\end{tabular}

\end{center}
\end{table*}

\begin{landscape}

\begin{table*}
\small
\begin{center}
\caption[The intensity integrated over the width of each line, and the optical depth  of the main line in each IRDC]{\small Linewidth,   the intensity integrated over the dispersion of each line (subtable I), and the integrated optical depth over the  velocity dispersion  of the strongest line among multiplet (subtable II). 
 \label{tab:lineprofile}}
\scalebox{0.9}{
\begin{tabular}{c|c|cc|cc|cc|cc}
\hline\hline
Species  &Freq$^{1}$       &\multicolumn{2}{c|}{G28.34\,S}   &\multicolumn{2}{c|}{IRDC\,18530}    &\multicolumn{2}{c|}{IRDC\,18306}  &\multicolumn{2}{c}{IRDC\,18308}\\
\hline
\multicolumn{10}{c}{I. Linewidth $\rm \Delta \upsilon~ (km~s^{-1})$ and integrated intensity $\rm \int T_B(\upsilon)d\upsilon~ (K km~s^{-1})$ in four IRDCs} \\
\hline

&(GHz)     &$\rm (km~s^{-1})$    &$\rm (K~km~s^{-1})$    &$\rm (km~s^{-1})$    &$\rm (K~km~s^{-1})$    &$\rm (km~s^{-1})$    &$\rm (K~km~s^{-1})$    &$\rm (km~s^{-1})$    &$\rm (K~km~s^{-1})$\\
 \hline
$\rm OCS$   & 85.1391   &$\rm 3.91\pm0.66$   &$\rm 0.26\pm0.04$    &$\times$    &$\times$    &$--$    &$\le$0.05    &$--$    &$\le$0.06\\
c-$\rm C_3H_2$   & 85.3389   &$\rm 2.97\pm0.10$   &$\rm 1.51\pm0.05$    &$\times$    &$\times$   &$\rm 1.99\pm0.12$   &$\rm 0.69\pm0.03$   &$\rm 1.76\pm0.23$   &$\rm 0.34\pm0.04$\\
$\rm c-CH_3C_2H^{(1)}$   & 85.4557   &$\rm 3.08\pm0.45$   &$\rm 0.31\pm0.04$    &$\times$    &$\times$   &$\rm 2.21\pm0.42$   &$\rm 0.18\pm0.03$   &$\rm 2.81\pm0.81$   &$\rm 0.22\pm0.06$\\
$\rm c-CH_3C_2H^{(2)}$   & 85.4573   &$\rm 2.44\pm0.42$   &$\rm 0.26\pm0.04$    &$\times$    &$\times$   &$\rm 1.08\pm0.55$   &$\rm 0.08\pm0.03$    &$--$    &$\le$0.08\\
$\rm NH_2D$   & 85.9263   &$\rm5.92\pm1.04$   &$\rm0.41\pm0.05$   &$\rm2.86\pm0.44$   &$\rm0.17\pm0.02$   &$\rm8.95\pm1.16$   &$\rm0.32\pm0.04$   &$\rm7.36\pm1.53$   &$\rm0.21\pm0.04$\\
$\rm H^{13}CO^+$   & 86.7543   &$\rm 3.07\pm0.06$   &$\rm 1.32\pm0.02$   &$\rm 2.01\pm0.06$   &$\rm 0.55\pm0.02$   &$\rm 2.05\pm0.10$   &$\rm 0.50\pm0.02$   &$\rm 2.09\pm0.09$   &$\rm 0.59\pm0.02$\\
$\rm SiO$   & 86.8470   &$\rm 4.83\pm0.39$   &$\rm 0.55\pm0.03$    &$--$    &$\le$0.03    &$--$    &$\le$0.25    &$--$    &$\le$0.03\\
$\rm HN^{13}C$   & 87.0909   &$\rm 2.99\pm0.07$   &$\rm 1.12\pm0.02$   &$\rm 1.71\pm0.08$   &$\rm 0.47\pm0.02$   &$\rm 2.24\pm0.10$   &$\rm 0.41\pm0.02$   &$\rm 2.08\pm0.11$   &$\rm 0.43\pm0.02$\\
$\rm HNCO$   & 87.9244  &$\rm2.68\pm0.09$   &$\rm0.83\pm0.02$   &$\rm1.75\pm0.14$   &$\rm0.25\pm0.02$   &$\rm2.13\pm0.18$   &$\rm0.25\pm0.02$   &$\rm2.11\pm0.35$   &$\rm0.17\pm0.02$\\
$\rm HCO^+$   & 89.1885   &$\rm 2.44\pm0.08$   &$\rm 5.27\pm0.13$   &$\rm 3.74\pm0.05$   &$\rm 4.74\pm0.05$   &$\rm 3.11\pm0.28$   &$\rm 1.72\pm0.14$   &$\rm 3.37\pm0.06$   &$\rm 5.52\pm0.08$\\
$\rm HNC$   & 90.6636   &$\rm 2.59\pm0.05$   &$\rm 5.57\pm0.08$   &$\rm 3.10\pm0.03$   &$\rm 4.01\pm0.03$   &$\rm 2.94\pm0.12$   &$\rm 1.85\pm0.07$   &$\rm 2.83\pm0.04$   &$\rm 3.95\pm0.05$\\
$\rm HC_3N$   & 90.979   &$\rm 3.36\pm0.16$   &$\rm 0.70\pm0.03$   &$\rm 2.22\pm0.23$   &$\rm 0.20\pm0.02$   &$\rm 2.34\pm0.22$   &$\rm 0.25\pm0.02$   &$\rm 3.03\pm0.87$   &$\rm 0.14\pm0.03$\\
$\rm ^{13}CS$   & 92.4943   &$\rm 3.06\pm0.39$   &$\rm 0.22\pm0.02$    &$--$    &$\le$0.03   &$\rm 2.28\pm0.42$   &$\rm 0.11\pm0.02$    &$--$    &$\le$0.03\\
$\rm H_2CO$   & 218.222   &$\rm 3.73\pm0.42$   &$\rm 0.98\pm0.09$    &$--$    &$\le$0.15    &$--$    &$\le$0.15    &$--$    &$\le$0.14\\
$\rm C^{18}O$   & 219.56   &$\rm 3.70\pm0.07$   &$\rm 7.83\pm0.13$   &$\rm 2.42\pm0.06$   &$\rm 3.71\pm0.08$   &$\rm 1.95\pm0.06$   &$\rm 3.75\pm0.10$   &$\rm 1.94\pm0.07$   &$\rm 3.02\pm0.10$\\
$\rm ^{13}CO$   & 220.399   &$\rm 4.83\pm0.20$   &$\rm 21.18\pm0.62$   &$\rm 3.78\pm0.13$   &$\rm 16.24\pm0.47$   &$\rm 3.11\pm0.25$   &$\rm 9.79\pm0.69$   &$\rm 2.75\pm0.13$   &$\rm 14.15\pm0.63$\\
$\rm C^{17}O$   & 224.714   &$\rm 3.72\pm0.25$   &$\rm 2.35\pm0.13$   &$\times$   &$\times$   &$\rm 2.06\pm0.12$   &$\rm 1.26\pm0.07$   &$\rm 2.61\pm0.36$   &$\rm 0.70\pm0.09$\\

\hline

\multicolumn{10}{c}{II. Linewidth $\rm \Delta \upsilon (km~s^{-1})$ and integrated optical depth $\rm \tau_0\Delta\upsilon (km~s^{-1})$ of the main line in  four IRDCs} \\
\hline

   &$\rm (GHz)$   &$\rm (km~s^{-1})$   &$\rm (km~s^{-1})$   &$\rm (km~s^{-1})$   &$\rm (km~s^{-1})$   &$\rm (km~s^{-1})$   &$\rm (km~s^{-1})$   &$\rm (km~s^{-1})$   &$\rm (km~s^{-1})$\\
 \hline
$\rm  N_2H^+$ & 93.1738     &$\rm 2.94\pm0.01$     &$\rm 1.14\pm0.01$     &$\rm 1.84\pm0.04$     &$\rm 1.20\pm0.18$     &$\rm 1.88\pm0.03$     &$\rm 1.19\pm0.15$     &$\rm 2.02\pm0.02$     &$\rm 0.20\pm0.04$\\
$\rm  C_2H$   &87.3169   &$\rm   2.90\pm0.07$   &$\rm   4.00\pm0.59$   &$\rm   1.85\pm0.08$   &$\rm   0.91\pm0.48$   &$\rm   1.73\pm0.07$   &$\rm   3.10\pm0.66$   &$\rm   1.76\pm0.11$   &$\rm   0.18\pm0.17$\\
$\rm  HCN$ & 88.6318     &$\rm 2.33\pm0.06$     &$\rm 0.23\pm0.10$     &$\rm 3.23\pm0.05$     &$\rm 0.32\pm0.04$     &$\rm 2.60\pm0.19*$     &$\rm 25.80\pm7.50^*$     &$\rm 3.15\pm0.10$     &$\rm 2.43\pm0.83$\\
$\rm  H^{13}CN$ & 86.3402     &$\rm 2.96\pm0.12$     &$\rm 0.54\pm0.19$     &$\rm 2.25\pm0.27^*$     &$\rm 0.23\pm5.61^*$     &$\rm 1.82\pm0.64^*$     &$\rm 0.62\pm1.77^*$     &$\rm 1.69\pm0.34^*$     &$\rm 0.17\pm7.08^*$\\

\hline
 \hline
  \multicolumn{10}{l}{}\\
  \multicolumn{10}{l}{{\bf Note.} 1. Transition numbers are listed in Table \ref{irdc:tab:line};}\\
  \multicolumn{10}{l}{~~~~~~~~~~2. Values are obtained from  the Gaussian/hyperfine multiplets fit program based on CLASS;}\\
   \multicolumn{10}{l}{~~~~~~~~~~3. Uncertainties on the measured intensities are typically $\le10\%$;}\\
  \multicolumn{10}{l}{~~~~~~~~~~4. For species that are not detected, an upper limit equal to the $\rm 3\sigma$ rms is given;}\\
   \multicolumn{10}{l}{~~~~~~~~~~5. ``$--$'' denotes  $\rm <3\sigma$ detection in our dataset, and the intensity upper limit comes from the  $\rm 3\sigma$ rms of the entire region}\\
   \multicolumn{10}{l}{~~~~~~~~~~~~    by integrating three channels in total around the $\rm V_{lsr}$;}\\
    \multicolumn{10}{l}{~~~~~~~~~~6. $\rm ``\times$'' denotes no observations in our dataset;}\\
 \multicolumn{10}{l}{~~~~~~~~~~7. ``$^*$'' denotes a Hyperfine multiplet that is not well fitted.}\\
\end{tabular}
}
\end{center}
\end{table*}

\end{landscape}

\newpage
\begin{landscape}
\begin{table*}
\small
\caption[]{Velocity integration range for each line imaged in Figure~\ref{irdc:molint}. }\label{irdc:tab:intrange}
\begin{center}
\scalebox{0.9}{
\begin{tabular}{llcccc}\hline\hline

Freq. & Transition  &\multicolumn{1}{c}{G28.34\,S}  &\multicolumn{1}{c}{IRDC\,18530}        &\multicolumn{1}{c}{IRDC\,18306}   &\multicolumn{1}{c}{IRDC\,18308}\\
(MHz) &                        &($\rm km\,s^{-1}$\text{--}\,$\rm km\,s^{-1}$)                 &($\rm km\,s^{-1}$\text{--}\,$\rm km\,s^{-1}$)          &($\rm km\,s^{-1}$\text{--}\,$\rm km\,s^{-1}$)       &($\rm km\,s^{-1}$\text{--}\,$\rm km\,s^{-1}$)      \\    
\hline
{\color{black}85139.103}        &$\rm OCS$               &77.84--81.75   &$\times$  &53.29--56.31~$^*$   &76.69--79.71~$^*$\\
{\color{black}85338.894}     &c-$\rm C_3H_2$         ($\rm 2_{1,2}\rightarrow1_{0,1}$)   &76.92--79.89   &$\times$   &53.51--55.49   &76.91--78.49\\           
{\color{black}85457.300}     &$\rm CH_3C_2H$         ($\rm 5_0\rightarrow4_0$)              &77.98--80.42   &$\times$   &54.95--56.05   &76.84--79.36\\
85926.278     &$\rm NH_2D$           ($\rm 1_{1,1,0s}\rightarrow1_{0,1,0a}$)    &75.19--81.61   &74.47--77.33   &52.77--57.23   &74.45--81.95\\
86340.163     &$\rm H^{13}CN$          ($\rm J=1\rightarrow0, F=2\rightarrow1$)         &76.92--79.88   &73.31--76.69   &52.94--54.66   &76.16--77.84\\
86754.288     &$\rm H^{13}CO^+$    ($\rm J=1\rightarrow0$)                               &76.86--79.94   &74.89--76.91   &53.77--55.83   &77.16--79.25\\
86846.960     &$\rm SiO$                 ($\rm J=2\rightarrow1$)                        &75.98--80.81   &74.39--77.41~$^*$   &53.29--56.31~$^*$   &75.49--78.51~$^*$\\
87090.850       &$\rm HN^{13}C$     ($\rm J=1\rightarrow0$)                             &76.91--79.89   &74.84--76.56   &53.88--56.12   &77.16--79.24\\
87316.898       &$\rm C_2H$                     ($\rm N=1\rightarrow0, J=3/2\rightarrow1/2, F=2\rightarrow1$)       &76.95--79.85   &74.97--76.83   &53.53--55.27   &76.12--77.88\\
87925.237               &$\rm HNCO$                     ($\rm 4_{0,4}\rightarrow3_{0,3}$)           &77.06--79.74   &74.12--75.88   &53.73--55.87   &75.41--78.59\\
88631.848               &$\rm HCN$                      ($\rm J=1\rightarrow0,F=2\rightarrow1$)         &77.23--79.56   &73.4--78.4   &53.5--56.1   &76.62--79.78\\
89188.525     &$\rm HCO^+$              ($\rm J=1\rightarrow0$)                         &77.18--79.62   &73.4--78.4   &53.24--56.35   &76.52--79.89\\
90663.568     &$\rm HNC$                        ($\rm J=1\rightarrow0$)                         &77.11--79.69   &74.28--77.52   &53.33--56.27   &76.78--79.61\\
90979.023     &$\rm HC_3N$              ($\rm J=10\rightarrow9$)                                &76.72--80.08   &74.79--77.01   &53.63--55.97   &76.74--78.86\\
92494.308     &$\rm ^{13}CS$            ($\rm J=2\rightarrow1$)                         &78.47--81.53   &74.39--77.41~$^*$   &54.43--55.57   &77.49--80.51~$^*$\\
93173.700       &$\rm N_2H^{+}$         ($\rm J=1\rightarrow0, F1=2\rightarrow1$)               &76.93--79.87   &74.98--76.82   &53.86--55.74   &77.19--79.21\\
218222.192    &$\rm H_2CO$              &77.69--80.31   &75.28--76.52~$^*$   &54.18--55.42~$^*$   &77.58--78.82~$^*$\\
219560.354      &$\rm C^{18}O$          &76.55--80.25   &74.69--77.11   &53.83--55.77   &77.23--79.17\\
220398.684      &$\rm ^{13}CO$          ($\rm J=2\rightarrow1$)                         &75.98--80.81   &74.01--77.79   &53.24--56.35   &76.83--79.58\\
{\color{black}224714.385}       &$\rm C^{17}O$          ($\rm J=2\rightarrow1$)                          &76.54--80.26   &$\times$   &53.77--55.83   &76.89--79.5\\

\hline \hline
\multicolumn{6}{l}{{\bf Note.} 1.``$\rm \times$'' denotes no observations in our dataset.}\\
\multicolumn{6}{l}{~~~~~~~~~ 2. ``$*$'' denotes a line having $\rm <4\sigma$ detections;  the integration range is given from three channels in total around the systematic $\rm V_{lsr}$ at its rest frequency.}
\end{tabular}
}
\end{center}
\end{table*}

\end{landscape}

\newpage
\begin{landscape}
\begin{table*}
\small
\begin{center}
\caption[Average column densities and abundances for molecules with HFS with respect to $\rm H_2$ in each IRDC]{\small Average column densities and abundances for molecules with hyperfine multiplet lines 
in a square region from [20\arcsec, 20\arcsec] to [-20\arcsec, -20\arcsec] of four IRDCs observed by IRAM 30\,m.
}\label{irdc:column-HFS}
\scalebox{0.82}{
\begin{tabular}{c|cc|cc|cc|cc}
 \hline\hline
Species   &\multicolumn{2}{c|}{G28.34\,S}   &\multicolumn{2}{c|}{IRDC\,18530}   &\multicolumn{2}{c|}{IRDC\,18306}   &\multicolumn{2}{c}{IRDC\,18308}\\
&$\rm T_{ex,\alpha}~(K)$   &(12 K)   &$\rm T_{ex,\alpha}~(K)$   &(9 K)   &$\rm T_{ex,\alpha}~(K)$   &(8 K)   &$\rm T_{ex,\alpha}~(K)$   &(12 K)\\
\hline
\multicolumn{9}{c}{I. Molecular Column density [$\rm x\pm y (z) =(x\pm y) \times 10^z cm^{-2}$]}\\
\hline
$\rm N_2H^+$   &$\it 1.76\pm0.05(16)$   &$\rm 1.46\pm0.02(15)$   &$\it 2.30\pm0.57(17)$   &$\rm 2.94\pm0.43(15)$   &$\it 7.46\pm1.32(17)$   &$\rm 4.24\pm0.51(15)$   &$\it 5.11\pm0.45(14)$   &$\rm 2.59\pm0.49(14)$\\
$\rm C_2H$   &$\it 4.93\pm0.80(14)$   &$\rm 2.81\pm0.42(15)$   &$\it 1.59\pm0.73(14)$   &$\rm 5.50\pm2.16(14)$   &$\it 3.63\pm0.84(14)$   &$\rm 1.11\pm0.23(15)$   &$\it 3.09\pm4.83(13)$   &$\rm 1.20\pm1.26(14)$\\
$\rm HCN$   &$\it 4.40\pm2.83(13)$   &$\rm 4.18\pm1.88(13)$   &$\it 7.15\pm0.41(13)$   &$\rm 6.31\pm0.82(13)$   &$\it 1.68\pm0.31(17)$   &$\rm 5.50\pm1.60(15)$   &$\it 3.06\pm0.11(15)$   &$\rm 4.36\pm1.49(14)$\\
$\rm H^{13}CN$   &$\it 9.27\pm0.28(14)$   &$\rm 1.01\pm0.35(13)$   &$\it 5.70\pm62.60(14)$   &$\rm 4.58\pm114.17(13)$   &$\it 6.94\pm6.64(15)~^\dag$   &$\rm 8.11\pm23.23(14)$   &$\it 3.91\pm77.43(14)~^\dag$   &$\rm 1.51\pm63.62(13)$\\
\hline

\multicolumn{9}{c}{II. Molecular Abundances with respect to $\rm H_2$    [$\rm x\pm y (z) =(x\pm y) \times 10^z$]}\\
\hline
$\rm N_2H^+$   &$\it 1.33\pm1.49(-7)$   &$\rm 1.11\pm0.18(-8)$      &$\it 1.34\pm1.15(-6)$   &$\rm 1.72\pm0.53(-8)$      &$\it 3.91\pm4.29(-6)$   &$\rm 2.22\pm1.11(-8)$      &$\it 8.35\pm9.49(-9)$   &$\rm 4.24\pm2.04(-9)$\\
$\rm C_2H$   &$\it 3.72\pm4.98(-9)$   &$\rm 2.12\pm0.67(-8)$      &$\it  9.28\pm15.50(-10)$   &$\rm 3.21\pm1.90(-9)$      &$\it 1.90\pm3.13(-9)$   &$\rm 5.81\pm3.59(-9)$      &$\it 5.06\pm16.10(-10)$   &$\rm 1.96\pm3.05(-9)$\\
$\rm HCN$   &$\it 3.32\pm6.28(-10)$   &$\rm 3.16\pm2.10(-10)$      &$\it 4.18\pm5.04(-10)$   &$\rm 3.69\pm1.07(-10)$      &$\it 8.78\pm13.90(-7)$   &$\rm 2.88\pm2.08(-8)$      &$\it 5.00\pm6.00(-8)$   &$\rm 7.12\pm4.77(-9)$\\
$\rm H^{13}CN$   &$\it 7.01\pm8.29(-9)$   &$\rm 7.64\pm4.17(-10)$      &$\it 3.33\pm45.60(-9)$   &$\rm 2.68\pm76.66(-10)$      &$\it 3.63\pm9.50(-8)~^\dag$   &$\rm 4.24\pm17.63(-10)$      &$\it 6.40\pm166.00(-9)~^\dag$   &$\rm 2.48\pm130.00(-10)$\\

\hline\hline
\multicolumn{9}{l}{{\bf Note.} 1. Column densities and abundances in italic font are obtained  at the temperatures listed in Table~\ref{irdc:tex}, under CTEX assumption;}\\
\multicolumn{9}{l}{~~~~~~~~~~2. Column densities and abundances in roman font  are obtained at mean $\rm T_{ex,CO}$,  with LTE assumption;}\\
\multicolumn{9}{l}{~~~~~~~~~~3. Uncertainties on the measured values are  determined from  hyperfine multiplet fit to $\rm \tau_0\Delta \upsilon$ and $\rm T_{ex,\alpha}$ or $\rm T_{ex,CO}$;}\\
  \multicolumn{9}{l}{~~~~~~~~~~4. ``$\dag$'' denotes a  species whose distribution is less extended than the 30\,m primary beam (filling factor $\it f\rm <0.5$).}\\
\end{tabular}
}

\end{center}
\end{table*}

\newpage
\begin{table*}
\small
\begin{center}
\caption[Average column densities and abundances for other molecules with respect to $\rm H_2$ in each IRDC]{\small Average column densities and abundances for molecules without hyperfine multiplet lines 
in a square region from [20\arcsec, 20\arcsec] to [-20\arcsec, -20\arcsec] of four IRDCs observed by IRAM 30\,m.
}\label{irdc:column-other}

\scalebox{0.77}{
\begin{tabular}{c|p{3cm}p{3cm}p{3cm}p{3cm}}
 \hline\hline
Species                          & G28.34\,S      &IRDC\,18530                  &IRDC\,18306   &IRDC\,18308 \\
                                           &(12 K)   &(9 K)    &(8 K)   &(12 K)\\
\hline
\multicolumn{5}{c}{I. Molecular Column density [$\rm x\pm y (z) =(x\pm y) \times 10^z cm^{-2}$]}\\
\hline
$\rm OCS$    &$\rm 2.31\pm0.34(13)$   &$\times$   &$\rm \le 5.62(12)$   &$\rm \le 5.32(12)$\\
c-$\rm C_3H_2$    &$\rm 1.10\pm0.03(13)$   &$\times$   &$\rm 3.61\pm0.17(12)$   &$\rm 2.48\pm0.32(12)$\\
$\rm c-CH_3C_2H^{(1)}$    &$\rm 5.01\pm0.69(13)$   &$\times$   &$\rm 1.53\pm0.29(14)~^\dag$   &$\rm 3.52\pm0.92(13)$\\
$\rm c-CH_3C_2H^{(2)}$    &$\rm 2.23\pm0.33(13)$   &$\times$   &$\rm 6.48\pm2.08(12)$   &$\rm \le 7.09(12)$\\
$\rm NH_2D$    &$\rm 5.53\pm0.61(12)$   &$\rm 6.60\pm0.80(12)~^\dag$   &$\rm 5.94\pm0.69(12)$   &$\rm 2.79\pm0.52(12)$\\
$\rm H^{13}CO^+$    &$\rm1.45\pm0.03(12)$  &$\rm5.13\pm0.14(11)$  &$\rm4.39\pm0.19(11)$  &$\rm6.51\pm0.23(11)$\\
$\rm SiO$    &$\rm 1.11\pm0.07(12)$   &$\rm \le 5.39(10)$   &$\rm \le 4.34(11)$   &$\rm \le 6.00(10)$\\
$\rm HN^{13}C$    &$\rm 2.55\pm0.05(12)$   &$\rm 9.13\pm0.35(11)$   &$\rm 7.45\pm0.33(11)$   &$\rm 9.69\pm0.45(11)$\\
$\rm HNCO$    &$\rm 1.75\pm0.05(13)$   &$\rm 6.50\pm0.44(12)$   &$\rm 7.24\pm0.56(12)$   &$\rm 3.60\pm0.46(12)$\\
$\rm HCO^{+~\bf*}$ &$\rm \bf 4.42\pm0.11(13)$     &$\rm \bf  3.82\pm0.04(13)$     &$\rm \bf  2.32\pm0.18(13)$ &$\rm \bf  4.05\pm0.06(13)$  \\

$\rm HNC^{~\bf*}$  &$\rm \bf  6.52\pm0.09(13)$     &$\rm \bf  4.02\pm0.03(13)$     &$\rm \bf  2.49\pm0.09(13)$ &$\rm \bf  3.97\pm0.05(13)$\\
$\rm HC_3N$    &$\rm 3.77\pm0.15(12)$   &$\rm 1.58\pm0.14(12)$   &$\rm 2.42\pm0.20(12)$   &$\rm 7.42\pm1.43(11)$\\
$\rm ^{13}CS$    &$\rm9.95\pm1.08(11)$  &$\rm\le 1.25(11)$  &$\rm4.66\pm0.82(11)$  &$\rm\le 1.37(11)$\\
$\rm H_2CO$    &$\rm 7.36\pm0.64(13)~^\dag$   &$\rm \le 9.15(11)$   &$\rm \le 9.67(11)$   &$\rm \le 7.81(11)$\\
$\rm C^{18}O$    &$\rm 4.48\pm0.07(15)$   &$\rm 2.50\pm0.05(15)$   &$\rm 2.78\pm0.07(15)$   &$\rm 1.73\pm0.06(15)$\\

$\rm ^{13}CO^{~\bf*}$   &$\rm \bf  3.41\pm0.10(16)$     &$\rm \bf  3.39\pm0.10(16)$     &$\rm \bf  2.01\pm0.09(16)$ &$\rm \bf  4.61\pm0.32(16)$\\
$\rm C^{17}O$    &$\rm1.33\pm0.07(15)$  &$\times$  &$\rm9.55\pm0.50(14)$  &$\rm3.97\pm0.54(14)$\\


\hline
\multicolumn{5}{c}{II. Molecular Abundances with respect to $\rm H_2$    [$\rm x\pm y (z) =(x\pm y) \times 10^z$]}\\
\hline
$\rm OCS$   &$\rm 1.75\pm0.55(-10)$   &$\times$   &$\rm \le 2.94(-11)$   &$\rm \le 8.70(-11)$\\
c-$\rm C_3H_2$   &$\rm 8.31\pm1.53(-11)$   &$\times$   &$\rm 1.89\pm0.75(-11)$   &$\rm 4.05\pm1.64(-11)$\\
$\rm c-CH_3C_2H^{(1)}$   &$\rm 3.79\pm1.17(-10)$   &$\times$   &$\rm 8.03\pm4.70(-10)~^\dag$   &$\rm 5.75\pm3.27(-10)$\\
$\rm c-CH_3C_2H^{(2)}$   &$\rm 1.69\pm0.54(-10)$   &$\times$   &$\rm 3.39\pm2.59(-11)$   &$\rm \le 1.16(-10)$\\
$\rm NH_2D$   &$\rm 4.18\pm1.16(-11)$   &$\rm 3.85\pm1.09(-11)~^\dag$   &$\rm 3.11\pm1.52(-11)$   &$\rm 4.56\pm2.18(-11)$\\
$\rm H^{13}CO^+$   &$\rm1.09\pm0.19(-11)$    &$\rm3.00\pm0.52(-12)$    &$\rm2.30\pm0.90(-12)$    &$\rm1.06\pm0.31(-11)$\\
$\rm SiO$   &$\rm 8.35\pm1.82(-12)$   &$\rm \le 3.15(-13)$   &$\rm \le 2.27(-12)$   &$\rm \le 9.81(-13)$\\
$\rm HN^{13}C$   &$\rm 1.93\pm0.33(-11)$   &$\rm 5.33\pm1.00(-12)$   &$\rm 3.90\pm1.54(-12)$   &$\rm 1.58\pm0.48(-11)$\\
$\rm HNCO$   &$\rm 1.32\pm0.24(-10)$   &$\rm 3.79\pm0.84(-11)$   &$\rm 3.79\pm1.66(-11)$   &$\rm 5.89\pm2.38(-11)$\\
$\rm HCO^{+~\bf*}$   &$\rm \bf  3.34\pm0.59(-10)$   &$\rm \bf  2.23\pm0.35(-10)$   &$\rm \bf  1.21\pm0.54(-10)$   &$\rm \bf  6.63\pm1.74(-10)$\\
$\rm HNC^{~\bf*}$   &$\rm \bf  4.58\pm0.76(-10)$   &$\rm \bf  2.18\pm0.33(-10)$   &$\rm \bf  1.56\pm0.60(-10)$   &$\rm \bf  6.49\pm1.69(-10)$\\
$\rm HC_3N$   &$\rm 2.85\pm0.56(-11)$   &$\rm 9.25\pm2.26(-12)$   &$\rm 1.26\pm0.56(-11)$   &$\rm 1.21\pm0.59(-11)$\\
$\rm ^{13}CS$   &$\rm7.52\pm2.06(-12)$    &$\rm\le 7.30(-13)$    &$\rm2.44\pm1.39(-12)$    &$\rm\le 2.25(-12)$\\
$\rm H_2CO$   &$\rm 5.56\pm1.39(-10)~^\dag$   &$\rm \le 5.35(-12)$   &$\rm \le 5.06(-12)$   &$\rm \le 1.28(-11)$\\
$\rm C^{18}O$   &$\rm 3.38\pm0.57(-8)$   &$\rm 1.46\pm0.24(-8)$   &$\rm 1.46\pm0.54(-8)$   &$\rm 2.83\pm0.80(-8)$\\
$\rm ^{13}CO^{~\bf*}$   &$\rm \bf  2.58\pm0.47(-7)$   &$\rm \bf  1.98\pm0.35(-7)$   &$\rm \bf  2.41\pm1.04(-7)$   &$\rm \bf  3.28\pm0.98(-7)$\\
$\rm C^{17}O$    &$\rm1.01\pm0.21(-8)$    &$\times$    &$\rm5.00\pm2.02(-9)$    &$\rm6.50\pm2.69(-9)$\\

\hline\hline
\multicolumn{5}{l}{{\bf Note.} 1. ``$\times$'' denotes a species that we did not observe in a certain source;}\\
\multicolumn{5}{l}{~~~~~~~~~2. For species that are not detected, an upper limit derived from  $\rm 3\sigma$ rms is given;}\\
\multicolumn{5}{l}{~~~~~~~~~3. Column density and abundances are obtained at mean $\rm T_{ex}$ of CO isotopologues with LTE assumption;}\\
\multicolumn{5}{l}{~~~~~~~~~4. Species with ``*'' and values in bold face are likely obtained from optically thick lines and  for which we did the optical depth correction;}\\
\multicolumn{5}{l}{~~~~~~~~~5. Uncertainties on the measured values are determined from $\rm T_{ex,CO}$, partition function $\rm Q(T_{ex,CO})$, and Gaussian fit to  $\rm \int T_B(\upsilon)d\upsilon$;}\\
    \multicolumn{5}{l}{~~~~~~~~~~6. ``$\dag$'' denotes a species whose distribution is less extended than the 30\,m primary beam (filling factor $\it f\rm <0.5$).}\\

\end{tabular}
}
\end{center}
\end{table*}

\end{landscape}

\newpage
\setcounter{figure}{0}
\renewcommand{\thefigure}{A\arabic{figure}}
\begin{figure*}
\begin{center}
\includegraphics[angle=0,scale=0.8] {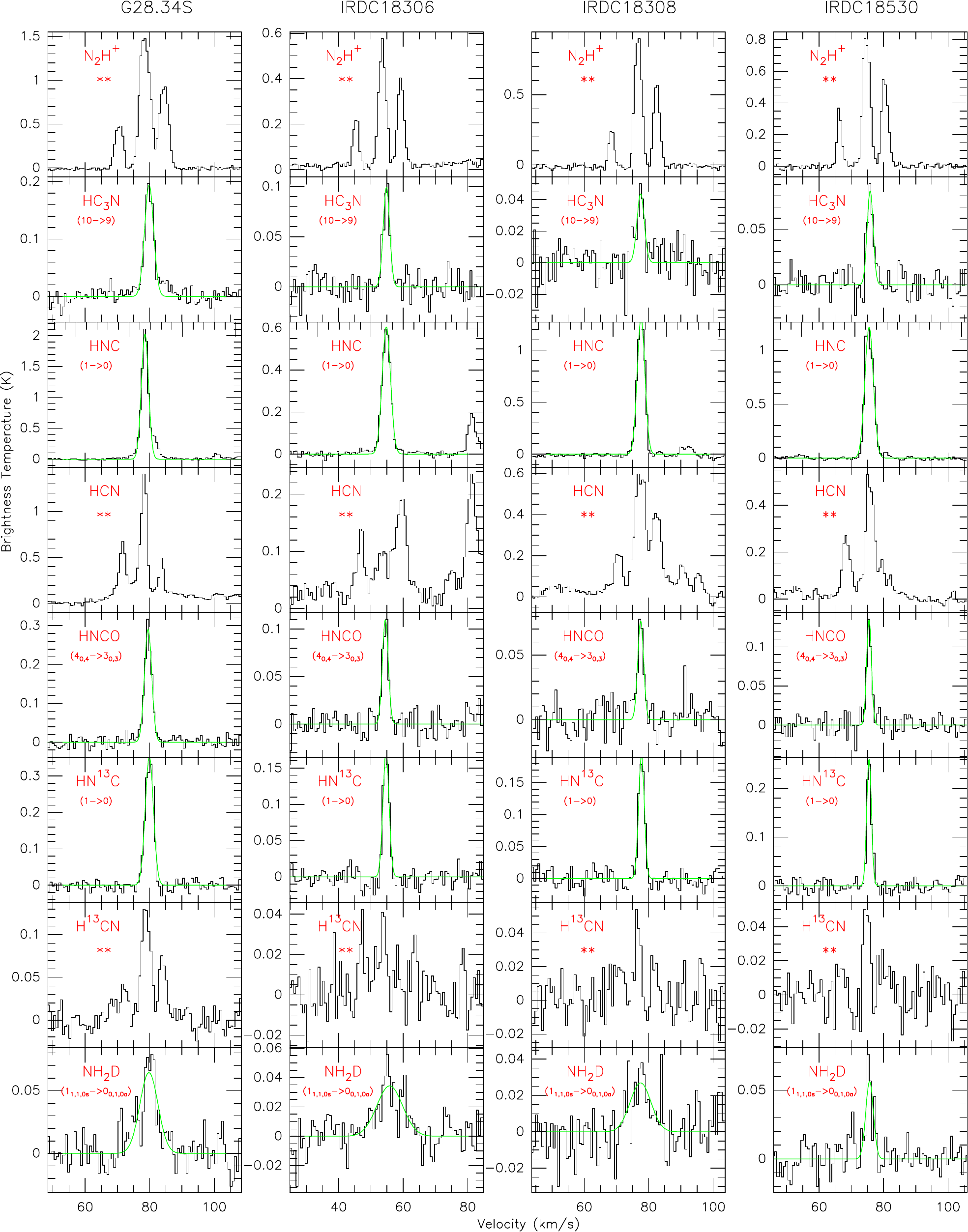}
\end{center}
\caption[Line profiles of identified species toward G28.34\,S, IRDC\,18530, IRDC\,18306 and IRDC 18308]{Line profiles of identified species toward each IRDC, {ordered in  groups as described in Section \ref{distribution}.}
The green line is from  a single-Gaussian fit. ``**'' denotes a species with hyperfine multiplets whose HFS fits are plotted in Figure~\ref{irdc:hyperfit}. Profiles in blue denote the lines we detect from the ``extra band" where the rms are higher than the others owing to less observation time.

}\label{irdc:velpro}
\end{figure*} 

\begin{figure*}
\ContinuedFloat
\begin{center}
\includegraphics[angle=0,scale=0.8] {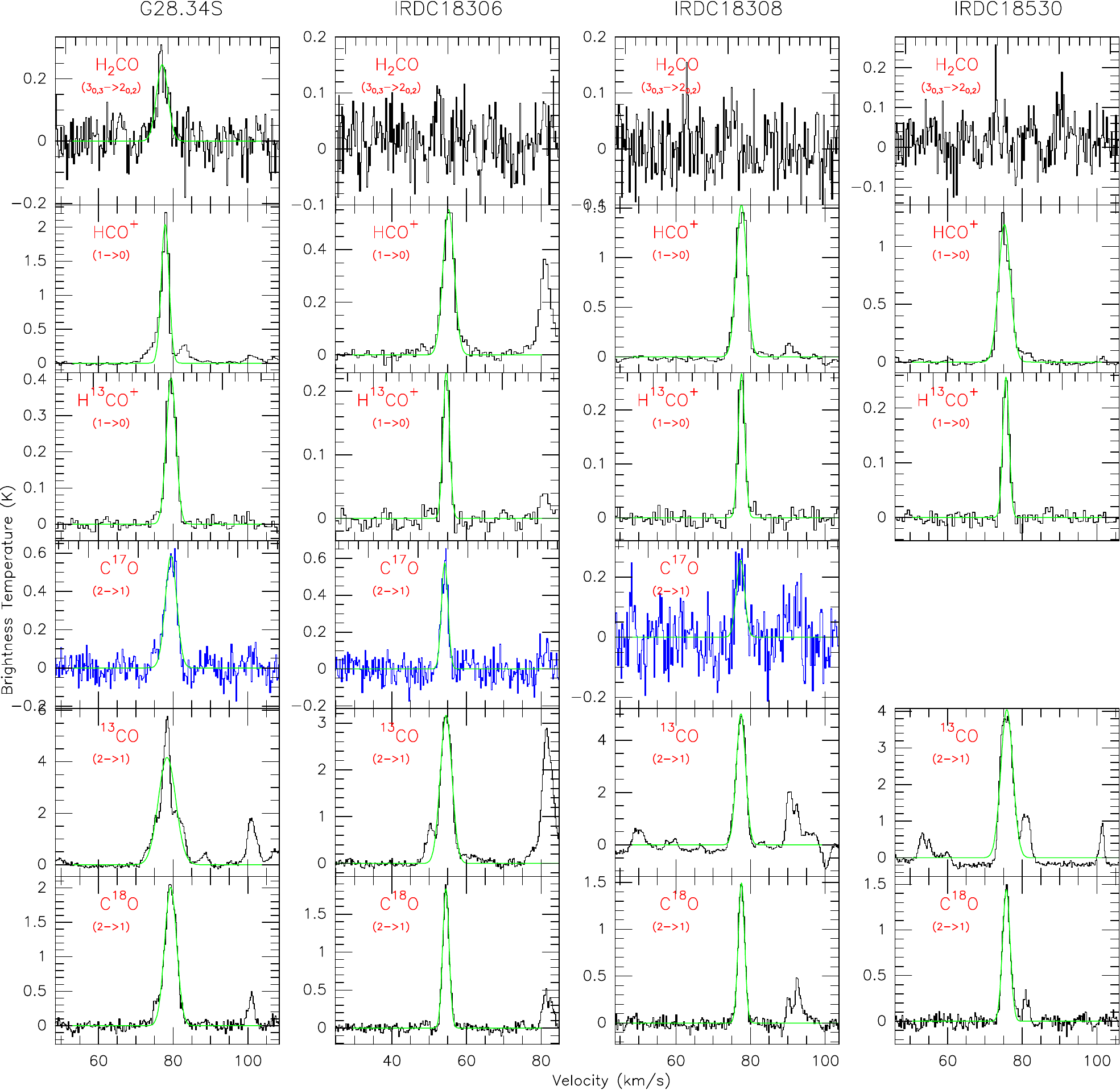}
\end{center}
\caption{(continued)}
\end{figure*} 

\begin{figure*}
\ContinuedFloat
\begin{center}
\includegraphics[angle=0,scale=0.8] {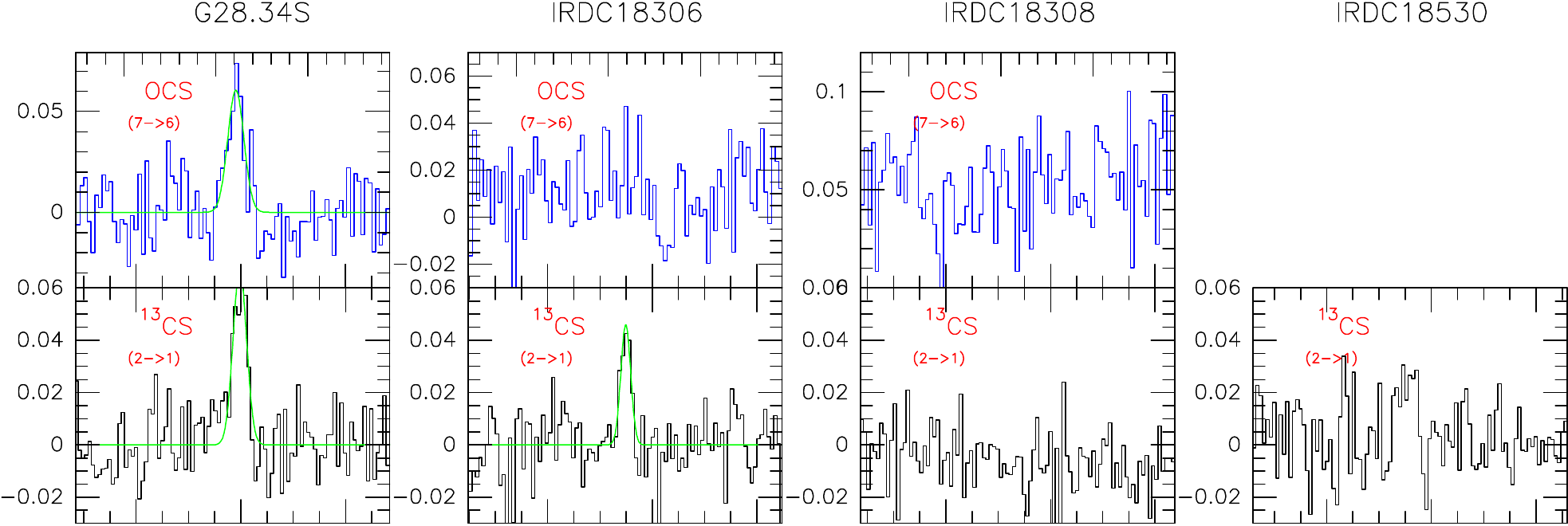}
\vspace{1em}
\includegraphics[angle=0,scale=0.8] {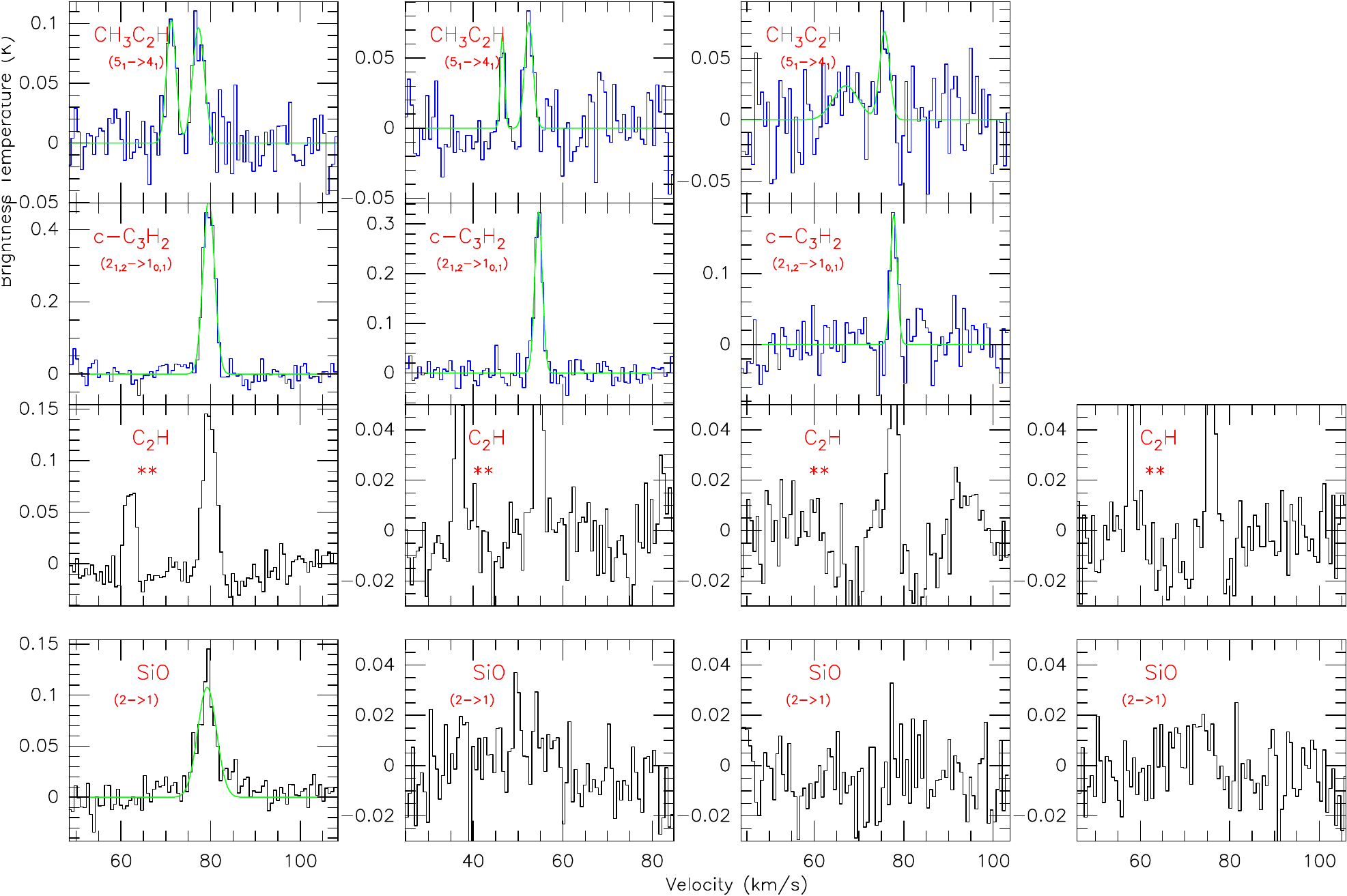}

Note: Here  a part of the multiplets of $\rm C_2H$ are shown; the complete set of  multiplets is in Figure \ref{irdc:hyperfit}\\
\end{center}
\caption{(continued)}
\end{figure*} 
\normalsize

\end{document}